# Persuasive Teachable Agent for Intergenerational Learning

## [A Book Draft]

**By**

**Su Fang Lim**

27 Jan 2016

# Abstract

Teachable agents are computer agents based on the pedagogical concept of learning-by-teaching. During the tutoring process, where students take on the role of the tutor to teach a computer agent tutee, learners have been observed to gain deeper understanding of the subject matter. Teachable agents are commonly used in the areas of science and mathematics learning where learners are able to learn complex concepts and deep reasoning by teaching the teachable agent through graphic representation such as concept maps.

Literature review on teachable agents as well as observations during field studies conducted by the researcher, have shown that many current teachable agents lack the interaction abilities required to keep learners engage in learning tasks. The result of this is learners deviating from the teaching process, and thus the learners are unable to benefit fully from learning with the teachable agent. The applications of teachable agents are restricted to the learning of academic subjects such as mathematics and science. The rapidly aging global population however has resulted in increased interest in expanding the applications of teachable agents, in order to utilize it as a tool to encourage intergenerational bonding as well as to benefit wider group of users.

In this thesis, we have proposed the Persuasive Teachable Agent (PTA), a teachable agent based on the theoretical framework of persuasion, computational and goal-oriented agent modelling. Consequently, in this thesis we ask how the PTA can be used to encourage intergenerational learning among learners from different age groups. We argue that the PTA, an autonomous agent, capable of encouraging attitude and behavioural change can offer a more meaningful and engaging learning experiences for learners from different age groups. Based on the findings from our research we argue that persuasive feedback actions generated by the PTA provide significant influence over learner’s decision to participate in intergenerational learning. The PTA plays a crucial role in the development of future persuasive technologies in artificial intelligent agents.

# Table of Contents

# Table of Figures

# 1 Introduction

Teachable agents are pedagogical agents that encourage students to learn through teaching the agents. The development of teachable agents is based on the learning-by-teaching pedagogy on the notion that students learn much better for themselves when they teach others. Teachable agents focus on life-like interaction between learners the tutoring process through verbal and visual simulation in order to motivate deeper learning, another area of research in developing agent architecture to support learning (Biswas & Leeawong, 2005). Teachable agents allow for learners to engage in tutoring of an artificial computer agent, and in the process learners are able to learn for themselves (Blair, Schwartz, Biswas, & Leelawong, 2006). Thus, with teachable agents, learners are able to benefit from the ability to acquire inquiry skills through asking questions, and obtain motivation through the cultivation of a sense of responsibility towards their peer tutee. Research in teachable agents has shown positive results in a learners ability to reflect on complex concepts while they teach the teachable agents, learners have also been observed to be receptive towards taking on the responsibility of teaching the teachable agent and therefore learn better in the process (Chase, Chin, Oppezzo, & Schwartz, 2009).

Teachable agents are applied in areas of education and learning in various domains, such as science and mathematics, have demonstrated positive effects in helping the students gain motivation, increasing self-regulated learning behaviours, and improving learning gains (Chase et al., 2009; Matsuda, Keiser, Raizada, Tu, et al., 2010; Pareto, Haake, Lindström, Sjödén, & Gulz, 2012).

However, in our literature review of teachable agent systems, presented in Chapter 2, it is shown that existing teachable agents lack social interaction causing frustrations among student users (Agneta Gulz, Haake, Silvervarg, Sjödén, & Veletsianos, 2011). As a result, they are not able to fully engage in learning with the teachable agents. Moreover, current teachable agents developed to date are targeted towards students in schools, and are applied in specific areas of education and learning domains, mainly in the areas of mathematics (Carlson, Keiser, Matsuda, Koedinger, & Rosé, 2012) and science learning (Biswas & Leeawong, 2005). Therefore, there is a limit in the areas in which the teachable agents can be applied.

The aim of this research is to improve on existing teachable agent by resolving the issues such as lack of social interaction between learners and teachable agents. As well as expanding the application of teachable agents to benefit a wider group of learners.

In order improve teachable agents in the way they interact with its users and to expand its application to a wider group of users, the Persuasive Teachable Agent (PTA) is proposed in this research. The PTA is a novel teachable agent developed based on persuasion theory. By introducing persuasion theory into existing teachable agent, the improved PTA will be able to social interact with different groups of users, such that users from different generations are able to come together in an environment to share and learn from the knowledge taught to the teachable agent. In this way, a wider range of population can benefit from the positive learning effects from teaching the teachable agent.

We have designed the PTA, a teachable agent with persuasion capabilities to support intergenerational learning. The PTA is a teachable agent that understands the learning motivation and abilities of individual users in the process of tutoring the agent. Improvements are also made to the PTA to encourage social interaction with its users. The result is the PTA that provides autonomous persuasive feedback, encouraging learners in teaching the agent.

The application of the PTA have also been extended to support intergenerational learning, which focus on building bonds and fostering interaction between the young and older generations. The learning by teaching pedagogy allows the inclusion of multiple tutors, and thus provides a way to motivate cross-generation learning, and thus helping them to lead a more fulfilling life by learning and sharing their knowledge with learners form other generations. This also helps build strong relational bonds between the different generations of learners there are transferable resources with many social benefits that can be passed on between generations of learners.

Another area in which this research, the PTA, can improve intergenerational learning is through the use of technology. There is a potential digital divide between the younger and the older users. Younger generations who are born in the digital world coined the "Digital Natives" are comfortable with the use of technology, their grandparents or parents the "Digital Migrants" may not be able feel uneasy about adopting the new technology (Prensky, 2001). We reason that through collaborative use, technology can be used to improve the quality time families member spend with their children.

To improve intergenerational relationships through technology and close the gap in the digital divide, teachable agents such as the PTA can be implemented into games that designed to support intergeneration bonding. The "Digital Migrants" who may be adults or the elderly will be motivated through game play, learning and teaching, to participate in intergenerational learning activities with the younger "Digital Natives"

The PTA is designed based on teachable agents and persuasion theory. Teachable agents have been used in educating young learners with well-established learning-by-teaching pedagogy. Moreover, traditional teachable agents have been widely studied in previous research, and demonstrating positive results in encouraging self-regulated learning, the Protégé Effect, and is already used for a number of both competitive and cooperative games (Chase et al., 2009; Matsuda et al., 2012; Pareto et al., 2012). Teachable agents such as the PTA can be used to motivate intergenerational learning among the different generations.

In this thesis, the researcher hopes to contribute to a future where the use of technology such as the PTA can help to encourage children to spend more time with their parents and grandparents, as well as encourage the younger generation to interact with the older cohort.

## 1.1 Research Objectives

As the population ages and with extended life expectancy, there is an urgency to develop systems to cater to the learning needs of the growing elderly population. One of the aims of this research is to support communities in the transmission of social capital, and continuing of lifelong learning across generations. The global demographic shift towards a growing elderly population has driven our research in the application of teachable agents to cater to aging learners.

Based on the literature review, there are limitations in current teachable agents systems such as the lack of social interaction between teachable agents and the inability for the teachable agents to assess the learners. At the same time, the literature review has also shown that teachable agents have been known to improve learning in young learning in science and mathematics learning domains. There is however, a research gap in the application of teachable agents in intergenerational learning scenarios. Hence, the following are the research objectives within the scope of this research.

1. **Overcome existing issues in teachable agents.**

The first and foremost objective of this research is to address the current issue in teachable agents, where the literature review on current teachable agents have revealed that teachable agents the lack capability to keep learners interested in teaching and learning with the agent. Therefore, the focus is to improve the teachable agent, so as keep the learners interested in learning and teaching the teachable agent.

2. **Design of PTA to infuse persuasion theory in teachable agents to encourage learning.**

The next objective following the identification of issues in teachable agent systems is the design of a teachable agent system that is able to understands user's learning ability and motivation, so that PTA will be able to influence learner's attitude and behaviour in learning and teaching. With this aim in mind, the theory of persuasion is examined as it provides a framework to provide the necessary influence in improving the learning experience of the learners when they interact with the agent.

3. **Implementation of PTA for intergenerational learning.**

Current teachable agent systems are applied in specific learning domains such as math and science. This research aims to expand the application of PTA to areas of informal learning, so as to benefit wider groups users. At the same time, the PTA provides a platform in which knowledge from different age groups can be shared in an intergenerational learning setting.

4. **Testing and evaluation of PTA for intergenerational learning**

The objective of this research also covers the testing and evaluation of the PTA in terms the effects of intergeneration learning to benefit the future works on teachable agent research.

## 1.2 Research Questions and Hypotheses

Given the background of this research, the following are the research questions and hypotheses that will be tested in this thesis.

**Question 1:** *What are intergenerational learning experiences during the interaction with the PTA?*

**Hypothesis 1:**

In this hypothesis, it is predicted that the participants from different age groups will benefit from intergenerational learning events such as taking on various roles such as teacher, student, leader and follower in the course of teaching the PTA. Some other possible intergenerational learning experiences include skills and knowledge areas such as computer skills, school related issues or academic knowledge.

**Question2:** *How do different age groups perceive the PTA during intergenerational learning?*

**Hypothesis 2:**

Teachable agents have shown to benefit learners by producing the *Protégé Effect*, where they are motivated to learning better for themselves by taking on the role of the teacher and teaching a novice learner. With the PTA, we predict that the difference in effort towards learning for different age groups will be greater for the older age group.

**Question 3:** *Does PTA improve learning across different age groups?*

**Hypothesis 3:**

Our hypothesis is that intergenerational learning increases the time spent teaching the PTA, and allow for greater interaction between participants from different age groups.

**Question 4:** *How can the findings from the study help in the future design of PTA for intergenerational learning?*

**Hypothesis 4:**

The findings from the study in this thesis will be useful to the future development of teachable agents, particularly in expanding the application of teachable agents to other aspects of learning such as informal learning and non-academic subjects, as well as towards the application of teachable agents for different age groups.

## 1.3 Organization of Report

In the first section of the report we have introduced the rationale for the research in PTA for intergenerational learning, which is first and foremost to overcome the existing issues in teachable agents and to expand its application from classroom learning topics such as science and mathematics domain, to informal learning towards building intergenerational learning relationship.

The following Chapter 2 of this paper introduces artificial intelligence and agent technology, followed by a literature review on teachable agent as a pedagogical agent introducing related work on teachable agents in supporting classroom learning that compares the key capabilities of different teachable agent systems, highlighting issues of different teachable agent systems in current research. The second part of the literature review examines the various persuasive theories that are used in social psychology to influence attitudes. And discuss on how persuasion can be used in computer interaction to generate positive attitudes in users of systems. The final part of the literature review will touch on intergenerational learning with case studies which introduce programmes initiated by various agencies to promote learning among the different generations to encourage bonding.

The Chapter 3 of the paper focuses on the current work in the initial design and development of PTA. A preliminary experiment was conducted to test the initial design of PTA and issues and challenges regarding our initial design will be discussed.

In Chapter 4 of this report, the actual implementation of the PTA in the latest improved version Virtual Singapura (VS) Saga, a 3D virtual learning platform will be introduced. A user study on the effectiveness PTA for intergenerational learning and the results and findings of the study will be discussed.

Lastly, Chapter 5 is the conclusion highlighting the contributions in the current work on PTA, and the report ends with the future work in implementation of PTA for intergenerational learning.

# 2 Literature Review

The first section of the literature review introduces the works on the pedagogical foundation of teachable agents. The development of teachable agents is traced from the root of computer aided learning with intelligent learning systems. Different teachable agent systems are also compared in terms of the domain in which they are applied and analysis of the capabilities and drawbacks of each teachable agent is included.

The literature review will also examine various persuasion theories in social psychology used to change individual opinion and attitudes on specific subject matter. The discussion section will describe how the persuasion theories aim to counter the drawbacks in current teachable agent systems.

Lastly, the literature review looks into the works on intergenerational learning, a relatively new domain which proposes that the teachable agent can be applied to benefit a wider user group of different ages.

## 2.1 Teachable Agents

Teachable agents are pedagogical agents through which students engage in learning through teaching the agents. The development of teachable agents is based on the learning-by-teaching pedagogy on the notion that students learn much better for themselves when they teach others.

There are several areas of research developments in teachable agents, the first area focuses on life-like interaction between learners and agent during the tutoring process through verbal and visual simulation in order to motivate deeper learning. Another area of research is the development of an agent architecture to support learning (Biswas & Leeawong, 2005). Teachable agents allow for student learners to engage in tutoring of an artificial computer agent, and in the process student learners are able to learn for themselves (Blair et al., 2006). Thus, with teachable agents, learners are able to benefit from the ability acquire inquiry skills in asking questions, and motivate learners by cultivating a sense of responsibility towards their peer tutee.

### 2.1.1 Development of Teachable Agents

Computers have a history of being used to support learning. Computer Aided Instruction Systems (CAI systems) which are used to scaffold learning has its roots traced to Pressey's 1925 instructional system, which was made up of a punch-board device and multiple-choice machine (Mann, 2008). The disadvantage of the traditional CAI system is that it requires learners to memorize learning materials repeatedly, which decreases student interest and motivation in learning (Arnold, 2000). This led to the emergence of Artificial Intelligence in Education (AIED) and the development of the intelligent learning environment (ILE) where Artificial Intelligence (AI) are used in CAI systems to support learning, which improving the interaction experience between learners and system learning (Blandford, 1994).

Within the ILE, there is the Intelligent Tutoring Systems (ITS) which tends to imitate the process of human to human tutoring by replacing human tutors with computer (Chan & Baskin, 1990). There are also advocates of alternate learning models (Self, 1990) whom have inspired researchers to investigate alternative methods of teaching such as mimicking the teaching process in traditional classroom within the ITS. For example, Chan and Baskin (1990) used the learning-by-teaching model to develop an ITS where multiple agents takes on the role of the teacher and learning companion (Chan & Baskin, 1990).

DENISE system (Nichols, 1993) is an ITS that was developed to support the learning of Economics using the learning-by-teaching pedagogy. The DENISE system aims to improve the cognitive and motivational aspects of learning by allowing learners to construct causal relations using question and answer dialogue sequences where student learners takes on the role of the tutor teaching the computer. However, the lack of natural interaction caused learner's frustration (Nichols, 1994).

To simulate a more naturalistic interaction between learners and system, a branch of research has focused on developing learning companion agents to increase learner's attention and keeping human learners motivated in learning (Chan & Baskin, 1990). EduAgent is one such learning environment that incorporates an artificial learning companion in learning Mathematics developed to accompany learners with different levels of competencies and to engage in meaningful conversation with learners (Hietala &

Niemirepo, 1998). With the development of systems such as EduAgent, there was an interest in studying the motivation for student to learn with learning companions. Thus, it is suggested that not only does the level of competency of the learning companion matters to how well the collaboration between human learner and learning companion, but that the personality and ability of individual learners should also affect the way the learning companion communicates with the human learner in keeping them motivated in learning (Hietala & Niemirepo, 1998; Uresti, 2000).

At MIT Lab, ”Learning Companion” was built to recognize a learner’s emotion and to provide intervention in order to guide students towards productive learning tasks (Cooper, Brna, & Martins, 2000; Kapoor, Mota, & Picard, 2001). Virtual agents that are based on learning-by-teaching pedagogy were integrated into Computer Aided Instruction System (CAI system) to help learners to learn more effectively (Fumiaki, Hiroshi, & Hidekazu, 2000).

Countering the drawbacks of monotonous learning in CAI systems and to maximize the potential benefits of learning-by-teaching pedagogy, teachable agents are developed to improve student learners learning experience when tutoring the embodiment of an artificial intelligence learning companion. Artificial Intelligence (AI) allows computers and machines to make human-liked decisions, solve problems accordingly, therefore enabling them to perceive reason and act humanly and rationally (Russell & Norvig, 2009). An intelligent agent is an autonomous computer agent that acts or operates in an environment with reasoning and learning based on observation through sensors to achieve certain goals.

With the integration of Artificial Intelligence (AI), teachable agents are developed to improve student learning by providing adaptable learning environment, tailored to an individual learner’s needs and capability. In a teachable agent learning environment, student learners are expected to explicitly teach the agent to carry out complicated learning tasks (Biswas, Schwartz, & Bransford, 2001).

Betty is a teachable agent developed by G. Biwas and colleagues at “AAA lab”, Stanford University and “Teachable Agent Group” at Vanderbilt University (Biswas & Leeawong, 2005; Blair et al., 2006) to be used in the domain of science topics. Learners teach the Betty teachable agent by modifying concept maps (Novak, 1990). By linking relations between concepts, visual representations of knowledge and reasoning mechanisms. The central idea of science topics is made clear to the learners with the use of concept maps. With the Betty teachable agent, learners are able to evaluate the progress of Betty's learning by observing the agent's explanation to a problem.

Within the Betty's Brain system, the "Protégé Effect" (Chase et al., 2009) was observed where learners take on responsibility for their teachable agent, and therefore, believing they were learning for their tutees, learners were willing to put significantly more effort towards learning and benefited from the enhanced Betty.

The Betty system learns using hidden Markov models (HMM) (Rabiner & Juang, 1986) generated from student behaviour activity logs to measure self-regulated learning based on learning behaviour in a learner's associated activities. According to established studies in cognitive science, self-regulatory and meta-cognitive skills are beneficial to students beyond classroom learning (Zimmerman, 1990), learners become better students for future learning, displaying autonomy and control even when they no longer have access to self-regulation learning environment (Leelawong & Biswas, 2008). The results conducted on the Betty's Brain system demonstrate that learners show better learning performance and meta-cognitive behaviours than those who only learn for themselves. With meta-cognitive feedback from the mentor agent while teaching the teachable agent, learners display advance focus and monitoring behaviours (Biswas, Jeong, Kinnebrew, Sulcer, & Roscoe, 2010).

SimStudent is another teachable agent developed at Carnegie University. The original goal of SimStudent was to facilitate authoring of cognitive tutors, a learning-by-tutored problem system which simulates classroom instructional environment in teaching mathematics. The system acts out the role of the teacher providing learners with information (Matsuda, Cohen, Sewall, Lacerda, & Koedinger, 2007). The first and foremost reason for setting up a teachable agent in SimStudent is that it would take up a lot time for students to wait for their turn to teach and learn from each other. A teachable agent would allow students to teach the agent and simulate the learning-by-teaching environment. Secondly, in actual person to person tutoring, the tutees may not benefit as much as the tutor (Matsuda, Keiser, Raizada, Tu, et al., 2010). SimStudent is incorporated into APLUS (Artificial Peer Learning environment), where learners are able to tutor SimStudent the artificial peer learner in a learning-by-teaching environment (Matsuda, Keiser, Raizada, Stylianides, et al., 2010).

SimStudent is a machine learning agent that has the potential to predict a human learner's performance. When learners are faced with new problems, SimStudent is able to observe the cognitive skills used by the learners, so as to extract a cognitive model, which it uses to explain solutions to learners. The concept of the teachable agent environment in SimStudent is to build a platform to support the learning-by-teaching pedagogy, while keeping the cost of setting up such an environment low to counter the drawbacks of conducting peer tutoring in classrooms.

Subsequently, the developments of pedagogy agents have been directed towards studies to explore the cognitive, affective and social aspects of interaction between learners and agents (Kim & Baylor, 2006). This led to the incorporation of personality traits in visual representation and conversational ability in teachable agents. Using EnALI (Enhancing Agent-Learner Interaction) (Veletsianos, Miller, & Doering, 2009) as a guiding framework and an existing teachable agent learning environment, Sjödén and colleagues developed a system which included teachable agent with added social and conversational abilities. The study on the system concluded the importance of agent context, learner knowledge and understanding the way learners communicate (Sjödén, Silvervarg, Veletsianos, Haake, & Gulz, 2011).

### 2.1.2 Progress in Teachable Agent Research

Improvements on Betty's Brain system has been made to explore the relationship between teachable agent feedback and learner's performance in the system. Agent prompts in Betty Brain are designed to encourage reflective interaction (Wu & Looi, 2008). It is shown that dialogue and action responsiveness has strong relations to the learning gain in the Betty's Brain system. Hence, cognitive and metacognitive models for effective self-regulated learning is compared to student learning strategy to derive feedback for students to overcome learning difficulties (James R. Segedy, John S. Kinnebrew, & Gautam Biswas, 2011).

On the other hand, SimStudent has included a natural language interaction between the learner and the teachable agent (Carlson et al., 2012). Text classification is used to train models in order to differentiate and categorize learner's feedback and enable interaction between learners and teachable agent.

Attempts have also been made to improve the believability of teachable agents by simulating emotions and modes of empathetic expression in immersive learning environments with the integration Affective Computing (Picard, 1997). Teachable agents that can express emotions as a result of a learner's interaction are often introduced as affective teachable agents. The purpose of an affective teachable agent is to evoke learner's interest in tutoring the agent.

Affective teachable agent was introduced in a Virtual Learning Environment (VLE), Virtual Singapura (VS) for Lower Secondary Science learning in Singapore designed with goal-oriented reasoning, categorizing a set of emotions based on the Ortony Clore Collins (OCC) model (Ortony, Clore, & Collins, 1990). OCC allows the teachable agent to generate appropriate emotional responses based on what the students have taught the agent (Ailiya, Shen, & Miao, 2011).

The Teachable Agent Arithmetic Game (TAAG) has been developed to help 5$^{th}$ grade students learn math in Sweden (Pareto et al., 2012). The competitive nature of teachable agents are explored as well, as the teachable agent was implemented in a game environment where students learn through collaboration or compete as a single player, or as two players, either with another student, a non-TA computer agent or other teachable agents (Sjödén, Tärning, Pareto, & Gulz, 2011).

Dynalearn is another project that consist a teachable agent that mimics a less capable learner that allows for learning-by-teaching pedagogy to be applied in an intelligent learning environment. The novelty of Dynalearn is in the qualitative reasoning model generated by the learner which can be compared to other learners as well as experts, thus allowing the system the automation to provide appropriate feedback and recommendations (Bredeweg et al., 2010).

More recently, the Tangible Activities for Geometry (TAG) is a teachable agent that was developed using digital augmented devices to help students learn geometry by solving problems in physical space, a different approach from other teachable agents in virtual environments (Muldner, Lozano, Girotto, Burleson, & Walker, 2013). In TAG, a robot named Quinn is constructed using the LEGO Mindstorm robot mounted with an iPod displaying facial expression. The students are to solve geometry problems by issuing command to Quinn where Quinn executes the command and check if the solution to the current problem is correct. Quinn is able to display emotions based on the TAG feedback to show the students how it feels through a gender neutral voice message. The current study on TAG focuses on the attributes impact of the interaction between Quinn and the students (Muldner, Girotto, Lozano, Burleson, & Walker, 2014).

The following Table 1 presents the comparison of the existing capabilities and features of the different teachable agent systems. The majority of teachable agents are able to conduct conversation with student learners. All of the teachable agents embody a learning companion. Artificial intelligence such as machine learning, decision making and affective computing are used in systems such as the Betty's Brain, SimStudent, Virtual Singapura and TAG. SimStudent and TAAG are teachable agents systems that are used in mathematics learning using competition and cooperation to engage student learning. The teachable agent in Virtual Singapura is the only teachable agent that is developed in a 3D virtual environment.

Table 1 Overview of existing capabilities and features of teachable agent systems

| **Teachable Agent** | **Artificial Intelligence** | | | | **HCI (Conversational Ability)** | **3D Virtual Environment** | **Competition / Cooperation** |
|---|---|---|---|---|---|---|---|
| | Machine Learning | Decision Making | Affective Computing | Embodiment | | | |
| Betty's Brain | | ✓ | | ✓ | ✓ | | |
| SimStudent | ✓ | | | ✓ | ✓ | | ✓ |
| Virtual Singapura | | | ✓ | ✓ | | ✓ | |
| TAAG | | | | | | | ✓ |
| DynaLearn | | | | ✓ | ✓ | | |
| TAG | | | ✓ | ✓ | | | |

### 2.1.3 Teachable Agents Related Issues

Teachable agents are based on the concept that students learn in the process of teaching others, the pedagogical concept of learning-by-teaching, aid learners to achieve deeper learning when they take on the role of the tutor (Bargh & Schul, 1980; Gartner, 1971). The key to learning-by-teaching is that during the tutoring process student learners are able to gain a more complete knowledge by refining their concepts through explaining and reorganizing their own knowledge (Coleman, Brown, & Rivkin, 1997).

During the tutoring process, learners are more prepared to teach their peer tutee as compared to when they are learning for themselves (Annis, 1983). Tutors also benefit from explanation and questioning interactions with tutees, thus, encouraging student tutors to reflect upon the problems to stimulate deeper reasoning (Craig, Sullins, Witherspoon, & Gholson, 2006). Developments in research on learning-by-teaching systems are based on the notion in education and cognitive science literature that learners benefited from tutoring as they learn as much as the tutee (Chi, Siler, Jeong, Yamauchi, & Hausmann, 2001). Besides improving understanding, there are considerable affective and cognitive gains during the tutoring process (Lublin, 1990; McNall, 1975). Low performing students are also motivated to perform better with teachable agents (Sjödén, Tärning, et al., 2011). As preparation to teach requires tutors to organize their knowledge, student tutors are observed to learn a great deal through assessment as well as reflection during the tutoring process (Biswas & Leeawong, 2005).

Despite the advantages of the amount of knowledge gained during tutoring, there are several drawbacks with learning with teachable agents. Firstly, novice student tutors are not able to gauge their own understanding of the topic that they are teaching. This may cause them to commit errors or make mistakes when judge other sources of information (Azevedo & Cromley, 2004). Secondly, learning-by-teaching requires the student learners to be self-monitored and self-regulated. Student learners who have little experience or low prior knowledge in teaching may not be able to handle tutee enquiries (Roscoe, Wagster, & Biswas, 2008). Moreover, student tutors also does not fully benefit from learning-by teaching if the tutee is too passive and does not provide adequate responses from the tutoring (Graesser & Person, 1994). In order for student tutors to fully benefit from tutoring the teachable agent, they have to be motivated to tutor the teachable agent (Uresti, 2000).

There are also several challenges highlighted by Gulz et al. (2011), with regards to the design of pedagogical agents in intelligent tutoring systems including satisfying "learners expectation of agent's knowledge and social profile", "dealing with learner's engagement in off-task conversation" and "managing potential abuse of the agent" (Agneta Gulz et al., 2011), especially with open-ended learning environment as in Betty's Brain (Segedy, Kinnebrew, & Biswas, 2012) and conversational SimStudent (Carlson et al., 2012). Therefore, there are suggestions to design specific "off-task" interaction that in particular help to build rapport and trust, provide breaks in between learning to improve engagement (A. Gulz et al., 2010).

The importance of understanding learner's ability in order to provide appropriate feedback is also echoed in earlier studies on SimStudent, which showed that "prior knowledge has a strong influence on the tutor learning"(Matsuda, Keiser, Raizada, Stylianides, et al., 2010). Segedy (2011) suggests the use of a combination of "metrics", such as "responsiveness" and "performance" as a benchmark to determine whether learners need more support or feedback (J. R. Segedy, John S. Kinnebrew, & Gautam Biswas, 2011) with the use of cognitive and meta-cognitive model derived from Hidden Markov model from learner's behaviour to understand student learning strategies, thus applying suitable level scaffolding and feedback.

The Teachable Agent Arithmetic Game (TAAG) was used to evaluate the student's performance, attitude and self-efficiency. The results showed positive improvements in conceptual arithmetic understanding and self-efficiency however does not support positive attitudes towards math (Pareto, Arvemo, Dahl, Haake, & Gulz, 2011).

More recent developments indicate the interest in understanding the benefits of social interaction of learners through tutoring and interacting with teachable agents (Ogan et al., 2012), as well as the motivation factors for learning-by-teaching in teachable agents (Matsuda et al., 2012).

The development of teachable agents demonstrates the benefits for applying learning-by-teaching pedagogy in intelligent tutoring systems. The significant portion of work on teachable agents are dedicated towards simulating peer-liked interactions between learners and teachable agent, providing support and feedback in open-ended teachable agent learning environment as well as encouraging competition and collaboration among student learners and teachable agents. There is also a need to look into ways to alter student's attitudes towards the subject domain, and expand the application of teachable agents.

Table 2 shows the comparison between the different teachable agent systems in terms of their innovations and application domains. The majority of teachable agents are applied in mathematics and science learning, other applications also include economics in the DENISE and environmental science in the DynaLearn teachable agent system. As can be seen from the literature review on teachable agents, none of the systems have been applied to intergenerational learning.

Table 2 Issues, Innovation and application domain of teachable agents systems

| Systems | Issues | Innovation | Application Domain |
|---|---|---|---|
| DENISE | Lack of natural interaction caused learner's frustration | Learning-by-teaching pedagogy | Economics |
| EduAgent | Learner's personality and ability, affects their preference for the strength of their teachable agent learning companion. | Artificial learning companion | Mathematics |
| Betty's Brain | Understanding learner's expectation of agent knowledge and ability. Dealing with open-ended conversation, off-task engagement. | Artificial Intelligence | Science |
| SimStudent | | Machine Learning | Mathematics |
| Virtual Singapura | Learners do not fully engage in learning with TA. | Affective Computing | Lower Secondary Science |
| TAAG | Students attitudes towards math does not improve after using TA, compared to group that does learning without TA. | Collaborative and Competitive Learning | Mathematics |
| DynaLearn | Lack of clarity and relevance of | | Environmental Science |
| | recommendations, insufficient support in learning and modelling processes. | | |
| TAG | More understanding on capitalizing social and affective elements in TA. | Social Robot | Mathematics |

## 2.2 From Persuasion Theories to Persuasive Technologies

Persuasion is a fundamental interaction between human beings, in the attempt to influence others by modifying attitudes, beliefs, intentions, motivations or behaviours (Gass & Seither, 2010) and widely applied in areas such as rhetoric, social psychology, communications, advertising and public relations. Simon (2001), defined persuasion as the "human communication designed to influence the autonomous judgment of others" (Simons, 2001).

In section 2.2 of the literature review, two closely related theories of persuasion that contribute to the development of the PTA, namely, the Heuristics-Systems of Model and the Elaboration Likelihood Model (ELM) of persuasion will be introduced. Following that, the literature review examines the development of persuasive technology, looking at the various persuasive frameworks and models that have emerge from persuasion theories and how they have been used to address issues in user interactions in systems design. At the end of this section of the literature review, various applications of persuasive technologies in healthcare and education will also be reviewed.

### 2.2.1 Heuristic-Systematic Model

Heuristic and systematic views of persuasion regard that the person receiving the message is concerned with assessing the validity of the message's overall conclusion (Shelly Chaiken, 1980).

The Heuristic-Systematic model (HSM) involves heuristic processing and systematic processing. During heuristics processing, the message recipient uses little effort in processing and assessing the validity of the message. Instead, judgments are formed based on existing learned memory. Heuristic processing is useful in presenting messages that require little cognitive effort on the message receiver. On the other hand, systematic processing requires much more cognitive effort in performing the task and requires the recipients understand and evaluate the arguments in the persuasive message in order to accept the message's conclusion (Chen, Duckworth, & Chaiken, 1999).

The experiment conducted by Chaiken in 1980 on source and message concluded that the content of the message out-weight the likeability of the source. In terms of persistence change, content mediated opinion change lasted longer than source-mediated persuasion (Shelly  Chaiken, 1980). HSM is a dual process model that explains the way persuasive messages are perceived.

### 2.2.2 Elaboration Likelihood Model of Persuasion

The Elaboration Likelihood Model (ELM) of persuasion is another example of dualistic model developed by Petty, Cacioppo and colleagues (1986). Both ELM and HSM offer broad similarity in the notion that persuasion varying in the degree of issue related thought processes. ELM differs from other models as there is added attention towards retention of change in attitudes through the use of peripheral and the central persuasion routes. The central route of persuasion requires the recipient to take on in-depth consideration of the information presented, while the peripheral route requires little or negligible scrutiny (Petty & Cacioppo, 1986).

According to the ELM, attitudes formed under the central route are more persistent compared to the attitudes formed under the peripheral route, and behaviours are predictive and resistant to change until clear counter information is presented to alter the merits of the object. The ELM outlines a general theory of attitude change and the underlying framework for persuasive communications. The following Fig 1 is a diagram that shows the central and peripheral routes of persuasion in which attitude change can be achieved, according to the ELM theory of persuasion.

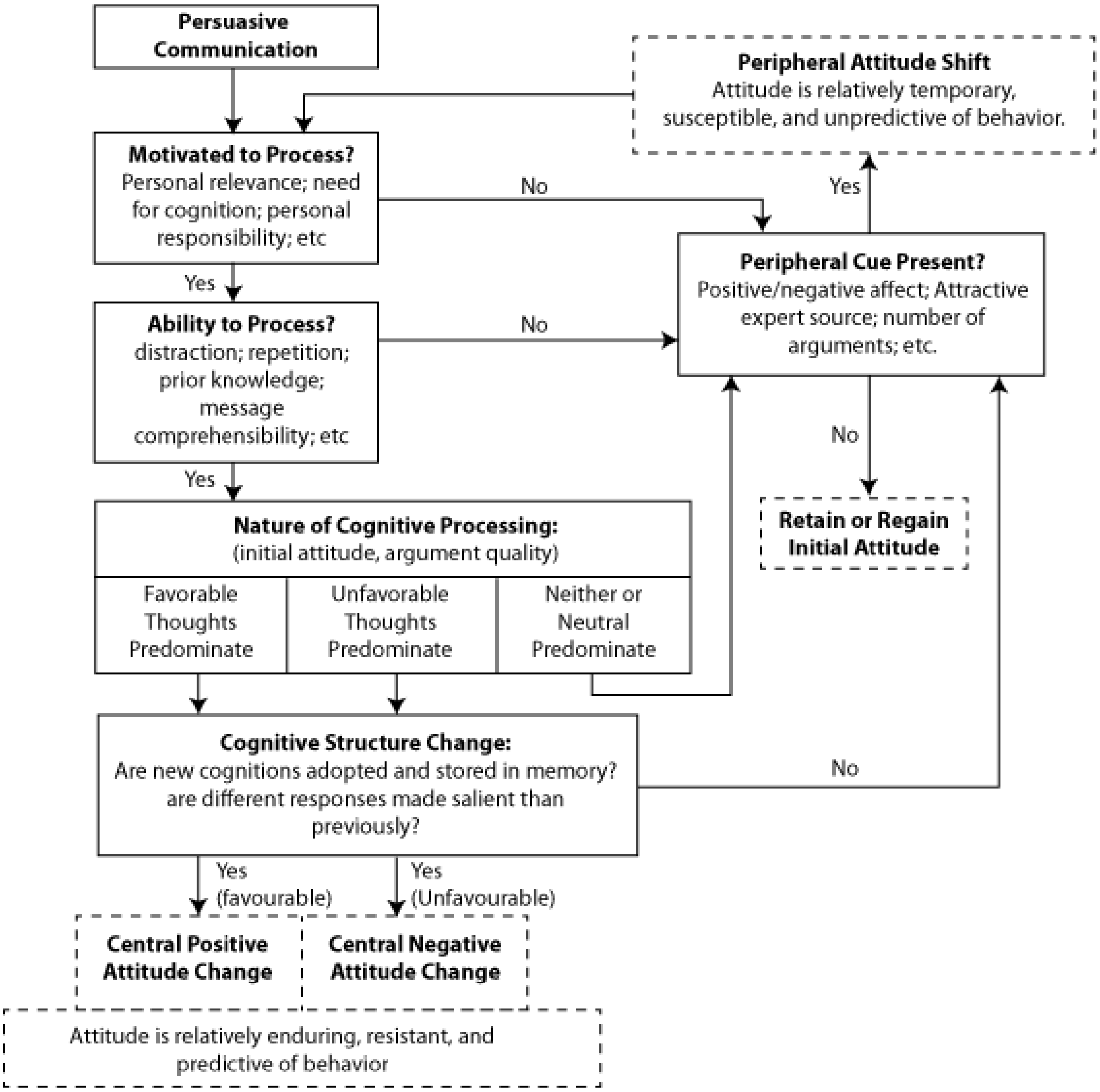

Figure 1 Central and Peripheral routes to Persuasion (Petty & Cacioppo, 1986)

***Elaboration***

The ELM recognizes that the characteristic of the recipient of a message varies in the degree of engagement in topic-related thinking in different situations. The term "elaboration" in ELM refers to the process of the degree of "issue – relevant thinking" required by the recipient of the persuasive message.

According to ELM, as the likelihood of elaboration increases, the quality of issue-relevant arguments becomes the determinant of persuasion. On the other hand, as the likelihood of elaboration decreases peripheral cues become more important.

Various factors have been identified to influence degree of elaboration. These factors that influence elaboration falls into two categories, one involves the motivation of the receiver and the other is the ability of the receiver to engage in elaboration. When the recipient is high in motivation and ability, able to process the content of the message, the central route is likely prevail, while peripheral route occurs when motivation and ability is low (Schwarz, Bless, & Bohner, 1991). Elaboration occurs when both ability and motivation is present (O'Keefe, 2002).

The following are variables that are can increase or decrease message elaboration under the central processing route, based on the amount of ability and motivation of the message recipient (Petty, Cacioppo, Sedikides, & Strathman, 1988).

***Factors affecting Elaboration Motivation***

PERSONAL RELEVANCE As messages become more personally relevant to the receiver, the receiver becomes more likely to engage in issue relevant thought. When personal involvement is high, individuals tend to process message content rationally. When personal involvement is low, the content of the message becomes less important, thus seeking the peripheral route rather than central route in message processing (Petty & Cacioppo, 1981, 1984; Petty, Cacioppo, & Schumann, 1983; Petty et al., 1988).

In ELM, research reports are labelled according to a receivers' level of "involvement" in topic related issues. In persuasion research, "involvement" covers a wide variety of relationship that the message recipient has with the topic, including judgment of importance and commitment to the issue. According to ELM, involvement is specific to the induction of personal relevance of the topic (O'Keefe, 2002).

Johnson and Eagly (1989), defined "involvement as motivational state induced by an association between an activated attitude and self-concept", this implies that the level of involvement affects the change in attitude. The message recipient is dependent on his or her self-concept that generates the sense of involvement. Johnson and Eagly (1989) distinguished three aspects of involvement, established by self-concept. Value-relevant involvement is activated by one's enduring value, outcome-relevant involvement depends on one's ability to obtain desirable outcome while establishing impression-relevant involvement depends on the impression one makes on others.

In their study involving value-relevant involvements and impression-relevant, subjects with high involvement are less persuaded compared to those low in involvement. With outcome-relevant involvement, subjects who are high-involvement are more persuaded than low involvement individuals (Johnson & Eagly, 1989).

Persuasion researches on involvement conditions have shown that under high involvement, the quality of arguments have greater impact compared to low involvement. Under conditions with low involvement, peripheral cues such as source expertise, and attractiveness of message source prevails, (Petty & Cacioppo, 1984; Petty, Cacioppo, & Goldman, 1981). Individuals put in much more cognitive effort to elaborate on issues when they are highly involved to take on a central route in information processing. Whereas under conditions of low involvement, they are more affected by simple heuristic cues such as acceptance and rejections based on source and message characteristics, choosing the peripheral route in processing information (S. Chaiken & Maheswaran, 1994; Petty et al., 1983).

PERSONAL RESPONSIBILITY Personal responsibility of an individual determines the amount of effort one is willing to put in evaluating the persuasive message. Petty, Harkins, and Williams (1980), showed that the quality of argument becomes less important with decreasing personal responsibility. With shared responsibility among group evaluators, weak messages are more favourable, whereas individual evaluators preferred strong messages. A mixed message containing both strong and weak arguments does not have effect on either shared or group evaluation (Petty, Harkins, & Williams, 1980).

NEED FOR COGNITION The receiver's need for cognition can influence the motivation for elaboration and indicate one's inclination towards engaging in cognitive activities. As Cacioppo and Petty (1982) describes, some people enjoy thought engaging activities while others do not (Cacioppo & Petty, 1982).

Findings by Cacioppo et al. (1983), shows that individuals with high need for cognition are more likely to discriminate between strong and weak arguments when evaluating the merits of supportive arguments. They are also more likely to be influenced by the quality of the message presented when forming impressions of the communicator, recall more message arguments regardless of cogency, and they account more cognitive effort compared to those with low need for cognition, (Cacioppo, Petty, Kao, & Rodriguez, 1986; Cacioppo, Petty, & Morris, 1983).

***Factors affecting Elaboration Ability***

MESSAGE REPETITION Message repetition is a characteristic that has been shown to influence the ability for recipients to elaborate on arguments. Cacioppo and Petty, (1989) have shown that repeated messages allow for more exposure and thus the recipients have more time to elaborate on messages. Compared to one exposure, three exposures to messages increased effectiveness of persuasion based on strong arguments but showed a counter effect on weak arguments (Cacioppo & Petty, 1989).

DISTRACTION Interruptions distract and have been shown to intervene with systematic message processing. The quality of argument becomes less important as there are fewer chances for issue relevant thinking. When the message recipient is highly motivated and has the ability to process the message, distraction becomes a bothersome disrupter to the ability to process information (Petty & Cacioppo, 1986). Recipients with good mood are less affected by distraction, suggesting they did not engage in message elaboration, compared to those who are in bad mood (Bless, Bohner, Schwarz, & Strack, 1990).

PRIOR KNOWLEDGE An individual ability to elaborate on a message is influenced by their prior knowledge on the topic suggesting that recipients hold salient self-schema that led them to act subjectively when evaluating persuasive messages. This in turn causes them to engage in increased issue relevant thinking to counter strong conflicting arguments to their beliefs (Cacioppo, Petty, & Sidera, 1982). Wood (1982), suggested similar notion that subjects who have less prior knowledge are more susceptible to persuasion as it reduces chances of counter-argument on the message (Wood, 1982).

## 2.3 Persuasive Technology

Technology that is designed to influence attitude and behaviour change are known as Persuasive Technology. Torning and Oinas-Kukkonen (2009) identified 4 areas of research in computer-based fields where persuasive systems and design can be observed, namely, human-computer interaction, computer-mediated communication, information systems and affective computing (Torning & Oinas-Kukkonen, 2009).

### 2.3.1 Persuasive Frameworks and Models

Information systems and computerized software that are designed to strengthen, modify attitudes and behaviours or both without coercion or deception are known as persuasive systems (H. Oinas-Kukkonen & M. Harjumaa, 2008).

The Persuasive System Design (PSD) model is a framework created to provide a guideline for persuasive design and software requirements (Harri Oinas-Kukkonen & Marja Harjumaa, 2008). The PSD model is made up of a persuasion context following a three phrase process, which consists of the *intent*, *event* and *strategy*.

According to the PSD model (Oinas-Kukkonen & Harjumaa, 2009), the *intent* comprises of the *persuader* and the *change type*. There are three types of *persuader*, the endogenous type, who is the creator of the technology, and the exogenous, those who distribute the technology, and lastly the autogenous type, who are the users of the technology. The *change type* of the *intent* defines the desired behaviour change. The event phrase defines the *use context*, *user context* and *technology context*. The *use* and *user context* help the system designer to understand the specific user issues and needs, while the *technological context* helps to evaluate the strength and weakness of the technology used of the system (Wiafe, Alhammad, Nakata, & Gulliver, 2012).

Lastly, the *strategy* phrase consists of the *message* and the *route*. The message considers the form to transform the intended change in behaviour. While the *route* refers to the selection of indirect or direct route or both approach to persuasion (Torning & Oinas-Kukkonen, 2009). In addition the PSD model examines the non-functional design principles in persuasive system design such as *primary task support*, *dialogue support*, *system credibility support* and *social support*.

Another key dimension to human computer persuasion is related to how users interpret persuasive messages from computers. This can be traced to social psychology research in persuasion. Social psychologist, Fishbein and Ajzen proposed the Theory of Reasoned Action (TRA) suggesting that the actions that a person intended to take in a situation can be predicted by their *attitude* and *subjective norm* towards the intended behaviour (Fishbein & Ajzen, 1975). TRA has been applied to the Technology Acceptance Model (TAM) in information system design, to anticipate user's acceptance to technology. Elaboration Likelihood Model (ELM) proposed by Petty, Cacioppo and colleagues, is a social psychology theory that explains how individuals perceive persuasive messages. The ELM is a dualistic model, proposing that attitude change takes on either a peripheral route or a central route of thought process. Under the peripheral route in the ELM, there is little thinking to process the information presented whereas the central route there is in-depth consideration involved (Petty & Cacioppo, 1986).

Together with the Persuasive System Design (PSD) framework used for developing and evaluating persuasive systems (Torning & Oinas-Kukkonen, 2009) and ELM model, the 3-Dimensional Relationship between Attitude and Behaviour (3-RAB) model is developed to analyse the *change type*, *user* and *use context* and *route* of persuasion to attitude and behavioural change (Wiafe et al., 2012).

The majority of research in persuasive technologies and persuasion in systems design are towards the applications of persuasive frameworks and the analysis and design persuasion in systems.

### 2.3.2 The Application of Persuasive Technology in Health Care and Education

The applications of persuasive technology can be seen far and wide. Persuasive technology has been applied include advertising, management, governance and public health. In view of the scope of this thesis, we have chosen to examine the areas of health and learning.

*Applications in Health Care*

In the area of health care, persuasive technologies are often used to encourage healthy lifestyle changes such as changing one's behaviour in making healthier food choices, and exercising more regularly (Pinzon & Iyengar, 2012). One such application involves "just-in-time" feedback to encourage healthier choices. This is synonymous with the concept of Kairos in persuasion, where information is provided at the right moment to influence one's decision. One example of the use of persuasive technology is the use of sensors in our environment to provide context-aware information. Smart devices are used to provide relevant information to make health related information. The MIT PlaceLab is a research facility that studies the possibilities of technologies to improve the wellbeing of residents in a smart home environment, where sensors and context-aware software are used to provide persuasive information to influence behaviours in a real life scenario (Logan & Healey, 2006).

Another example area in which persuasive technology can be used in health care is the smart pill box which reminds the elderly to take their medications. Such devices can be used to understand the reason why the elderly forget to take their medication. By analysing the behavioural and situational context information, appropriate persuasive information can be provided to remind them to take their medications (Lundell et al., 2006).

Wearable computing is another area in which persuasive technology can be applied. Such devices help users monitor their lifestyle and motivate them to adopt healthy lifestyle changes. Commercial devices include health monitors, pedometers and activity trackers. These wearable devices help users can keep a log on their food intake, to track their diet and workout routines (Ananthanarayan & Siek, 2012). These devices allow users to track and share their activity logs and share them with friends on social media to motivate them in changing their behaviours through healthy competition.

*Applications in Education and Learning*

The education field is an area that is fast changing. This requires educators to adapt to a generation of young learners who are comfortable with learning with technology. Digital narrative software with persuasion infused has been developed to motivate children to read and write, using persuasion principles such as similarity, tailoring and credibility to motivate literary learning (Lucero, Zuloaga, Mota, & Muñoz, 2006).

Persuasive strategies are also adapted in persuasive dialogue systems such as embodied conversational agents (ECA) designed for learning. An example of an ECA that provides supportive and expressive dialogue is the learning companion in the "How Was Your Day?" application, where the ECA provides affective persuasion that aims to influence the user's attitude positively (Cavazza et al., 2010).

Efforts such as the EuroPLOT project (persuasive learning objects and technologies) (Gram-Hansen, Schärfe, & Dinesen, 2012; Herber, 2011) aim to adapt persuasive design as a component in the pedagogical framework. Persuasive design is also considered in the development of the content design tools and design patterns in the EuroPLOT project to encourage active learning engagement.

The application of persuasive technology in educational setting can be demonstrated in the HANDS (Helping Autism-diagnosed teenagers Navigating and Developing Socially) project (Mintz & Aagaard, 2012). The HANDS project aims to assist autistic children to develop social skills using smartphone application with persuasive design. Drawing upon the concept of Kairos, namely delivering the right message at the correct timing and place, the mobile software application is user-centred design approach expressing the needs of teachers, children and parents. The user-centred design approach considers the user's limitations, needs and wants at every stage of the software development process (Abras, Maloney-Krichmar, & Preece, 2004). Teachers are able to develop intervention sequences using a flexible web-based toolkit to support each individual child in their learning.

The application of persuasive technology in educational and learning setting shows potential. With the integration of persuasive design principles drawn from multi-disciplinary research, there are possibilities to extend its applications.

## 2.4 Intergenerational Learning

According to the Encyclopaedia of the Sciences of Learning, intergenerational learning is defined as "the learning that happens naturally in the home between parents and children and includes the wider family and community," ("Intergenerational Learning," 2012).

The concept of intergenerational learning stems from the idea exchange of knowledge from the older generation to the young and vice versa. With the growing aging population, the idea of "productive aging" where the elderly leading meaningful and active life were promoted at the end of the 20th century (Brabazon & Disch, 1997; M. Kaplan, Kusano, Tsuji, & Hisamichi, 1998; Newman et al., 1997), where the elderly are highlighted as valuable resources instead of seen as a burden to the society for the growing cost of health care system and public pension (Gonyea, 1999).

## 2.5 Intergenerational Learning Programmes (ILP)

In 1997, Newman proposed the use of intergenerational programmes as an embodiment of intergenerational concept that "involved planned ongoing interactions between non-biologically linked children, youth and older adults" (Newman et al., 1997). Kaplan (1998), however, defined intergeneration initiatives as "activities, events and ongoing programs designed to increase cooperation, interaction or exchange between people between sixty years of age and older and twenty years of age and younger (M. Kaplan et al., 1998)."

Recognizing the changes in society that are occurring worldwide, the UNESCO Institute of Education published a monograph, highlighting a series of initiatives in various countries in intergenerational work that drives social and policy changes. The intergenerational learning programmes highlighted and identified various areas of impact in public policies and recommendations for development and implementation strategies.

Countries such as Germany, Japan, the Netherlands, the USA, and the UK have developed the following three most frequently described models of intergenerational programmes. These frequently described models are, firstly, involving older adults serving children. Secondly, children and youths serving older adults or the old, and lastly, the young serving the communities (Boström et al., 1999).

In 1999, the UNESCO Institute of Education (UIE) held a meeting in Dortmund among countries where researchers from China, Cuba, Germany, Japan, the Netherlands, Palestine, South Africa, Sweden, the United Kingdom and the United States came upon a definition for the concept of intergenerational programmes (Boström, 2003), "Intergenerational programmes are vehicles for purposeful and ongoing exchange of resources and learning among older and younger generations" (Boström, Hatton-Yeo, & Ohsako, 2000).

Intergenerational learning programmes are channels in which purposeful exchange of resources and learning among the older and younger generations can be achieved (Boström et al., 2000). The goal of ILP is to create opportunities, where interactions between the young and old promotes social growth and learning in informal settings (Newman & Hatton-Yeo, 2008). Informal learning occurs when people learn from one another, which includes values, attitudes knowledge and skills (Tuijnman & Boström, 2002).

M.S. Kaplan (2002) describes intergenerational programmes as those used in schools to which benefit student learning and the communities involved, particularly the elderly participants, where young people are given the opportunity to "interact, support, and provide care for one another" (M. S. Kaplan, 2002)

Boström (2003) also additionally observes the potentially beneficial social interaction between teachers, grandparents and the child which resulted from the increase amount of time the child spends in new social structures outside of the family such as day-care centre and schools (Boström, 2003).

Cross generational relationships are rewarding. However, due to the changes in modern lifestyles there is less time for bonding. This could be due to work or school commitments resulting in less time for the child to interact with their parents, or children moving away from their parents. As a result, the older generation has a lesser chance to spend time with their children or grandchildren.

Table 3 shows a series of programmes initiated by the European Commission aimed at examining the challenges of aging and looking at areas in which the elderly contribute to the society ("ICT For Seniors' And Intergenerational Learning - Projects funded through the Lifelong Learning Programme from 2008 to 2011," 2012).

Table 3 European Commission Intergenerational Programmes

| **European Commission Intergenerational Programmes 2008-2011** |
|---|
| A.L.I.CE - Adults Learning for Intergenerational Creative Experiences |
| BASIC-LIFE: Basic Web 2.0 Skills by Learning in Family Environment |
| Community Media Applications and Participation |
| Connect in Laterlife – Social Networking for Senior Citizens |
| CROSSTALK: Moving stories from across borders, cultures and generations |
| CT4P- CyberTraining -4-Parents |
| Detales: Digital Education Through Adult Learners' EU-Enlargement Stories |
| eScouts – Intergenerational Learning Circle for Community Service |
| Intergenerational ICT Skills |
| LEAGE – Learning Games for older Europeans |
| LIKE – Learning through Innovative management concepts to ensure transfer of Knowledge of Elderly people |
| Mix@ges – Intergenerational Bonding via Creative New Media |
| My Story – creating an ICT-based inter-generational learning environment |
| OWLE50+- Older Women Learning and Enterprise |
| PEER – Sapere aude! Dare to be wise! |
| SETIP – Senior Education and Training Internet Platform |
| SIGOLD – Turning the silver challenge into the golden opportunity |
| Silver: Stimulating ICT Learning for Active EU Elder |
| TKV – The Knowledge Volunteers |
| VinTAge – Valorisation of Innovative Technologies for Aging in Europe |
| W@ve 2.0 – Meeting Social Needs of Senior Citizens through Web 2.0 technologies |

### 2.5.1 Intergenerational Learning Programmes in Singapore

According to the Committee on Aging Report on Ageing Population in 2006, there are existing programmes organized by the community. This report recommend that more can be done to provide learning as well as greater intergenerational bonding initiatives opportunities to the seniors ("Committee on Ageing Issues Report on the Ageing Population," 2006).

More recently, Singapore seeks to provide more opportunities for the seniors to engage in lifelong learning so that they stay relevant and keep their mental awareness. One initiative is the Intergenerational Learning Programme (ILP) by the Council for the Third Age (C3A) setup in 2011, which provides an opportunity for the elderly to engage in lifelong learning where youths are matched with senior “teachers” in a classroom-based learning programme to encourage intergeneration bonds ("Intergenerational Learning Programme (ILP)," 2011). Family Services Central (www.family-central.sg) is a service appointed by the Council for the Third Age (C3A) as the main organizer in conducting these ILPs in Singapore. Photography, wellness, web design IT, floral arrangement and photo editing courses such as “Basic Photoshop for Photographers” are organized regularly for seniors above 50 years of age. These courses are initiated by school clubs, student groups as service learning initiatives. The aim is to provide an opportunity for youths to learn important life lessons and values through interaction with the elderly, while the seniors gain skills and knowledge from their younger counterparts ("Intergenerational Learning Programme ", 2011).

### 2.5.2 Intergenerational Games – Research

Intergenerational learning can be achieved through playas as the younger and older generations can benefit from the exchanges which take place during intergenerational play (Davis, Larkin, & Graves, 2002). Research has been done in the area of interactive games which aim at keeping families members engaged through intergenerational play. Various intergenerational games have been developed to facilitate this research. Table 4 shows the list of intergenerational games that are reviewed.

Table 4 List of Intergenerational Games in Research

| Year | Game Title | Innovation | Description |
|---|---|---|---|
| 2006 | Curball | Encouraging play activity among the elderly, and child | Prototype collaborative game based on bowling game which uses tangible devices, sensors and augmented reality components, the game can be played between an older person and a child. |
| 2008 | Age invaders (AI) | Adaptable game parameters to suit simultaneous gaming of elderly and young, thus compensating for elderly disadvantages | An interactive intergenerational social – physical game that the elder can play with children in physical space, in real-time or remotely through the internet. |
| 2010 | Family Quest | Study of the nature of intergenerational play using active theory as a conceptual and analytical framework | A multi-user 3D educational game where parents and children ages 9-13 come to together to play Quest Atlantis. |
| 2012 | Xtreme Gardener | The study showed challenges of paired social interaction between younger and older players | This game requires cooperative game play while players to tend to a virtual garden with silhouette interaction. |

Curball is one of the applications that allows for interactive play between the elderly and their grandchildren. The concept of the game is based on the Curling and Bowling game. The players in the Curball game are required to “throw” a tangible ball embedded with sensors in order to manoeuvre a virtual ball through augmented reality to the end of the board. The Curball game encourages communication between the older person and the other player e.g. the grandchild (Kern, Stringer, Fitzpatrick, & Schmidt, 2006).

Age invader (AI) is a multi-player game where players interact in physical space or participate in the game over the internet.  This is a physical game, with additional games that requires puzzle solving. AI is a game that stimulates cognitive, physical health, encourage collaboration and bonding among the seniors and young (Khoo, Cheok, Nguyen, & Pan, 2008).

Family Quest is study based on a multiplayer 3D game. The game reveals the dynamics of intergenerational play between Players ages 9-13 and their parents (Siyahhan, Barab, &

Downton, 2010). Family Quest is analysed using Active Theory. The components of activity are broken down into items such as tools, goals, norms and rules are constantly evolving in relation as a result of the activity system (Cole, 1996; Greenberg, 2001).

Xtreme Gardener is a game which requires the cooperation between a pair of players to nurture a set of virtual garden plates using silhouette interaction as modes of control, the study highlights the challenge of designing intergenerational bonding games (Rice, Yau, Ong, Wan, & Ng, 2012).

## 2.6 Case Studies

In this section, two case studies on educational gaming platforms that the researcher was involved in which have the potential to be developed for intergenerational learning will be reviewed.

The first case study is the Family Pet is an educational game which allow for users to engage in communication with family members. The second case study is the Virtual Singapura (VS) learning environment, which is developed as an educational game platform for science learning.

### 2.6.1 Case Study: Family Pet

Family Pet is an educational game designed for the Apple iOS, with the goal of promoting family bonding (Anthony, 2013). In the Family Pet, there is a virtual family pet that simulates real-life pet. Family members take on the responsibility to ensure the well-being of a virtual pet. In the process, they can participate in educational and casual activities together and thus foster strong bonds.

During the Family Pet gameplay, the young players take on the main role of taking care of the family pet and decisions are made on the way the virtual pet is raised such as the type of food the virtual pet consumes and the activities that the pet engage in.

The older family members are informed of the choices made by the younger players had made through the pet. The Family Pet game platform then prompts discussions between the generations to create an opportunity for the family to communicate with each other.

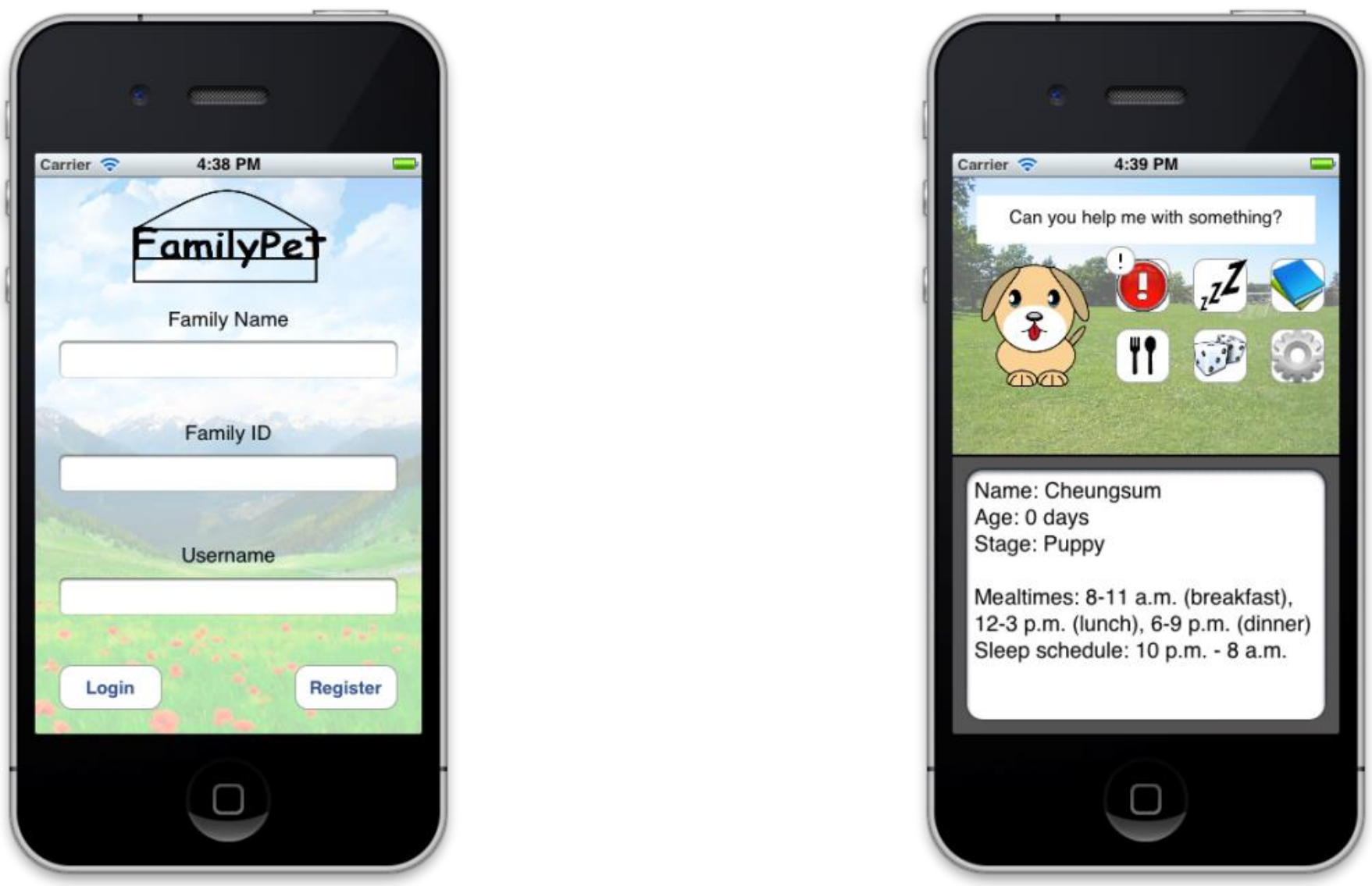

Figure 2 Family Pet Login Screen Shot

Fig. 2 shows the login screen of Family Pet where user can choose to login using an existing account or register as a new user. The registered user can view the status of the pet including the name, age and stage of as well as the daily activities of the pet. Besides naming the pet, the platform allows the user to personalize the background of the application and to adjust the font sizes of the text in Family Pet as shown in Fig. 3

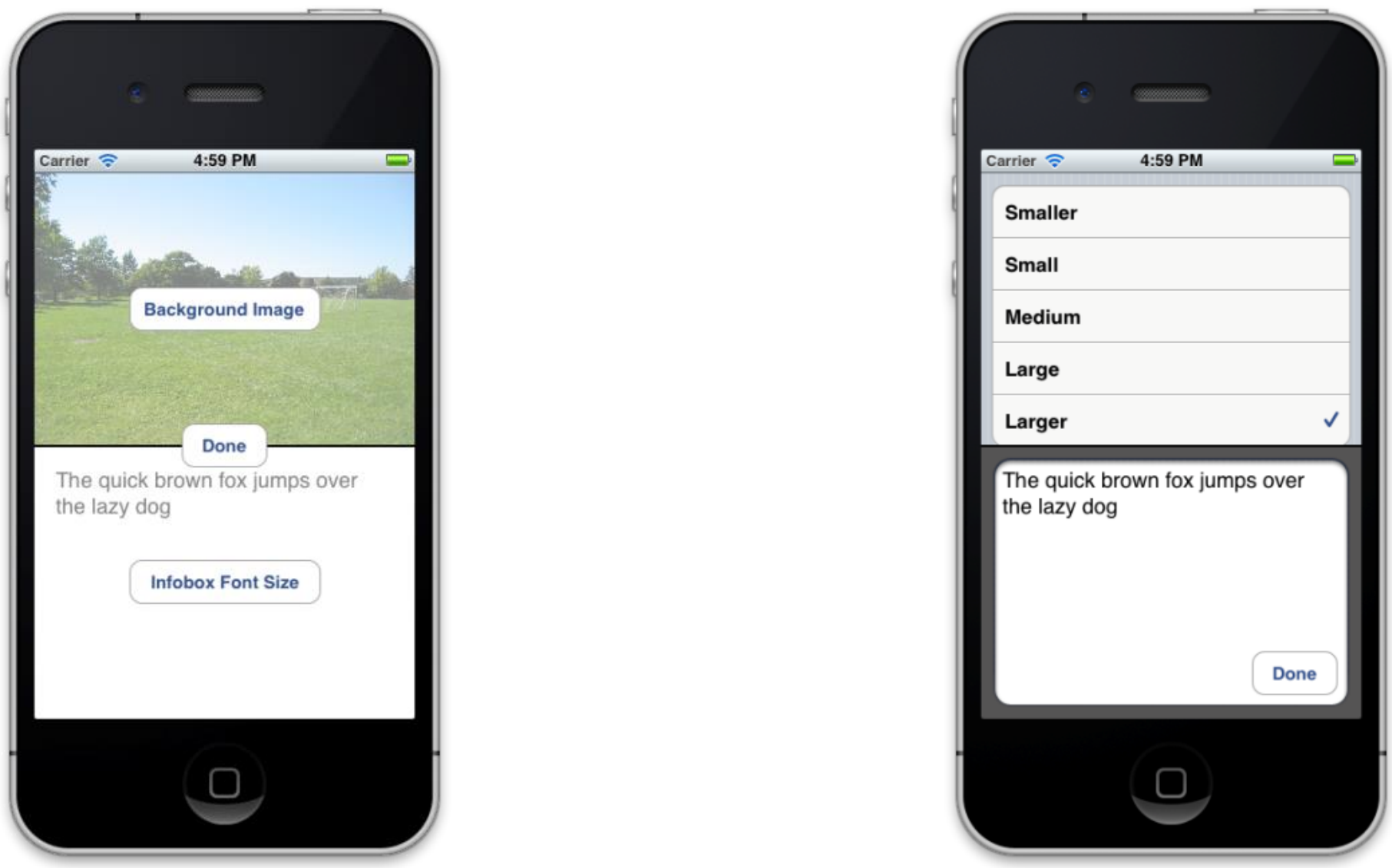

Figure 3 Personalization of Platform

The notifications screen in Fig 4 shows the current status of the pet that the pet owner needs to address. For example, upon the notification as shown in the screenshot, the user is required to participate in studying activity with the virtual pet. Upon choosing the activity to participate, the user can select from the various subjects such as mathematics, science, social for discussions which the user can participate together with the pet.

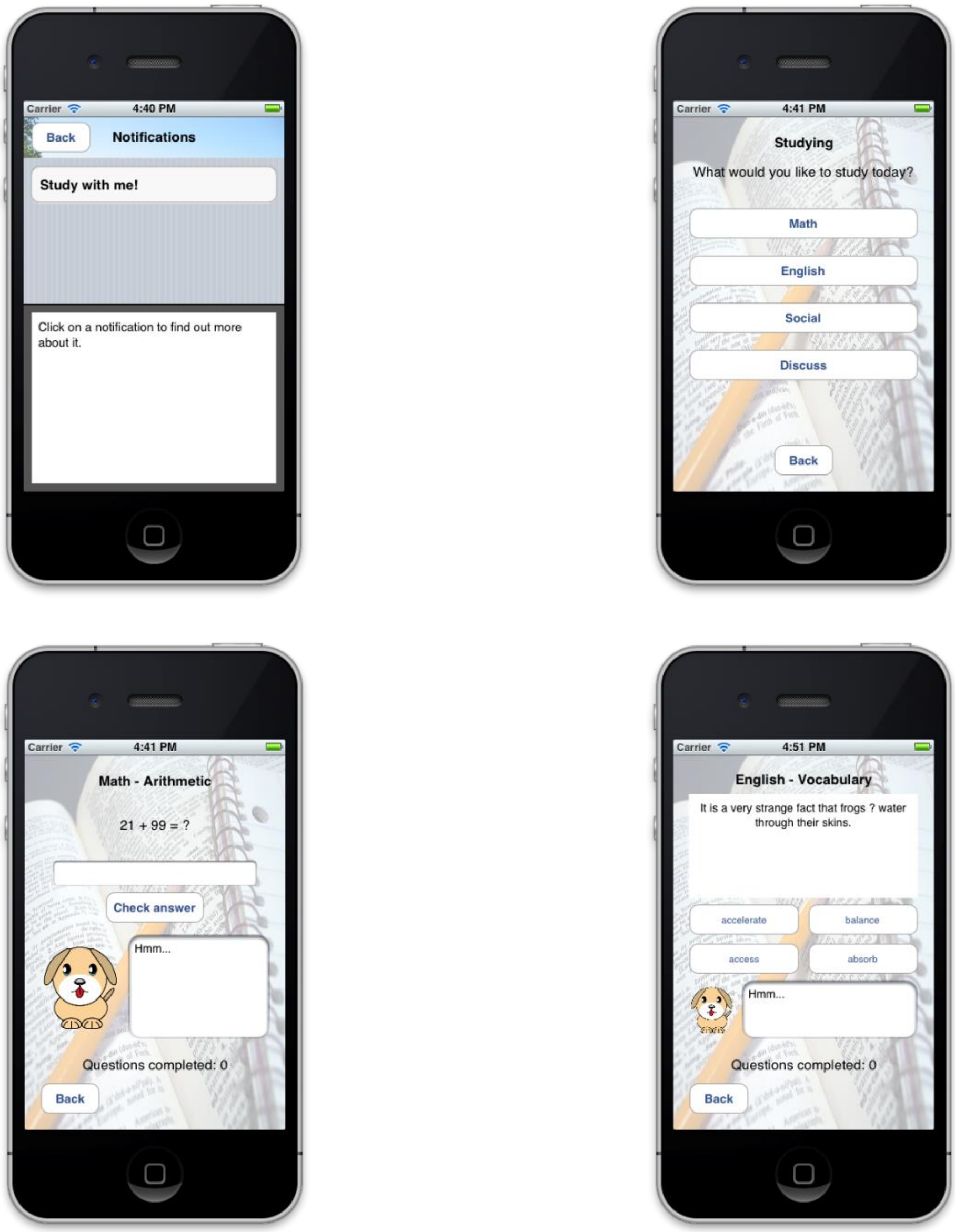

Figure 4 Notifications and Activities Screen Shot

In addition to studying with their pet, users can feed or choose to play with their pet as shown in Fig. 4.

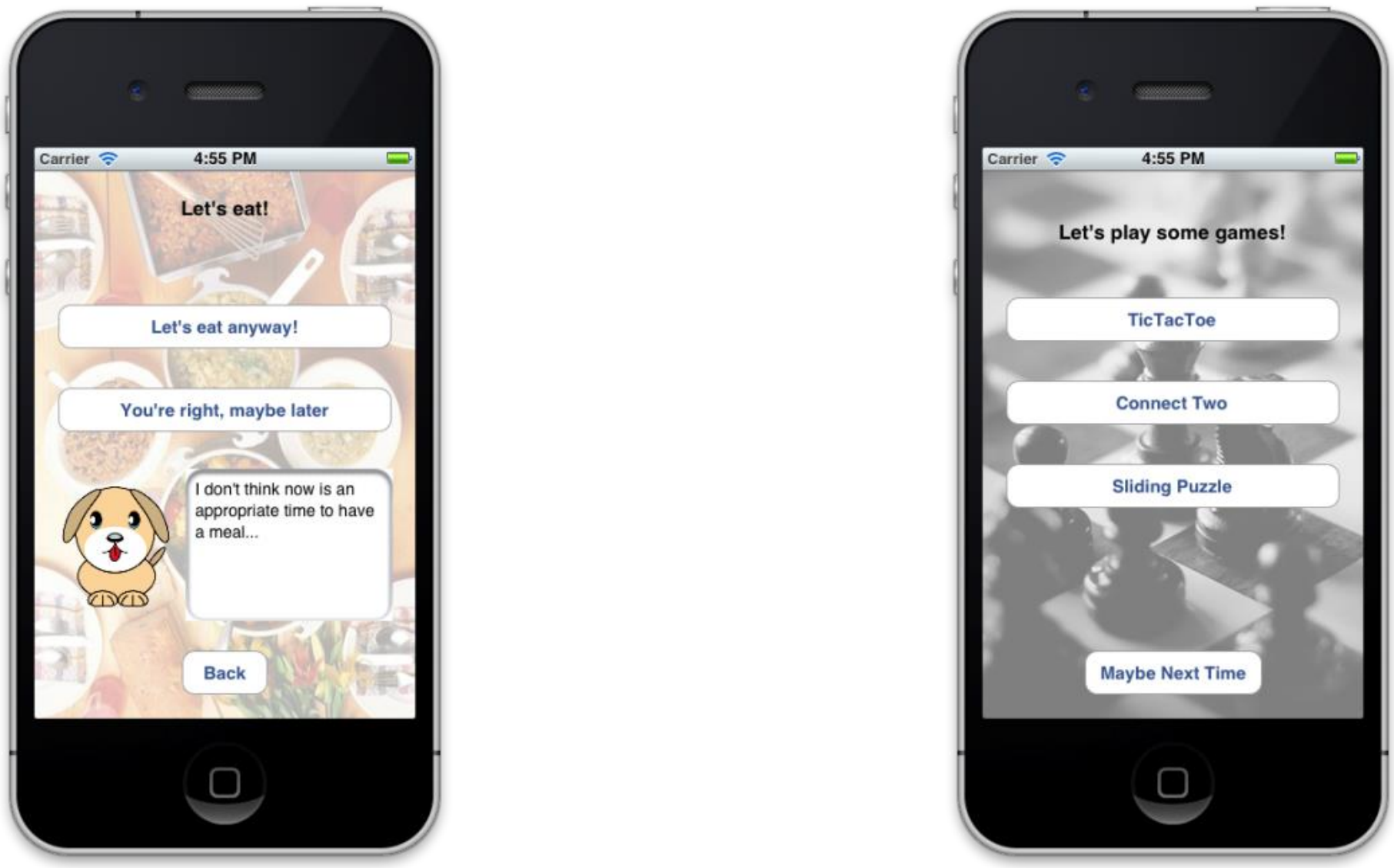

Figure 5 Feeding and Playing with Family Pet

With care and attention given to the pet, it will evolve in terms of growth. The application will also unlock new study topics and games whereas if the state of the pet is not well taken care of the pet it will cause the state of pet to retrogress. Fig.5 shows some of the games in Family Pet, such as the tick-tack-toe, connect two and sliding game.

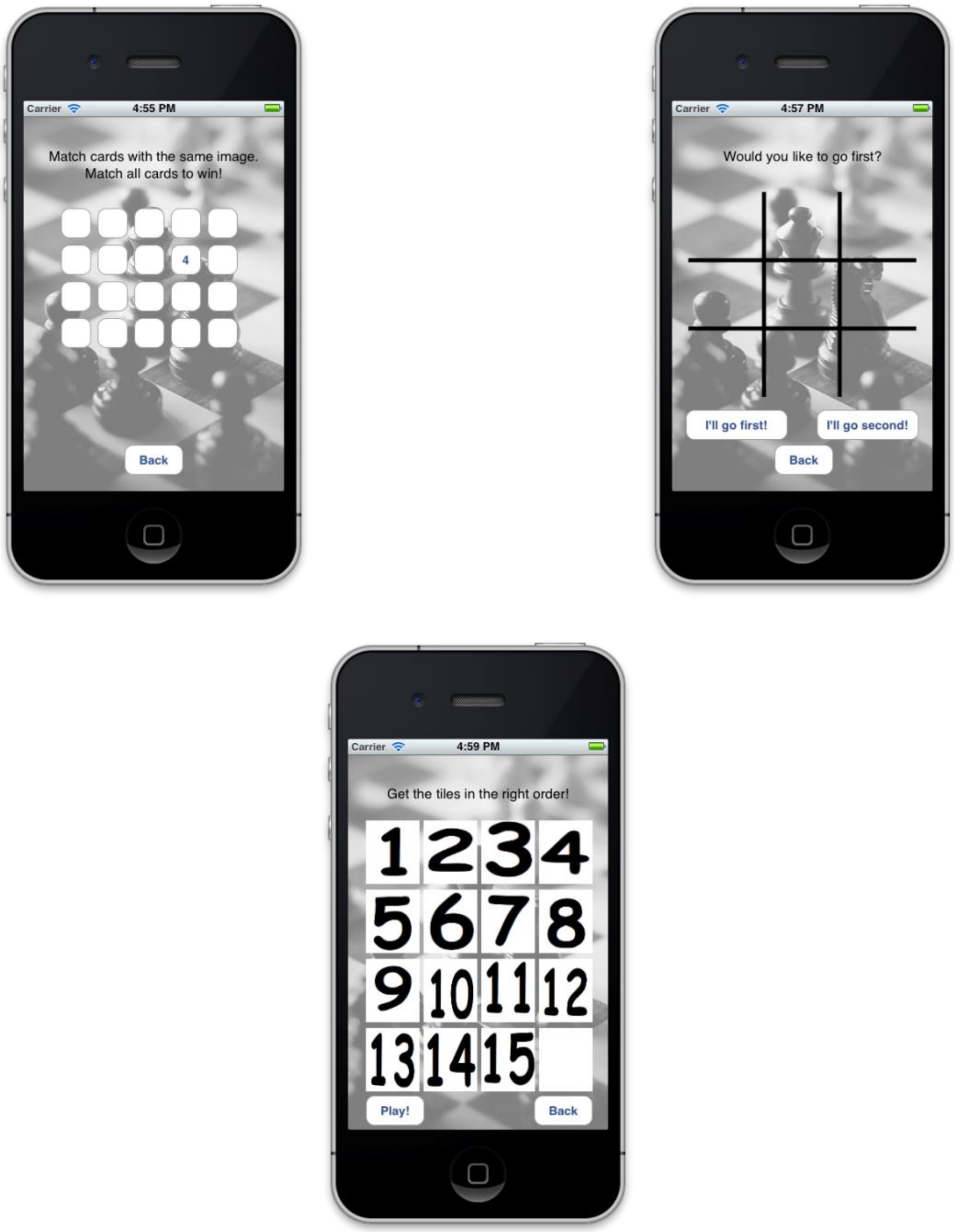

Figure 6 Screen shots of games in Family Pet

### 2.6.2 Persuasive Features in Family Pet

The Family Pet has adopted several persuasive strategies in the effort to motivate users of the application. One of the prominent persuasive features in Family Pet for example is operant conditioning. Coined by B.F Skinner, operant conditioning involves reinforcement and punishment as the consequences of behaviour (Skinner, 1938). In the case of family pet, when the user performs a correct behaviour, such as when they have taken good care of the pet in Family Pet, they will be rewarded with additional games and study topics as well as the growth in their pet. If the user does not take care of their pet properly, the state of the growth of the pet will be affected negatively.

Another persuasive feature is the role of the virtual pet as a persuasive social actor in which the pet plays a social role of a family pet in the application. A social actor can bring about social responses from the users by providing social support (Fogg, 2002; Reeves & Nass, 1996). The social actor in Family Pet aims to persuade its users by creating a relationship with the pet, thus bridging the communication between the child user who takes care of the pet and adult users who monitors the application.

Social cues such as physical attractiveness of the pet, designed to appeal to young users were also considered and incorporated. Other social cues in the form of psychological expression of empathy and personality through the language of the pet also attributes to the building of relationships with the user. There are also social dynamics such as turn taking in the games that the user plays with the virtual pet and positive responses from the pet when the user has completed the designated task. As well as social role in which the user plays the role of a friend towards the family pet and reciprocity in game play.

### 2.6.3 Case Study: Virtual Singapura

Virtual Singapura (VS) is a 3D virtual learning environment (VLE) developed to help learners from lower secondary level understand concepts of science topics and gain inquiry skills by completing learning tasks through interaction with teachable agents in the VLE. The teachable agent in VS requires students to teach the teachable agent by completing a concept map to simulate the learning-by-teaching process. The students are evaluated based on how well their teachable agents are taught.

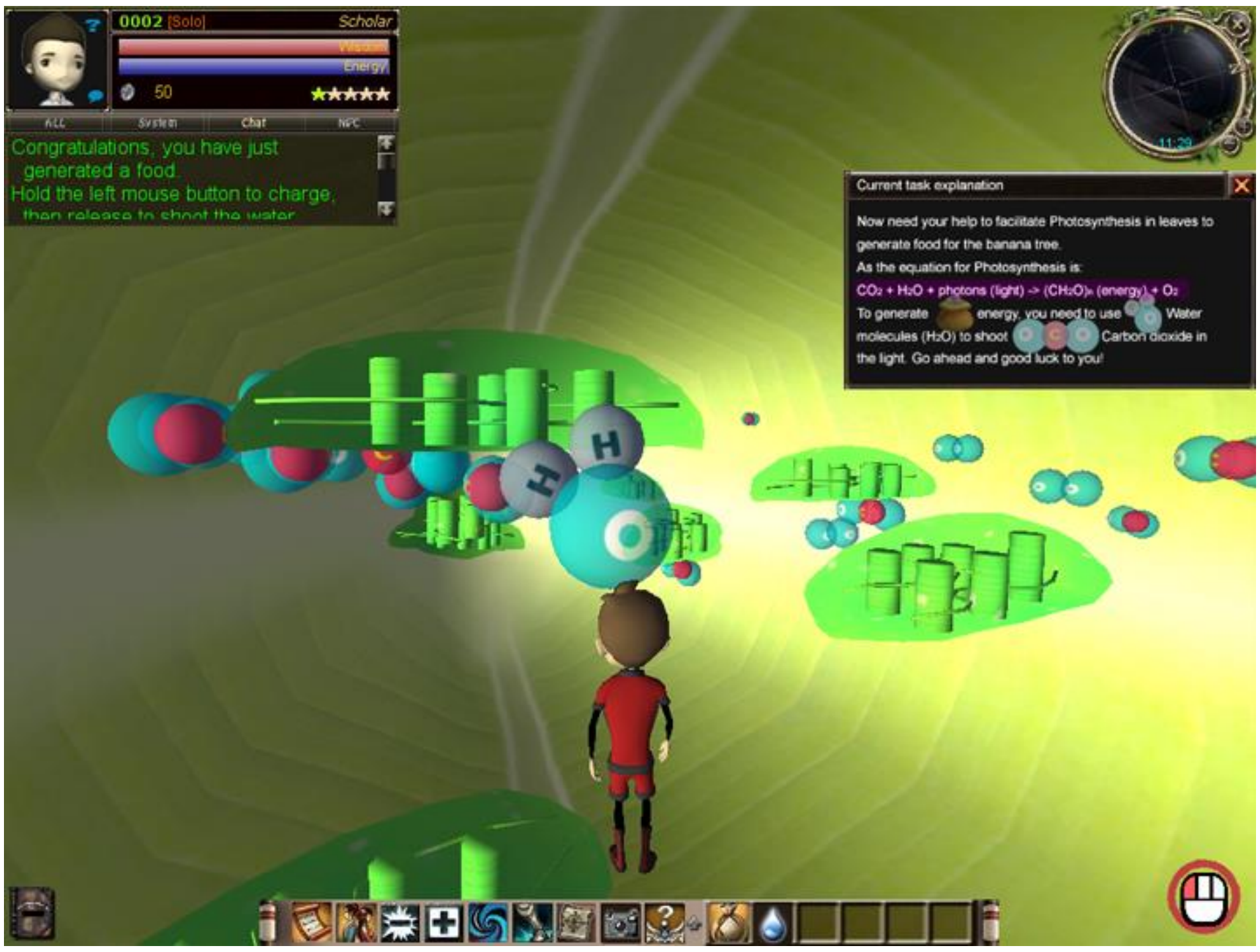

Figure 7 Screenshot of the mini game in VS project.

During the field study conducted in multiple lower secondary schools in Singapore, it was observed that student's learning experiences had been satisfactory (Yu et al., 2009). The students enjoyed the ability to fly, float and walk on water, actions which were only possible in the immersive virtual environment. Students also enjoyed participating in mini games shown such as the "shooting game", where they were required to generate "food" with other team players collaboratively by combining carbon dioxide molecules with water molecules as shown in Fig. 7. Based on the qualitative data collection during the field study however, it was observed that most of the activities that the students enjoyed were not related to learning with the teachable agent.

This case study has prompted us to improve the existing teachable agent so that students will find interacting with teachable agents more enjoyable.

## 2.7 Discussion

Based on the case studies reviewed, to our best knowledge, there is a current research gap in applying persuasion to teachable agents. So far, none of the teachable agents reviewed have the ability to apply persuasion for the users to learn by teaching.

Although learning-by-teaching have been proven to benefit the young (Chase et al., 2009; Sjödén, Silvervarg, et al., 2011). There is currently no research applying the theory in the context of intergeneration learning, since the majority of intergeneration games research is focused on engaging players from different generations through game play.

# 3 Persuasive Teachable Agent (PTA)

Intergenerational learning involves people from all ages to working together in meaningful learning activities where they are able to learn from each another to allow for continuous transfer of knowledge. To accomplish this aim, the PTA behaviour has to be convincing enough so that learners will be motivate to teaching the PTA. By engaging in teaching the PTA, learners from different generations can come together and share their knowledge and benefit from the learning-by-teaching process.

In this section, the definition of the PTA is defined. The PTA architecture will be introduced. The persuasion reasoning will be presented to explain the relationship between the persuasion theory, agent and computation model that allows the PTA to generate the persuasive actions that the agent carries out to influence learners to teach the PTA.

## 3.1 PTA Architecture

The PTA consists of the teachable agent learning environment where the learner teaches the agent by constructing a concept map using the learning-by-teaching pedagogy. Fig. 8 is the diagram of the PTA architecture.

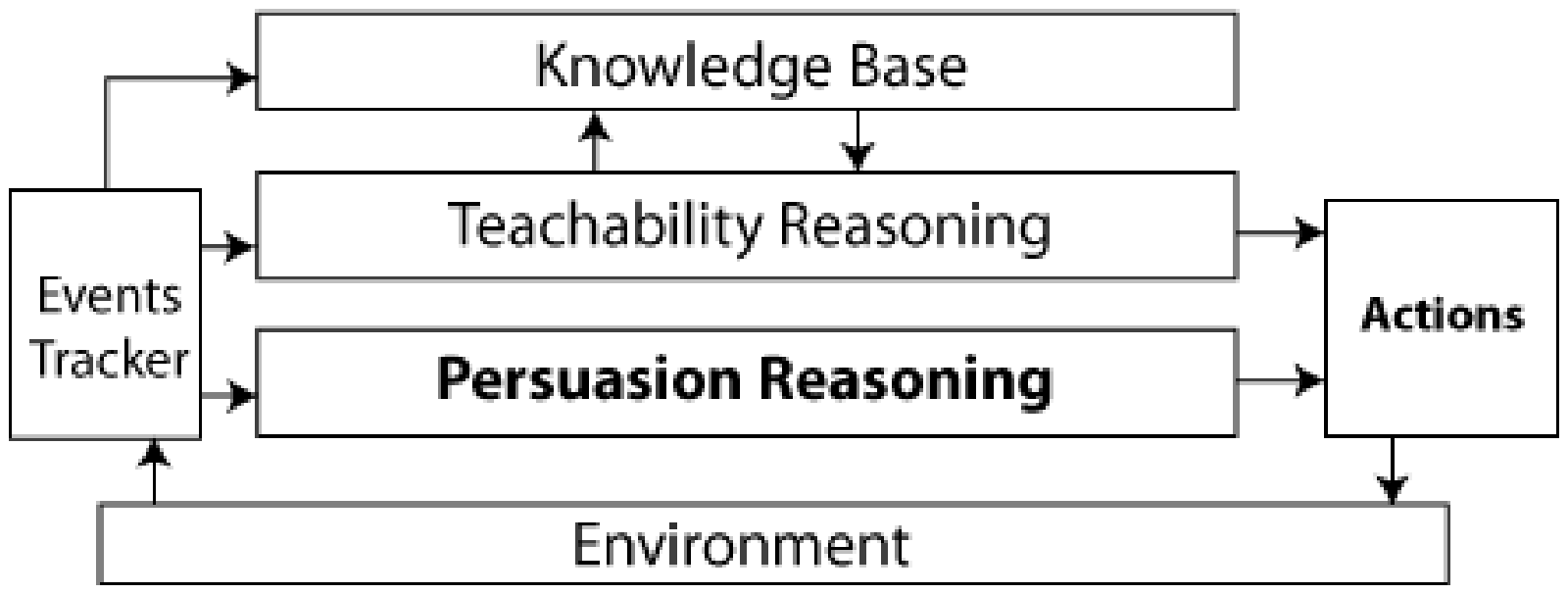

Figure 8 Diagram of persuasive teachable agent architecture

The PTA architecture in Fig. 8 can be expressed as a tuple in the form of:

$PTA = (E, P, K, Tr, Pr)$

- *E,* **Environment**

- The environment in which the agent interacts with

- *P,* **Events Tracker**
  - Precepts that the agent senses in the environment
- *K,* **Knowledge Base**
  - Stores the knowledge the agent has learnt from the learner
- *Tr,* **Teachability Reasoning**
  - Reasons and reflects on the knowledge taught
- *Pr,* **Persuasive Reasoning**
  - The reasoning model for the agent to derive persuasion

The events tracker receives the user learning activity data and the knowledge that the learners have taught the teachable agent from the environment. The knowledge base stores the PTA's knowledge that the agent learnt from the student tutor, while the teachability reasoning is made up of the learning cycle and the reasoning cycle (Ailiya et al., 2011). The learning cycle tracks the relationship links in the concept map where the learners are drawn in to the process to teach the teachable agent, whereas the reasoning cycle, provides the appropriate responses of the PTA according to the learner's inquiry of the learner and reflect the responses in the environment.

The persuasion cycle *PPrA* can be defined as:

- *P, **Perceive:*** Perceives the events that the learner participates in the environment.
- *Pr, **Persuasive Reasoning:*** Analyses the learner's motivation and ability towards teaching the PTA based on the participating events in the environment.
- *A, **Action:*** Generates actions to encourage change in learning attitudes and behaviours.

The diagram in Fig. 9 shows that the Persuasive Reasoning uses the Elaboration Likelihood Model (ELM) (Petty & Cacioppo, 1986) theory of persuasion as its theoretical basis. The Goal Net agent model (Zhiqi, Chunyan, Xuehong, & Robert, 2004) is used to model decision process in which the PTA selects the persuasive strategies it takes to influence the learners. The Fuzzy Cognitive Map (FCM) (Kosko, 1986) computes the learner's motivation and ability parameters.

In the following section, we will describe the theoretical, agent model and computational model that the PTA is based on.

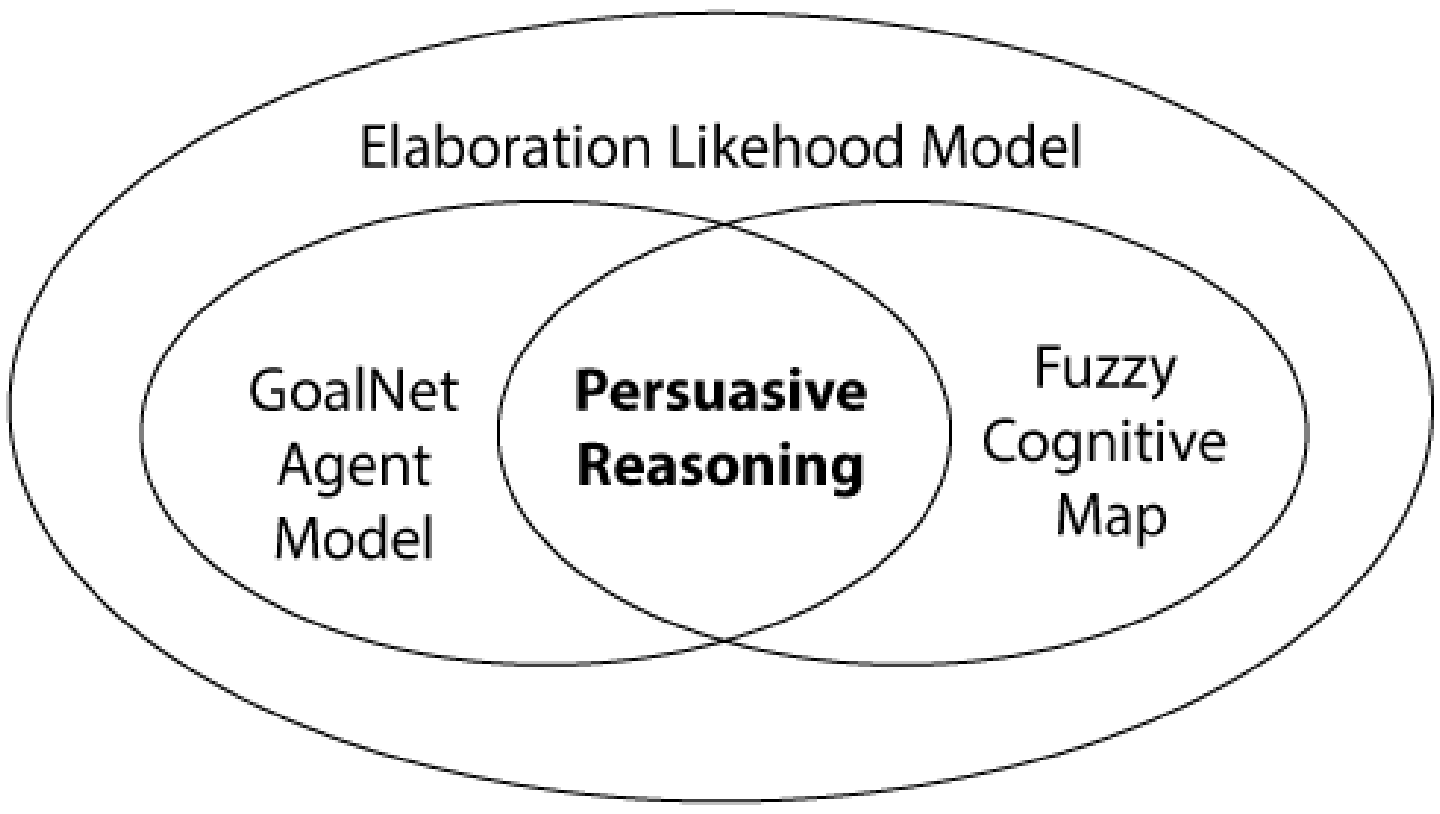

Figure 9 Theoretical, Agent and Computational Models in Persuasive Reasoning

### 3.1.1 Elaboration Likelihood Model of Persuasion

Elaboration Likelihood Model (ELM) is a well-established theories in the field of persuasion. ELM uses a dualistic approach, modelling attitude change through central and peripheral routes (Petty & Cacioppo, 1986).

According to ELM , if an individual is motivated and has the ability to perform a behaviour he or she is likely to take on the central route of persuasion, in which their change of attitude is more persistent. The individual is more likely to think deeply about the relevant issue. On the other hand, if one is less motivated and does not have the ability to perform a behaviour, one is likely to adopt the peripheral route of persuasion, where attitude change is less enduring and one is less interested in elaborating on issues.

The PTA integrates the ELM persuasion theory to instil lasting attitude change towards learning by applying appropriate persuasive strategies based on the learner’s motivation and ability levels. By categorizing events that occur in the environment, the PTA is able to determine the motivation and ability level based on the users’ choices to participate in events.

Factors influencing motivation and abilities are selected based on the ELM to match the events in the environment. The following Table 5 shows the factors influencing motivation and ability in the PTA.

Table 5 Factors Influencing Motivation and Ability based on ELM

| **Factors influencing Motivation and Ability** | |
|---|---|
| **Motivation** | Personal Relevance |
| | Personal Responsibility |
| | Need for cognition |
| **Ability** | Prior Knowledge |
| | Distraction |
| | Repetition |

Motivation is assessed according to the personal relevance, personal responsibility and need for cognition factors derived from the ELM persuasion theory. These motivation factors correlate to events that the user participates in within the environment. Personal relevance relates to how relevant the event is related to the learning topic for the learner. Personal responsibility relates to whether the event reflects the learner’s responsibility for his or her learning. And the need for cognition factor reflects whether the learner is enjoying the learning process.

Ability is assessed through the events that correlate to prior knowledge, distraction and repetition factors. Prior knowledge determines whether the learner has the knowledge resources required to teach the PTA. Events related to prior knowledge increase the learner’s knowledge so they can be better equip with the knowledge to teach the PTA. Distractions are events that reduce the learner’s ability and is evaluated based on the number of distracters present in the environment. Because repetition increases cognitive processing ability the number of attempts the learner teaches the PTA determines the learner's cognitive processing ability.

The ELM is a useful framework in understanding the learner specifically what drive the learners to learn and their ability to perform the task in teaching and learning with the PTA, With the ELM the PTA is able to conceptualize the change in attitude about learning in learners in terms of their motivation and ability.

### 3.1.2 Goal Net Agent Model

The persuasive reasoning component is based on a goal-oriented model for agent modelling known as Goal Net (Zhiqi, 2005; Zhiqi et al., 2004). The Goal Net is composed of *states* represented by nodes in which the agents need to complete to achieve its goal. Fig 10 shows an example illustration of the Goal Net model.

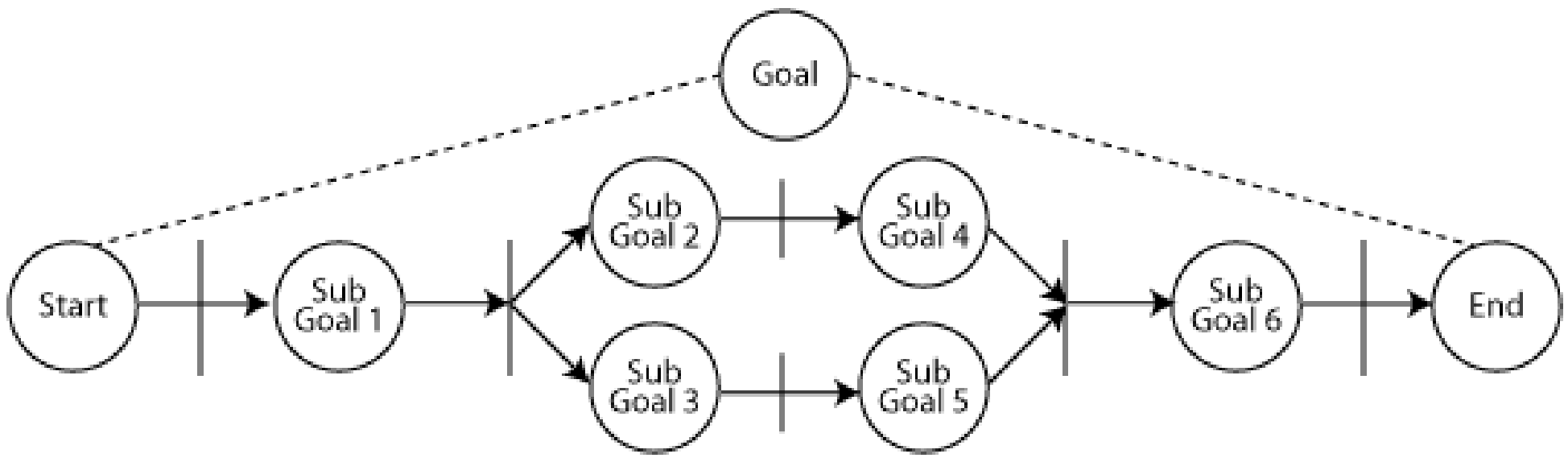

Figure 10 Example Illustration of Goal Net

The *transition* in the Goal Net model is represented by an *arc* with vertical bars, that associates the *task list* and *task function* the agent has to complete to move on to the next state. *Arcs* are represented by directed arrows that connect the relationship between *states* and *transitions*. A *composite state* (shaded node) can be broken down into *atomic states* (blank node) which consist of a single *state*, via *transitions*. An agent's goal within the Goal Net is to obtain an optimal state through the completion of sub-goals in a sequence.

There are four temporal relationships during the transition between input and output states: sequence, choice, concurrency and synchronization. The hierarchical structure of Goal Net allows for complex goal oriented models to be constructed using a composite of sub-goals.

**Sequence**: is a direct causal relationship from state *i* to state *i*+1.

**Concurrency**: is a simultaneous completion of tasks from state *i*. For example, state *i*+1 and state*i*+2 can be achieved at the same time.

**Choice**: is a selective connection from one state to other states. For example, state *i*+1 and state

**Synchronization**: is a convergent of multiple input states into a single output state. For example, state *i* and *i* +1 is converged to a single output state *i*+2.

There are three types of transitions which are, *direct*, *conditional*, and *probabilistic*.

**Direct Transition**: is represented by a vertical line, indicates that the input and output state is a fix action or sequence of actions where there is no selection mechanism involved.

**Conditional Transition**: is represented by a diamond that indicates a transition after the completion of rule-based reasoning and selects actions according to runtime conditions.

**Probabilistic Transition**: a hexagon represents a transition that uses probabilistic inference to select actions in an uncertain situation.
The following figure is a n example illustration of the Goal Net which shows the *transitions* between the sub-goals.

Goal Net has advantages in modelling autonomous, adaptive agents in dynamic situations. Therefore, complex goals can be achieved by combining multiple sub-goals. Mental models of the agent as well as its implementation can be modelled with Goal Net.

### 3.1.3 Fuzzy Cognitive Map Computational Model

FCMs were first proposed by Kosko (1986) to represent causal relationships with fuzzy feedback in dynamic systems. A collection of concepts are interconnected to reflect cause-effect relationships in the FCM dynamic system.

Fig. 11 is a diagram of a simple FCM. The casual relationships in the FCM are represented by links between nodes (concepts) and directed edges (causal relationships).

The weighted causal relationships in the FCM quantifies the strength of the causal effects (Groumpos, 2010). FCM is a widely accepted method to model interrelated causal relationships between different factors affecting decisions in a visual graph representation.

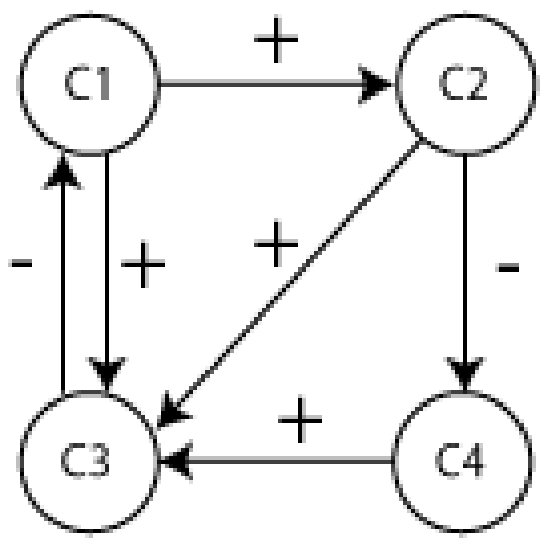

Figure 11 A Simple FCM, directed edge shows the causal flow between concept nodes.

The simplicity and flexibility of the FCM allows it to be applied in wide variety of applications. Some of the applications FCM include expert systems that solve decision problems (Taber, 1991), in finance to solve stock investment analysis (Lee & Kim, 1997), in medical field (Stylios, Georgopoulos, Malandraki, & Chouliara, 2008). The FCM is particularly useful in structuring virtual worlds, due to its dynamic nature that allows changes over time (Dickerson & Kosko, 1993).

In a simple FCM as shown in Fig 11, concept values C*i* are values in real numbers [-1, 1] and the causal edges are values in [-1, 0, 1]. The edge matrix for the simple FCM in Fig. 11 is presented as connection matrix **E** listing the causal links between the nodes as shown in Fig. 12:

$$E = \begin{array}{c} \\ C1 \\ C2 \\ C3 \\ C4 \end{array} \begin{array}{c} \begin{array}{cccc} C1 & C2 & C3 & C4 \end{array} \\ \begin{bmatrix} 0 & 1 & 1 & 0 \\ 0 & 0 & 1 & -1 \\ -1 & 0 & 0 & 0 \\ 0 & 0 & 1 & 0 \end{bmatrix} \end{array}$$

Figure 12 Connection matrix **E** shows the causal relationships in FCM.

The two main components of the FCM are *causal concepts* and *causal relationships*. *Causal concepts* are represented by nodes which can model, *events*, *actions*, *goals* or *lumped-parameter processes*. While *direct edges* represent the *causal relationships* between concepts, a positive sign "+" stands for causal increase, whereas a negative sign "-" indicates a causal decrease. Arbitrary numbers are assigned in weighted FCM to give value to the increase or decrease in causal strength. The *causal relationships* in the FCM listed as a connection matrix that **E** between which concept node in $C_i$ (Dickerson & Kosko, 1993). The strength of the edges $e_{ij}$ of concept $C_i$ ($i^{th}$ column) takes on the value [-1, 1] which indicates negative or positive causal relationships. For example, if edge $e_{ij}$ is negative, -1, $C_i$ decreases $C_j$. On the other hand, if edge $e_{ij}$ is positive, 1 $C_i$ increases $C_j$.

The advantage of the FCM is a quick and easy method to represent and acquire knowledge. Moreover, knowledge can be obtained from multiple expert sources and easily combined into a single FCM without restrictions to the number of concepts. FCM can be easily changed by adjusting the strength of the weights. It is also easy to modify changes in the model.

As such, FCM is suitable to be applied in situation where human behaviour is a major factor or where problems are complex due to the large number of factors and there is no right or wrong answers. The FCM is also useful when limited data is available, missing or incomplete (Özesmi & Özesmi, 2004).

## 3.2 Persuasive Reasoning in PTA

The persuasive reasoning component is based on a goal-oriented model for agent modelling know as Goal Net (Zhiqi et al., 2004). An agent's goal within the Goal Net is to obtain an optimal state through the completion of sub-goals in a sequence. The fuzzy cognitive map transition in the Goal Net structure is used to translate the external states of the world and user's activities into internal states that decide on the agents reactions in dynamic situations (Cai, Miao, Tan, & Shen, 2006).

The goal of a PTA is to direct learners to stay on the central route of ELM with attitude change, where the receiver's attitude changes towards an intended behaviour. In this case, the persuasive reasoning Goal Net is composed using the Goal Net, FCM and the ELM persuasion theory to persuade learners to teach the PTA.

### 3.2.1 Persuasive Reasoning Goal Net

The diagram in Fig. 13 shows the Persuasive Reasoning Goal Net. The Persuasive Reasoning Goal Net receives events from the events tracker once the agent encounters the learner. Once the events from the environment have been received, the Persuasive FCM is activated as a *transition* in the Goal Net.

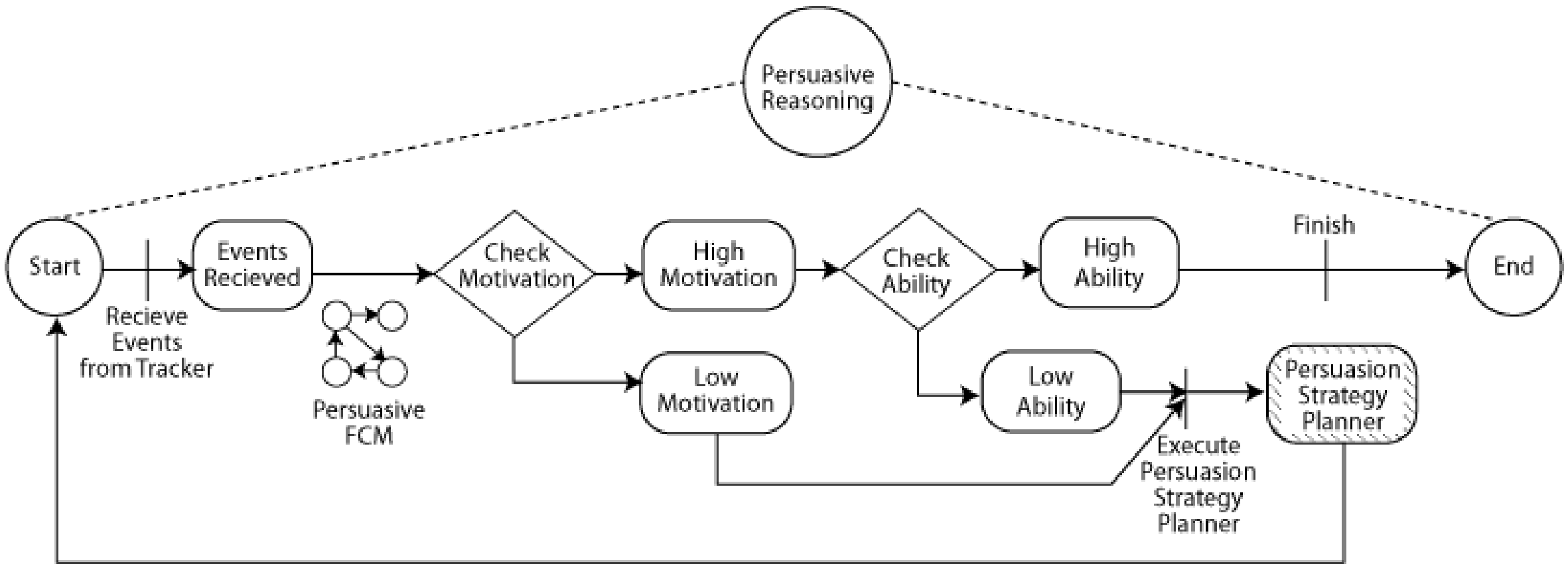

Figure 13 Persuasive Reasoning Goal Net.

In Fig. 13 the Persuasive FCM is shown as the fuzzy cognitive map transition in the Goal Net structure translates external states of the world and users activities into internal states that decides the agents reactions in dynamic situations (Cai et al., 2006). The Goal Net structure provides temporal control over events that occur within the dynamic environment while the fuzzy cognitive map handles the internal reasoning based on the influence of external environmental factors. The outcome is a fuzzy cognitive Goal Net reasoning that uses external events as a basis to model complex internal states of the agent. The agent then proceeds select and persuasion strategies and execute actions in the environment.

If the learner possess both motivation and ability to teach the agent the persuasive reasoning ends. If the learner has low ability and/or low motivation, the persuasion strategy planner will be activated. The appropriate peripheral cue is selected based on the student's motivation or ability and actions are carried out in the environment, once the actions are carried out the PTA persuasive reasoning returns to its initial state.

### 3.2.2 Persuasive FCM

External events influence the learner's motivation and ability in teaching the PTA. Thus, the PTA needs to understand the events in the environment and how each event influences the motivation and ability of the learners to learn and teach the teachable agent. The concept value of motivation and ability enables the PTA to select the appropriate persuasion strategy and action to act upon in the environment.

According to ELM persuasion theory, motivation is influenced by personal relevance to the issue, personal responsibility and one's need for cognition. The ability to elaborate on the issue depends on the presence of distracters. The learner's ability to participate in an event also depends on the repetition to the task and prior knowledge to the issue.

The Persuasive Fuzzy Cognitive Map (Persuasive FCM) is a fuzzy map based on the ELM persuasion theory that shows the causal relationships between external events. Factors affecting motivation and ability that influence the PTA's actions are based on the ELM theory of persuasion. The final persuasive FCM consists of the persuasive FCM with weighted edges. The initial Persuasive FCM was designed based on the case study on Virtual Singapura (VS).

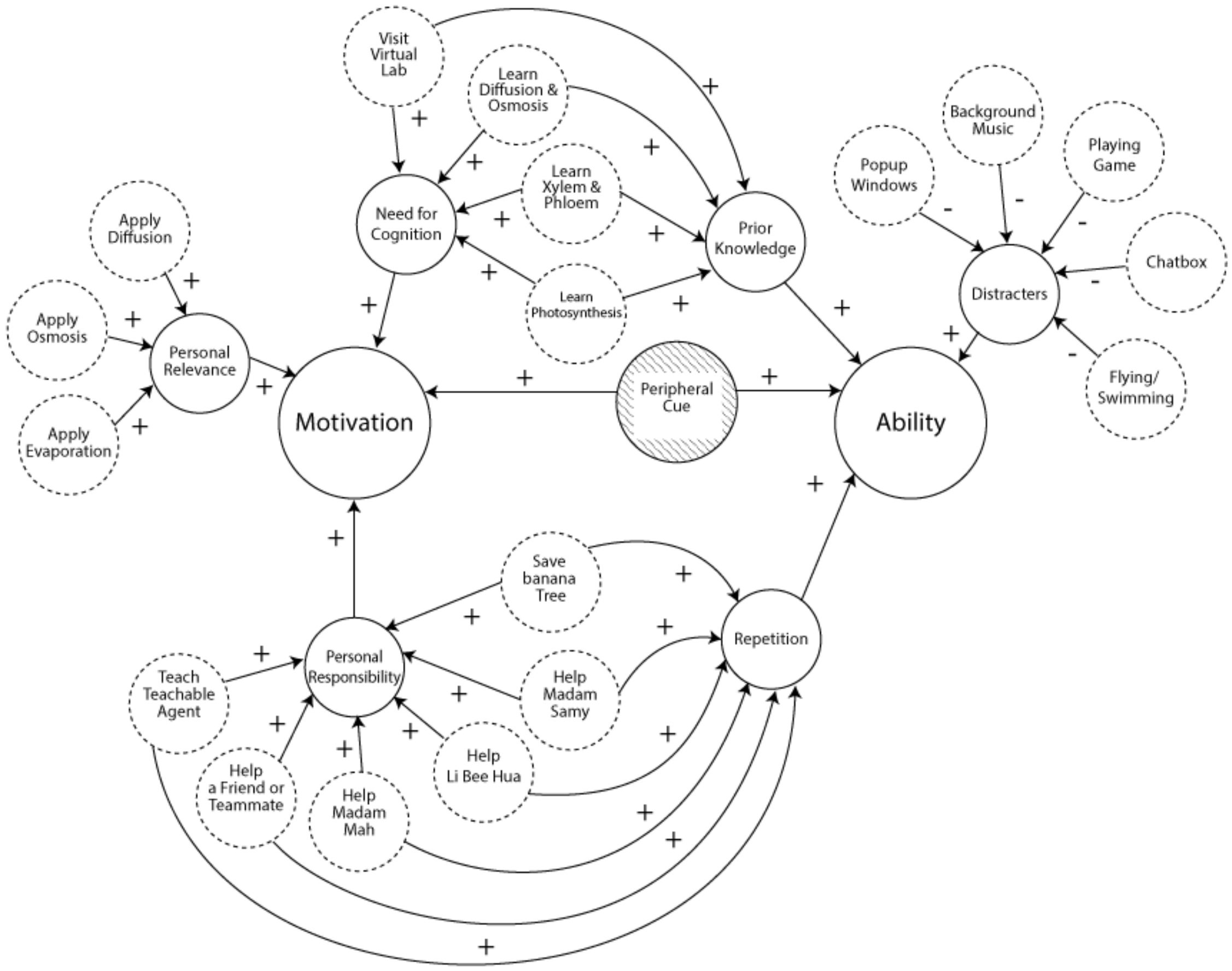

Figure 14 Causal relationships between external events and peripheral cue.

The learning motivation and ability to encourage students to teach the PTA are affected by the external events that the learners engage in within the VS learning environment, where the dotted nodes representing the external events that are causally linked with directed edges. The peripheral cues indicated with a shaded node in the Persuasive FCM represents the actions initiated by the PTA. Fig. 14 is the graphical representation of the Persuasive FCM that shows the causal relationships between the external events and the peripheral cue.

The Persuasive FCM is based on the findings during the field study of VS learning environment in local secondary schools. The causal relationships that are interlinked between external events and factors influenced both motivation and ability, and these relationships are crucial for the PTA to determine the types of peripheral cues that that the agent performs the environment, as some of the peripheral cues provided by the PTA influences the decision of students to learn and teach the PTA in the learning. The following is the definition of the Persuasive FCM.

The Persuasive FCM is defined as a set of $Concept = \{C_1, C_2, \dots, C_n\}$ where,

$$C_i \in \boldsymbol{Motivation} \cup \boldsymbol{Ability} \cup \boldsymbol{PeripheralCue}$$

The external events in the environment that student learner participates in that determines his or her motivation and ability. ***RL, RS*** and ***NC*** are event sets that are linked to **personal relevance**, **personal responsibility** and **need for cognition** respectively affecting an individual's motivation. And ***PK, DT***, ***RP,*** refers to **prior knowledge**, **distracters** and **repetition** event sets that affect an individual's ability to perform a task. Each external event *n* is given an initial state value [0], as the student learner participates in a particular event the state value of event *n* becomes [1].

Within each event sets there are n number of events,

$$\boldsymbol{RL} = \{RL_1, RL_2, \dots RL_n\}$$

$$\boldsymbol{RS} = \{RS_1, RS_2, \dots RS_n\}$$

$$\boldsymbol{NC} = \{NC_1, NC_2, \dots NC_n\}$$

$$\boldsymbol{PK} = \{PK_1, PK_2, \dots PK_n\}$$

$$\boldsymbol{DT} = \{DT_1, DT_2, \dots DT_n\}$$

$$\boldsymbol{RP} = \{RP_1, RP_2, \dots RP_n\}$$

$\boldsymbol{Motivation}$ , refer internal characteristics of the student learner that compels him or her to learn and seek out new knowledge. Motivation is defined as:

$$\boldsymbol{Motivation} \in \boldsymbol{RL} \cup \boldsymbol{RS} \cup \boldsymbol{NC}$$

Personal relevance, personal responsibility and need of cognition are factors in the ELM persuasion theory that influences an individual's motivation.

$\boldsymbol{Ability}$, refer to skill required by the student learner to teach the teachable agent. According to the ELM persuasion theory, an individual's ability is influenced by one's prior knowledge, the presences of distracters during processing of information and repetition to persuasive message presented. Ability is defined as:

$$\boldsymbol{Ability} \in \ \boldsymbol{PK} \cup \boldsymbol{DT} \cup \ \boldsymbol{RP}$$

The persuasion actions that the teachable agent performs in the environment are referred as ***PeripheralCues***. ***PeripheralCues*** include hints from ***ExpertSourceSet, AttractiveSourceSet*** and emotions from ***AffectSet.***

$$\boldsymbol{PeripheralCue} \ \in \boldsymbol{ExpertHintsSet} \ \cup \boldsymbol{AttractiveSourceSet} \cup \boldsymbol{AffectSet}$$

***ExpertHintsSet***, ***EH*** contains a set of hints from expert source that the teachable agent provides to help student learner overcome difficulties during the teaching process where,

$$\boldsymbol{EH} = \ \{EH_{1,} \, EH_2 \ \ldots EH_n\}$$

***AttractiveSourceSet***, ***AS*** is a set of popular individuals that the teachable agent will recall to provide support to the student learners.

$$\boldsymbol{AS} = \ \{AS_{1,} \, AS_2 \ \ldots AS_n\}$$

***AffectSet***, ***AF*** is a set of emotions that teachable display in the environment to influence student learner's learning motivation and ability.

$$\boldsymbol{AF} = \ \{AF_1, AF_2, \ldots AF_n\}$$

Each item, *n* from the ***ES, AS*** and ***AF*** takes on a value of [1, 0], and is activated when a peripheral cue have been selected and carried out in the virtual environment. The state value of motivation, $Mot^t$ at time *t* is the sums of RS, RL, NC state vectors, multiply by their respective weight value of *WRS*, *WRL*, and *WNC*. Similarly, state values of ability and peripheral cue, $Ab^t$ and $PeriCue^t$ are the result of the sum of individual state vectors multiply by their respective weight values.

$$Mot^t = RS * WRS + RL * WRL + NC * WNC$$

$$Ab^t = PK * WPK + DT * WDT + RP * WRP$$

$$PeriCue^t = EH * WEH + AS * WAS + AF * WAF$$

The state vector of the Persuasive FCM can be expressed as

$$A^t = [Mot^t, Ab^t, PeriCue^t]$$

A new value for each concept is updated due to the activation of a new event in the virtual environment can be computed as,

$$A^{t+1} = f(A^t * W + A^t)$$

Where $A^{t+1}$ is the new state vector of concepts $C_i$ and $C_j$, and $A^t$ is the previous state vector of $A^t$ multiply by the weight matrix $W$ of the persuasive FCM.

### 3.2.3 Persuasive Strategy Planner

In the Persuasive Strategy Planner sub-goal shown in Fig. 15, the PTA selects the persuasion strategy based on the motivation and ability values from the Persuasive FCM. Once the persuasive strategy to improve motivation or ability is selected, a peripheral cue is selected in the Persuasion Strategy Selection sub-goal. The peripheral cues are based on the ELM (Cacioppo et al., 1986) theory of persuasion, where individual under the peripheral route are influenced by attractive expert source, or strong arguments and affect.

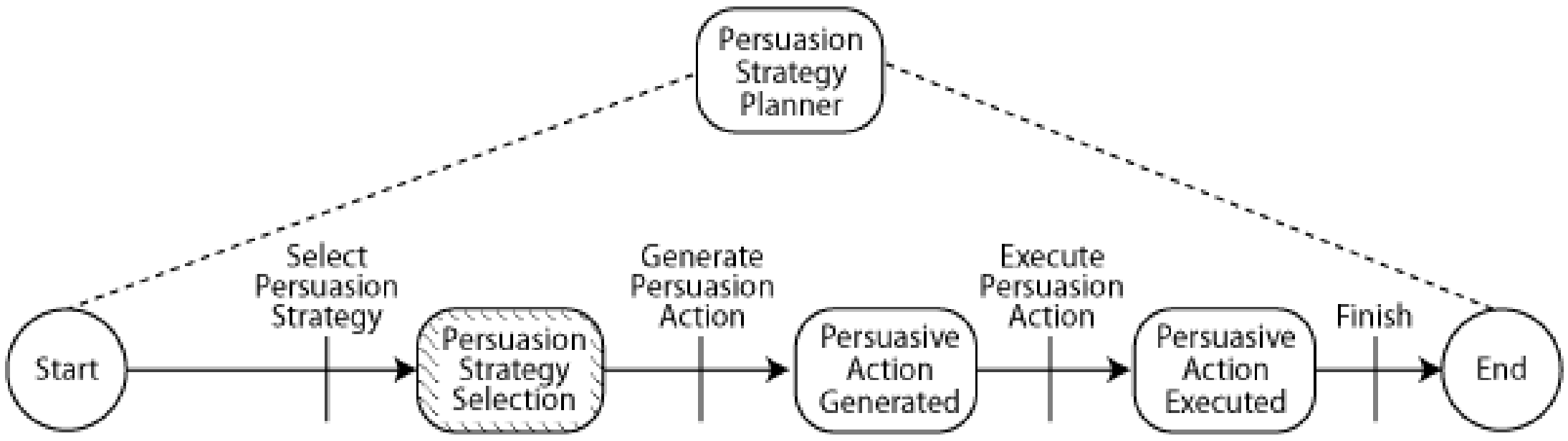

Figure 15 Persuasion Strategy Planner sub-goal of the Persuasive Reasoning Goal Net.

### 3.2.4 Persuasive Strategy Selection

In the Persuasive Strategy Selection sub-goal shown in Fig. 16, the persuasive strategy is selected based on the motivation and ability value. If either motivation or ability value is low, a suitable peripheral cue is selected. The persuasive action is generated and executed in the environment.

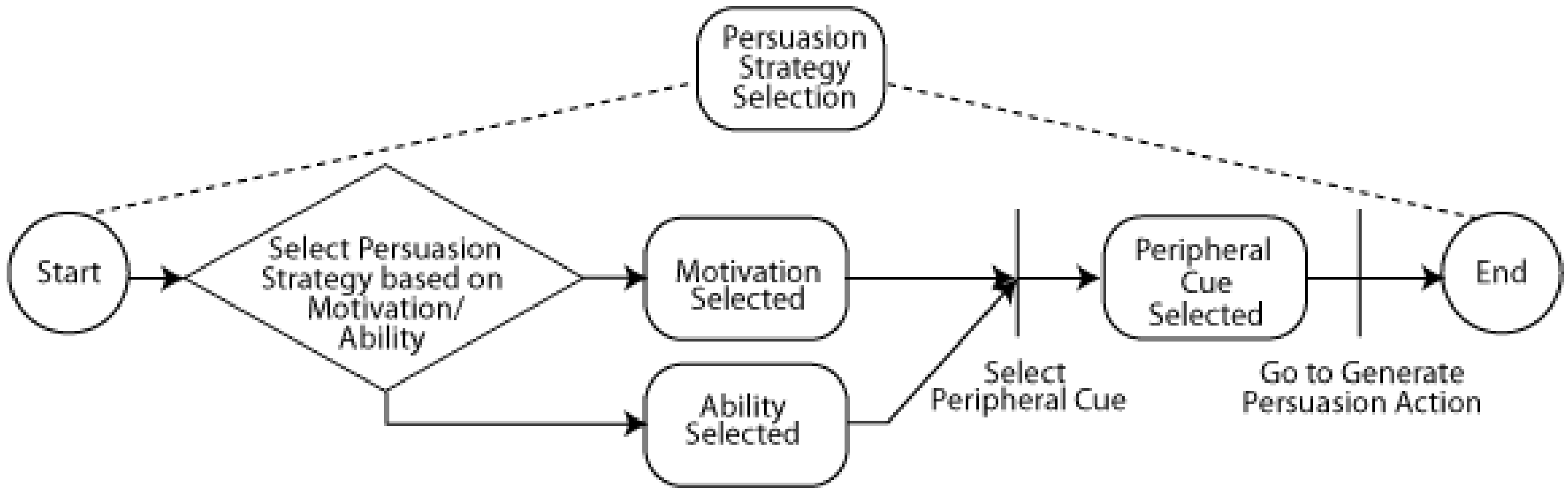

Figure 16 Persuasion Strategy Selection sub-goal of the Persuasion Strategy Planner

## 3.3 Preliminary Experimental Results

The persuasive FCM model based on the VS project is simulated in the following scenario examples that demonstrate the Persuasive FCM model. This simulation was conducted using the Fuzzy Cognitive Mapping software (FCMapper). The FCMapper is an Excel based FCM analysis tool freely available (Skinner, 1938).

*Experiment 1: Learner A engages in a series of events involving gradually change personal relevance by gradually applying diffusion, osmosis and evaporation.*

Table 6 Experiment Results from Persuasive FCM

| Concepts | Steady State | Apply Diffusion | Apply Diffusion and Osmosis | Apply Diffusion, Osmosis and Evaporation |
|---|---|---|---|---|
| Peripheral Cue | 0.87154658 | 0.8625956 | 0.86348113 | 0.86412577 |
| Motivation | 0.9692525 | 0.9230875 | 0.93050934 | 0.93593814 |
| Ability | 0.9454505 | 0.9139302 | 0.91399983 | 0.91405049 |

As shown in Table 6, the increase in concept value of the personal relevance corresponds to the increase in the concept value of motivation and ability. As the student learner's motivation and ability increases, the peripheral cue needed to stimulate interest decreases gradually. Based on the concept values of motivation, ability and peripheral cue from the Persuasive FCM iteration, the PTA will determine the persuasion strategy to execute in the sub-goal of the Persuasive Reasoning in the previous section.

### 3.3.1 Illustrative Example

The following section is a description on events that the learner can participate in, and is tracked using the Persuasive Reasoning Goal Net. Once an event is received, motivation, ability and peripheral cue values are extracted from Persuasive FCM. Actions based on the values derived from the Persuasive FCM are executed by the PTA with the aim to persuade the learner in teaching the PTA.

In the initial process, the PTA receives events that the learner participates in the learning environment.

**Receive Events:**

1. Flying/swimming in the environment
2. Learn diffusion and osmosis (Knowledge Farm)
3. Apply osmosis and diffusion (Virtual Laboratory)
4. Teach teachable agent (Save banana plant quest)

**Check Motivation:**

The result of the motivation values after each event is shown in Table 6. There is an increase in motivation value after each event.

**Check Ability:**

The result of the Ability values after each event.

**Execute Persuasion Strategy Planner:**

Peripheral cue values are extracted from the Persuasive FCM as shown in Table 6. In this case, a higher value of the peripheral cue indicates that the PTA needs to persuade the learner more. For example, if the peripheral cue value is high, the emotions of the Persuasive Teachable Agent will be activated from the *AffectSet*, *AF* which returns the “sad” emotions. Therefore, the persuasive actions for the PTA will be to execute a “cry” animation.

### 3.3.2 Discussion on Issues and Challenges

In comparison to the existing teachable agent, the improved PTA is able to respond more intuitively to suit the student’s learning requirements. Learners are motivated to teach the PTA. The improved PTA is able to respond better to individual needs of the learner based on their ability and motivation. At the same time, the PTA possesses the capability to provide strategies to encourage social interaction which will keep learners engaged while teaching the PTA.

Although this initial PTA model addresses the issues in teachable agent, namely, the lack of interaction between the teachable agent and the learners. However, this version of the PTA model still poses several limitations, as highlighted in the review of the initial of the PTA model (Zeng, 2015).

Firstly, the design of the initial PTA model focuses on the implementation of persuasion theory onto the teachable agent model, which improves the ability of the PTA to persuade the learners to teach it. However, the initial PTA model lacks integration of the agent’s ability to practice on the knowledge taught by the learner. Moreover, in terms of reusability, the initial PTA model adopts a quantitative model that was difficult to reuse in terms of computational processes. Lastly, the initial PTA model was largely based on theoretical a model which presents a gap in the practical implementation of the PTA model in a real world game engine.

## 3.4 Improvements to the PTA

Zeng (2015) further suggested the following improvements to the original version of the PTA model which includes a complete integration of the *Persuasion Reasoning, Teachability Reasoning and Practicability Reasoning.* The original *Persuasion Reasoning* is simplified and fully combined with the *Teachability Reasoning* and *Practicability Reasoning.* The *Teachability Reasoning* and *Practicability Reasoning* have been added to the improved

Persuasive Goal Net to reflect on the PTA's ability to be taught and practice the knowledge taught by the learner. An improved version of the PTA model allows for a reusable quantitative model where the *Persuasion Reasoning* model that could be applied to different contexts. This means that future applications of the *Persuasion Reasoning* can be reused in other knowledge domains and implementations in future work. And a refined version of the system architecture that describes the implementation of the reusable PTA control structures are proposed and developed in a 3D virtual environment.

### 3.4.1 Redefining PTA Model

The improved PTA model shown in Fig. 17 is proposed by, Zeng (2015) is redefined by a tuple*:*

*PTA = (E, ET, K, Rs, R, A)*

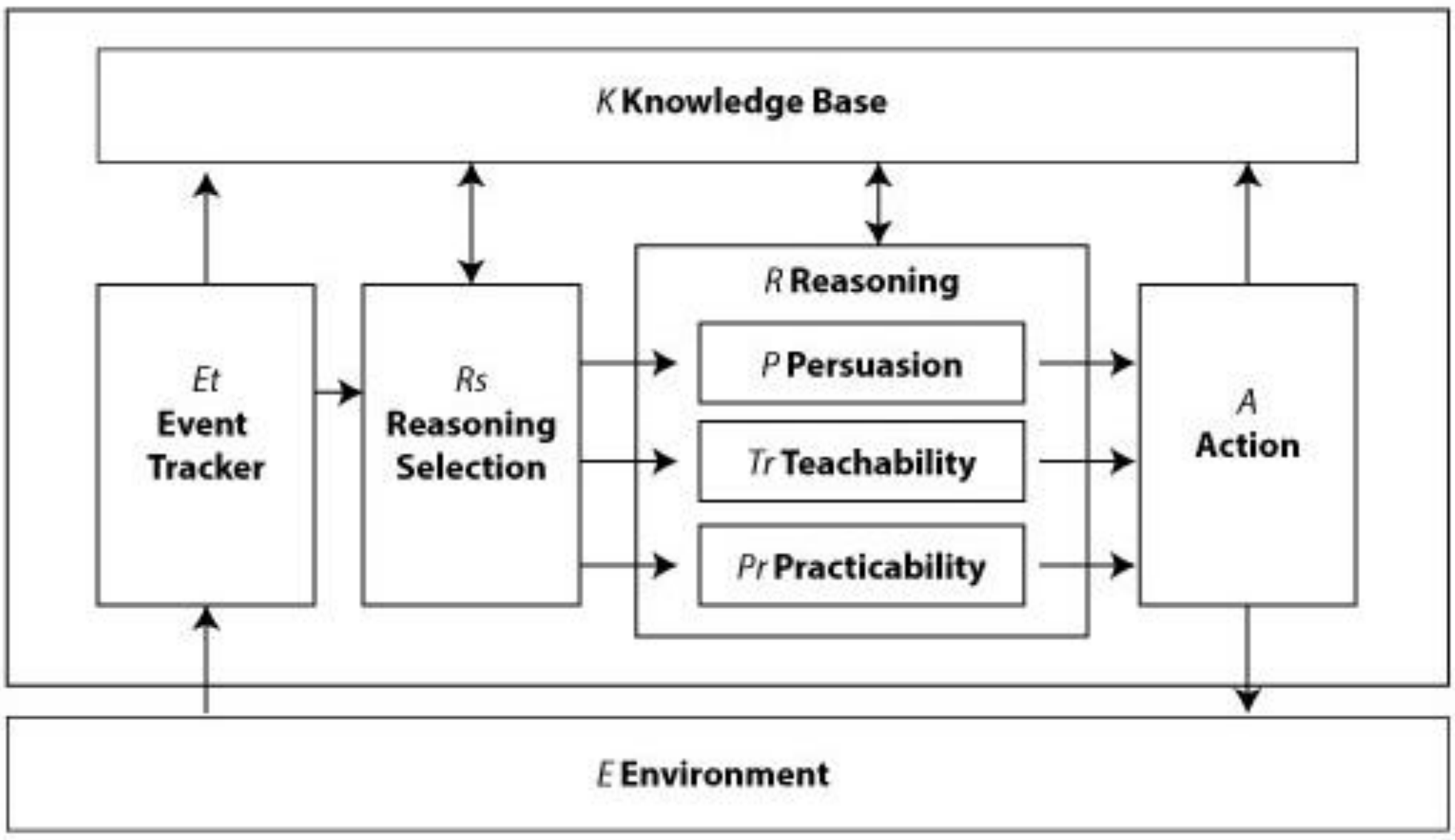

Figure 17 Improved PTA Model (Zeng, 2015)

- *E*, **Environment**
    - The set of environments states perceived by the agent.
- *ET*, **Precepts /Events Tracker**
    - The set of precepts or events that the agent perceives from the environment.
- *K,* **Knowledge Base**

- The knowledge base of the agent, which consists of the following subset.
  - *Goal Net Model,* is the set of Goal Net structures which controls the characteristics and drives the agent towards its goals pursuit
  - *FCM Model,* stores the FCM structure that derives the values of motivation and ability.
  - *Domain Knowledge,* stores the expert knowledge related to the learning topics.
  - *Learnt Knowledge,* stores the knowledge that the agent learnt from the knowledge taught
  - *Runtime Data,* stores data such as the events information and their history that is generated during runtime.

- *Rs,* **Reasoning Selection**
  - The selection mechanism set the agent has to adopt to select reasoning in *R* for each cycle.
- *R,* **Reasoning Set**
  - The set of reasoning for the agent to perform, which and can be further defined by the tuple: *R = (P, Tr, Pr)* where,
    - *P,* is the *Persuasion Reasoning* mechanism that enables the agent to persuade the learner.
    - *Tr,* is the *Teachability Reasoning* that allows the agent to learn from the knowledge taught.
    - *Pr,* is the *Practicability Reasoning* that allows the agent to practice on the knowledge learnt.
- *A,* **Actions**
  - The set of actions to be performed by the agent in the environment.

In addition, the improved PTA model enables the agent to through the following three different cycles:

- Persuasion Cycle: *EtRsP(KA)*
  1. *Perceive:* The agent's perceptions on the events in the environment

2. *Reasoning Selection:* The agent selects a suitable reasoning according to its perceptions

3. *Persuasion Reasoning:* The agent goes through reasoning to determinate the type persuasion based on the learner's level of motivation and ability. The agent then selects the appropriate persuasion cue according to its perception if persuasion is required.

4. *Knowledge Base:* The agent retrieves the persuasion cue from the knowledge base if persuasion is required.

5. *Action:* The agent executes the persuasion cue if persuasion is required.

- Teaching Cycle: *EtRsTrK*

   1. *Perceive:* The agent's perceptions on the events in the environment

   2. *Reasoning Selection:* The agent selects a suitable reasoning according to its perceptions

   3. *Teachability Reasoning:* The agent acquires the knowledge taught and through its teachability reasoning, interprets the knowledge into knowledge representations.

   4. *Knowledge Base:* The agent stores the interpreted knowledge representation in the knowledge base.

- Practicing Cycle: *EtRsPr(KA)*

   1. *Perceive:* The agent's perceptions on the events in the environment

   2. *Reasoning Selection:* The agent selects a suitable reasoning according to its perceptions

   3. *Practicability Reasoning:* The agent decides the appropriate responses to the learner's queries according to its knowledge. The agent chooses a response if correct response if the responses are pre-determined.

   4. *Knowledge Base:* The agent retrieves the response from the knowledge base if the response is pre-determined.

   5. *Action:* The agent response according to the learner's query.

The improved PTA model based on the earlier proposed PTA architecture offers a clearer view of the agent reasoning capabilities and flexibility for the PTA to select its reasoning functions according to its precepts of the environment, thus, allowing the PTA to be more autonomous.

### 3.4.2 Modelling Teachability and Practicability Reasoning with Goal Net

Significant additions to the PTA architecture are the *Teachability Reasoning* and *Practicability Reasoning* in the *R*, Reasoning Set cycle. The *Teachability Reasoning* and *Practicability Reasoning* is modelled with the Goal Net agent model and integrated in the Goal Net model of PTA (Zeng, 2015), which forms the highest level of the Goal Net in the PTA known as the main routine shown in Fig. 18.

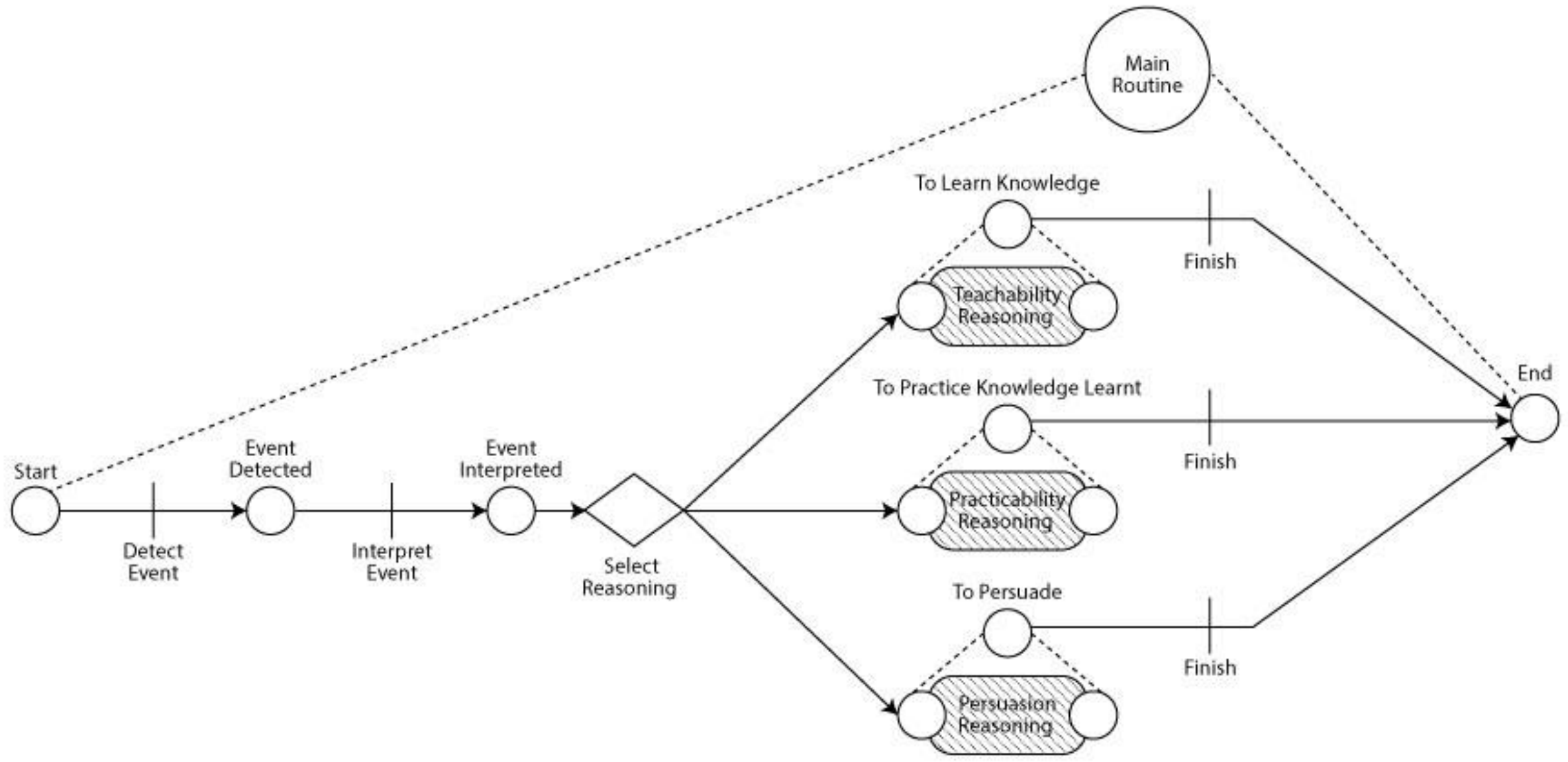

Figure 18 Goal Net Model of PTA Main Routine (Zeng, 2015)

The following Table 7 shows the PTA definition and the corresponding states and transitions in the main routine in Fig.18.

Table 7 PTA Definition corresponding to States/Transitions in Main Routine

| PTA Definition | States/Transitions |
|---|---|
| *Et*, **Event Tracker** | Detect Event, Event Detected, Interpret Event, Event Interpreted |
| *Rs*, **Reasoning Selection** | Select Reasoning |
| *Tr*, **Teachability Reasoning** | To Learn Knowledge |
| *Pr,* **Practicability Reasoning**<br>*A,* **Actions** | To Practice Knowledge Learnt |
| *P,* **Persuasion Reasoning** | To Persuade |

The main routine runs continuously as long as the agent continues to pursue its goal. The agent will check for an update of events in each cycle. If one or more events are detected, it will select the reasoning based on the events perceived. After each cycle, the agent will return to its original start node and begins a new cycle.

There are three sub-goals, namely, *Teachability Reasoning*, *Practicability Reasoning* and *Persuasion Reasoning* in the main routine. These sub-goals allow for the agent to select the appropriate reasoning goals according to the events in environment.

***Teachability Reasoning Sub-Goal***

The Teachability Reasoning sub-goal in Fig. 19 illustrates the agent's desire to learn new knowledge. In this sub-goal, the agent requires the learner user to teach itself. Once the response has been received by the agent, it checks if the request have been accepted or rejected. If the learner agrees to teach the agent, the goal net initiates teaching. Once teaching has been initialized the agent acquires the knowledge after it saves the stored the knowledge received and the cycle ends.

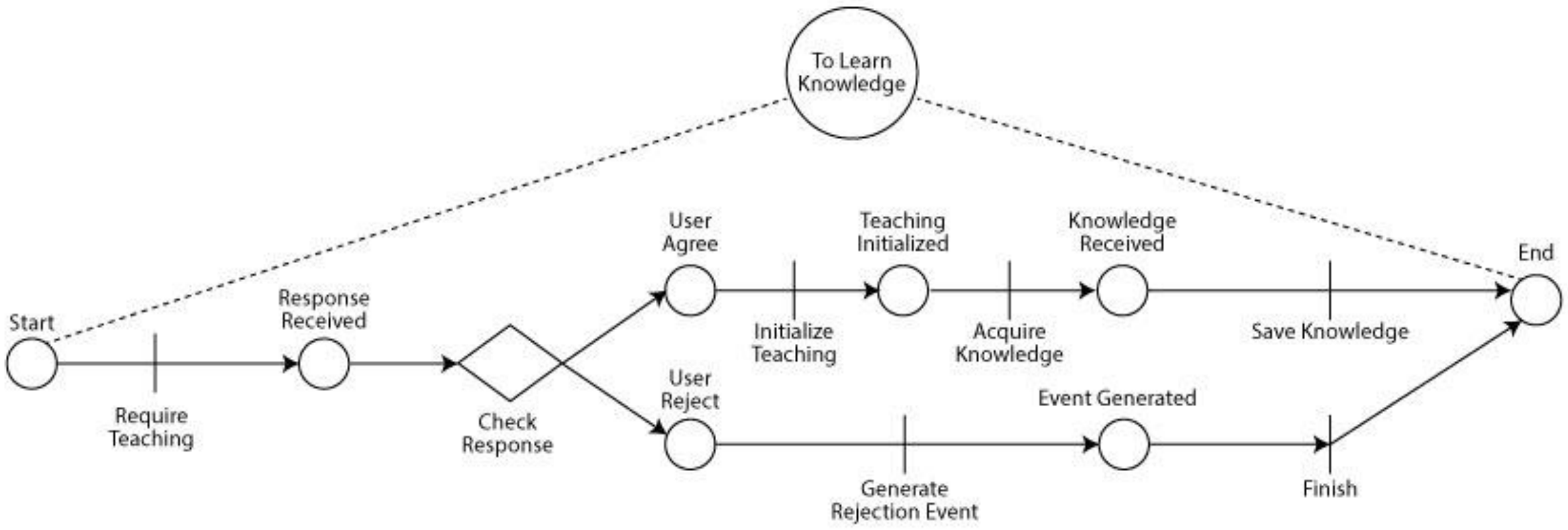

Figure 19 Teachability Reasoning Sub-Goal Net (Zeng, 2015)

***Practicability Reasoning Sub-Goal Net***

The Practicability Reasoning sub-goal in Fig. 20 starts with the agent's query in its knowledge base which retrieves the knowledge that it has learnt from the learner. Through its reasoning, the agent generates either a correct solution or a wrong solution based on the knowledge in its knowledge base.

This simulates the agent's capability to practice its knowledge and provides a means for the agent to reflect on the knowledge learnt through feedback on the learners' teaching. If the result from the reasoning is the correct solution, the agent will carry out the solution. On the other hand, if the result of the reasoning is a wrong solution event, upon the next cycle of the Goal Net of PTA, the agent will go through the Persuasion Reasoning Goal Net in the attempt to improve the learner's motivation.

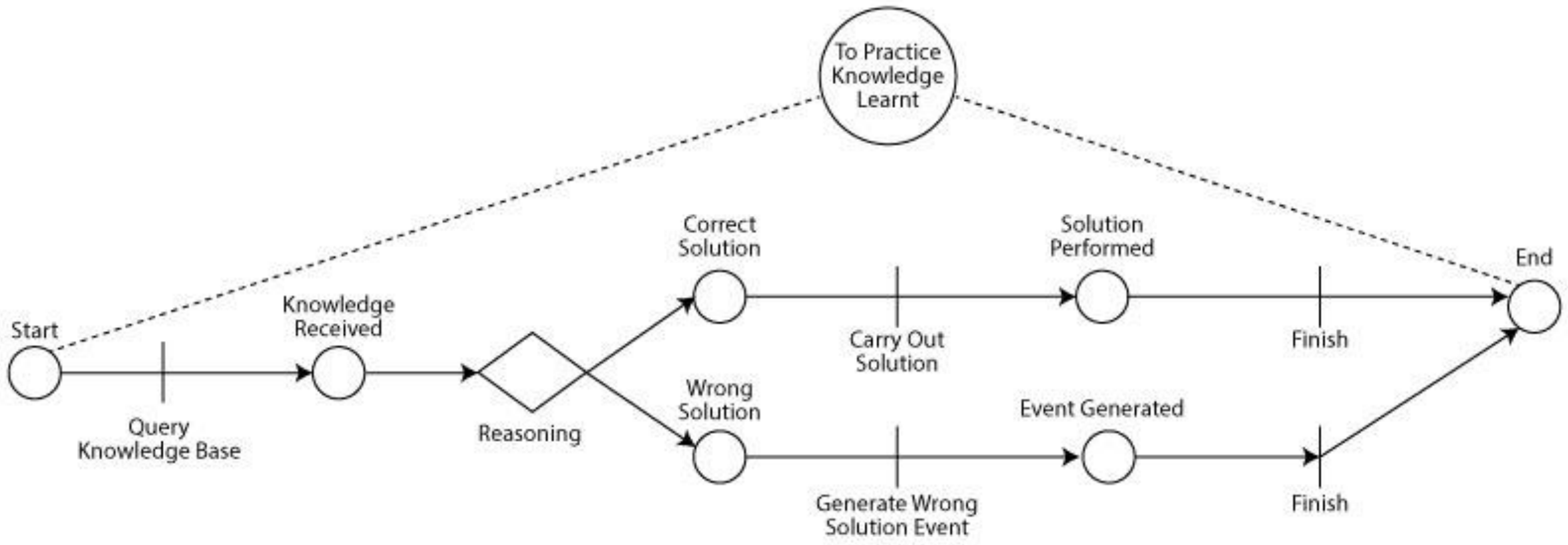

Figure 20 Practicability Reasoning Sub-Goal Net (Zeng, 2015)

***Persuasion Reasoning Sub-Goal Net***

The Persuasion Reasoning Sub-Goal shown in Fig, 21, is adapted from the Persuasive Reasoning Goal Net. In this sub-goal, the goal of the agent is to persuade the learner to improve their ability and motivation.

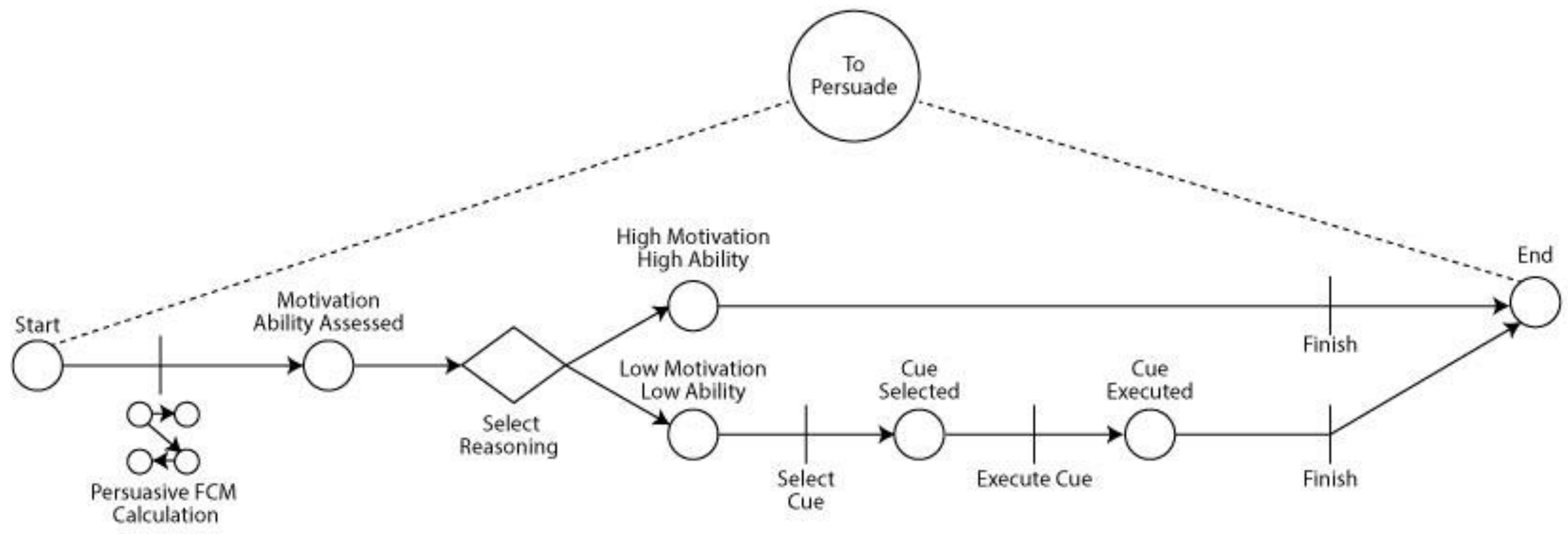

Figure 21 Persuasion Reasoning Sub-Goal Net (Zeng, 2015)

Similar to the Persuasive Reasoning Goal Net, the Persuasion Reasoning Sub-Goal starts with the calculation in the Persuasive FCM. Following that, learner's motivation and ability are assessed. The agent then selects a reasoning based on the level of motivation and ability by comparing the results to predetermine the baseline values. If both motivation and ability is high, the cycle ends and no further actions are required. If either motivation or ability is low, the agent will then select an appropriate persuasion cue that aims to improve the learner's motivation and ability and execute the cue. The Persuasion Reasoning Sub-Goal Net will end after the cues have been executed and return the next node in the main routine.

### 3.4.3 Quantitative Modelling of the Persuasive FCM

To facilitate the computational processes Zheng (2015) had suggested separating Persuasive FCM to form a Main FCM and the Sub FCM as shown in Fig. 22.

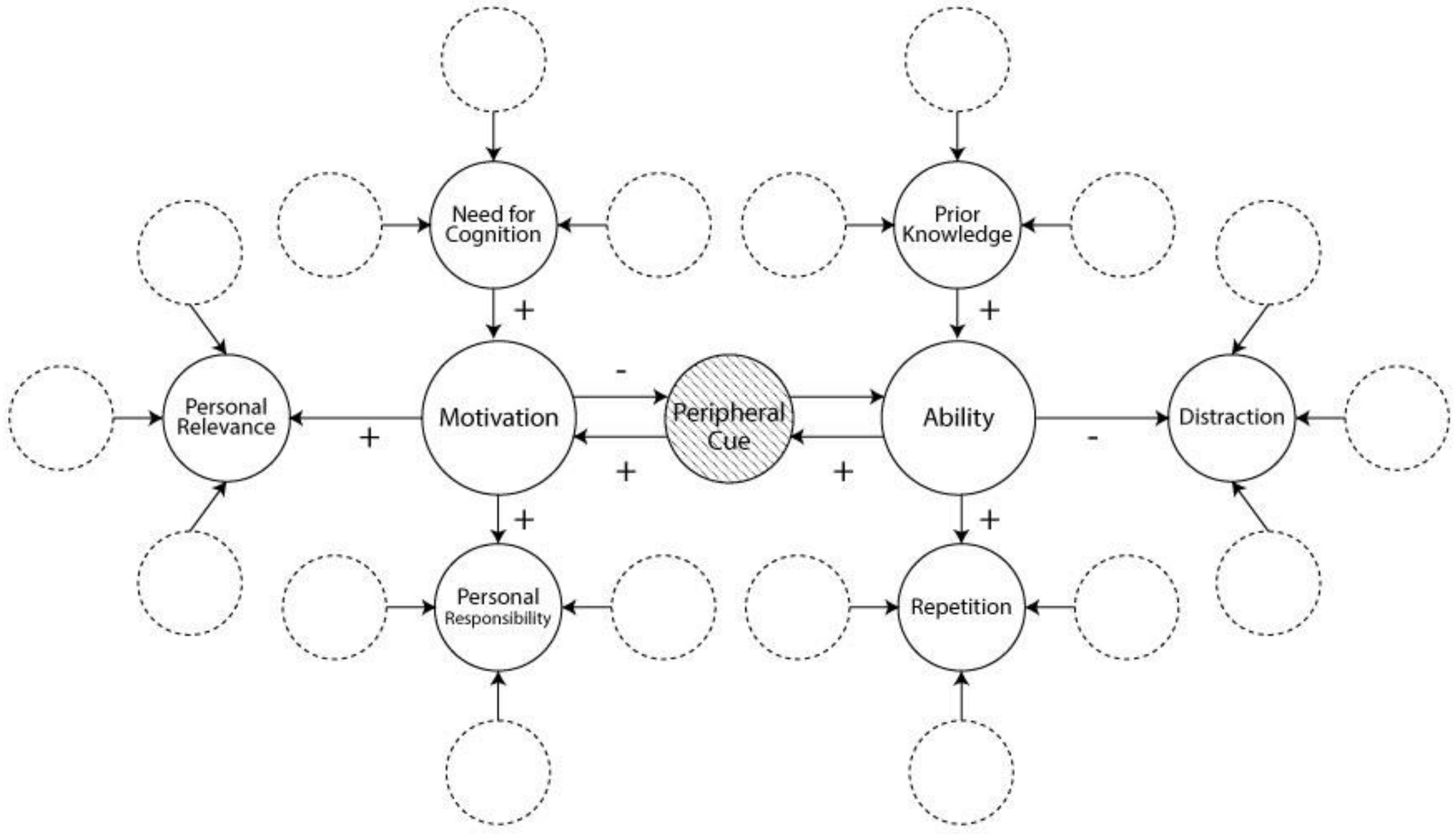

Figure 22 Remodelled Persuasive FCM (Zeng, 2015)

The Main FCM is formed by the stem nodes made up of the causal concepts namely, peripheral cue, as well as, motivation and ability which are connected to the factors that affect the input of motivation and ability forms the *stem nodes*. In addition to the Main FCM, the Sub FCM is connected to the *stem nodes*. Multiple are *leaf nodes* (dotted circles) are attached to the stem, each *leaf node* an event in the environment that has direct influence to the any of the input factors related to motivation and ability.

The improved Persuasive FCM model allows for greater flexibility and computational efficiency if future events in the environment were to be added, as these events could be included in the *leaf nodes*.

### 3.4.4 PTA System Architecture Design

To further close gap between the theoretical model and the implementation of the PTA, Zheng (2015), proposed an improved version of the PTA system architecture. The improved system architecture is more flexible as it is independent of the topics of learning. The embodiment of the PTA is not restricted and it can be deployed and implemented in other environments.

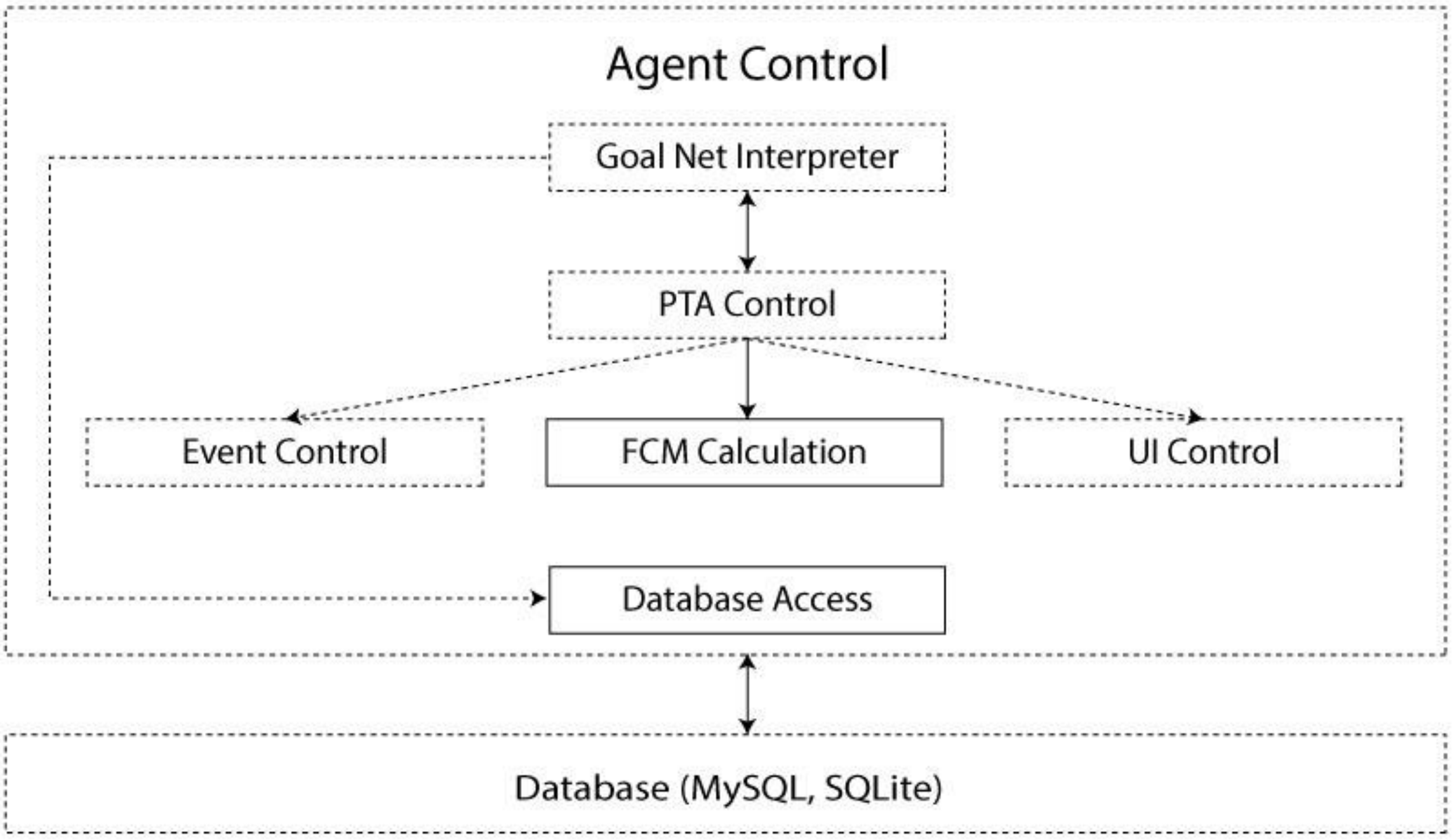

Figure 23 Improved PTA System Architecture Design (Zeng, 2015)

The following Table 8 shows the components in the PTA system architecture and its corresponding functionalities and roles.

Table 8 Functionalities and Roles of Components in PTA System Architecture

| Components | Functionality/Role |
|---|---|
| Goal Net Interpreter | Control the goal-oriented behaviours of the PTA. |
| PTA Control | Implement functions, coordinate PTA goal-oriented behaviours. |
| Event Control | Deals with the creation, logging, processing and removal. |
| FCM Calculation | Performs FCM computational processes. |
| UI Control | Control the interface elements in the PTA environment. |
| Data Access | Connects to the database and performs queries. |

***Goal Net Interpreter***

At the top level of the PTA system architecture is the Goal Net Interpreter. The purpose of the Goal Net Interpreter is to control the goal-oriented behaviours of the PTA through its reasoning cycle. The Goal Net models in the PTA are created in the Goal Net Designer (Han, Zhiqi, & Chunyan, 2007) and stored in the Goal Net database.

The main purpose of the Goal Net Interpreter is to load the saved Goal Net models from the Goal net database and interpret the Goal Net models dynamically in real-time. The Goal Net Interpreter also determines the tasks the PTA was to perform so as to pursue it goal, during the state transition in the main route of the PTA Goal Net. After the tasks are determined, the Goal Net Interpreter then updates the PTA Control to perform the tasks assigned.

***PTA Control***

The role of the PTA control is to implement the functionalities in the PTA. The PTA Control performs the task functions shown in Table 9 by acting upon the instructions from the Goal Net Interpreter. This is done by breaking down the tasks into tasks and sub-tasks and sends them to other components by calling on the corresponding modules in these components, so that the components are able to work together towards the goal of the PTA.

Table 9 Task Functions to be implemented by the PTA Control

| **Goal Net Model** | **Functions to be Implemented** | |
|---|---|---|
| Main Routine | DetectEvent<br>InterpretEvent | SelectReasoning<br>Finish |
| To Learn Knowledge | RequireTeaching<br>CheckResponse<br>InitializeTeaching<br>Acquire Knowledge | SaveKnowledge<br>GenerateRejectionEvent<br>Finish |
| To Practice Knowledge | QueryKB<br>Reasoning<br>CarryOutSol | GenerateWrongSolEvent<br>Finish |
| To Persuade | FCMCalculation<br>CheckMotAbi<br>SelectCue | ExecuteCue<br>Finish |

For the PTA to realize the three reasoning cycles in the PTA definition, the PTA Control has to implement the task functions that are associated with the transitions in the PTA Goal Net models. The following are the Goal Net models in the PTA and their associated functions to be implemented by the PTA Control.

***Event Control***

The Event Control corresponds to the **Event Tracker** *ET* in the PTA definition. It handles the lifetime of an event such as the creation, logging, processing and removal of events. The PTA Control also uses the Event Control to handle events in the PTA Goal Net model.

The Event Control keeps track of the events by checking for new events at predetermined intervals and log newly generated events in the Event Log in the Event Control. The Event Control prioritizes the events and determines the processing sequence of the events when the checking is performed as well as decides on the number of interaction events to process in batches.

The Table 10 below shows the event category and the examples of the few types of events that can be tracked in the PTA environment.

Table 10 Events Tracked in the PTA Environment

| **Event Category** | **Event** | **Example** |
| --- | --- | --- |
| Learning Behaviour | Dialogue event | • The learner refuses to teach the PTA during dialogue.<br>• The student is distracted by a non-player character. |
| | Location event | • The learner has arrived at the designated location. |
| | Time event | • The learner has been inactive for more than 5 minutes. |
| Learning Achievements | Collection of rewards | • The learner has been awarded a reward for completing the mission. |
| | Completion of mission | • The learner has completed the mission. |
| Knowledge Data | Make an error | • The learner make an error in teaching the PTA. |
| | Feedback event during teaching | • Generating the correct solution from the knowledge acquired from the learner. |

***FCM Calculation***

The FCM Calculation component processes the computational aspects in the Persuasion Reasoning in the PTA. The FCM model is loaded from the database and stored in the FCM Calculation component when the PTA is initialized, the PTA Control actives the FCM Calculation in the Persuasion Reasoning Cycle to compute the motivation and ability values.

***UI Control***

The UI Control component provides the module that enables the agent to perform the actions as defined in the PTA definition in the environment. This includes the embodiment of the PTA in the environment either as a 2D or 3D avatar. The UI Control allows for the learner user the graphical interface to control the PTA in the environment.

***Database Access***

The Database Access component allows for the other components to access the database. Multiple databases may be accessed through the Database Access component. For example the Goal Net Interpreter is able to access the state and transitions in the Goal Net database. Another example is when the FCM Calculation component loads the FCM model from the FCM database.

### 3.4.5 Discussion on the Improvements to the PTA

To summarise the improvements to the PTA, the reasoning component to the PTA is fully integrated with the combination *Persuasion Reasoning*, *Teachability Reasoning* and *Practicability*. The *Persuasion Reasoning* is simplified with a quantitative model allowing the PTA to be used in different contexts. This allows greater flexibility for the PTA to be deployed in other applications platforms in the future.

The improved system architecture provides better controls for the PTA to handle multiple functionalities and prepare the PTA for deployment in the applications such as the 3D game engine, which will be discussed further in the following section.

## 3.5 Implementation of PTA in Virtual Singapura (VS) Saga

In this section, the implementation of the PTA in a 3D game engine will be discussed in detail.

Firstly, the Unity 3D game engine in which the PTA is deployed will be introduced, followed by a walk-through of the Virtual Singapura (VS) Saga storyline. The walk-though of VS will serve as a guide on how the PTA will be used by the learner to simulate learning-by-teaching with the PTA.

### 3.5.1 Developing VS Saga using Unity 3D Game Engine

The VS Saga is developed in the Unity 3D game engine, which is an easy to use, multi-platform game engine. This allows for applications to be created for mobile devices, PC, consoles and websites. The PTA is implemented in the Unity 3D game engine using the PTA model and system architecture.

### 3.5.2 Virtual Singapura (VS) Saga Storyline

The VS Saga 3D virtual learning environment (VLE) is developed for lower secondary level science learning. Fig. 24 shows the screenshot of the main navigation interface of the VS saga.

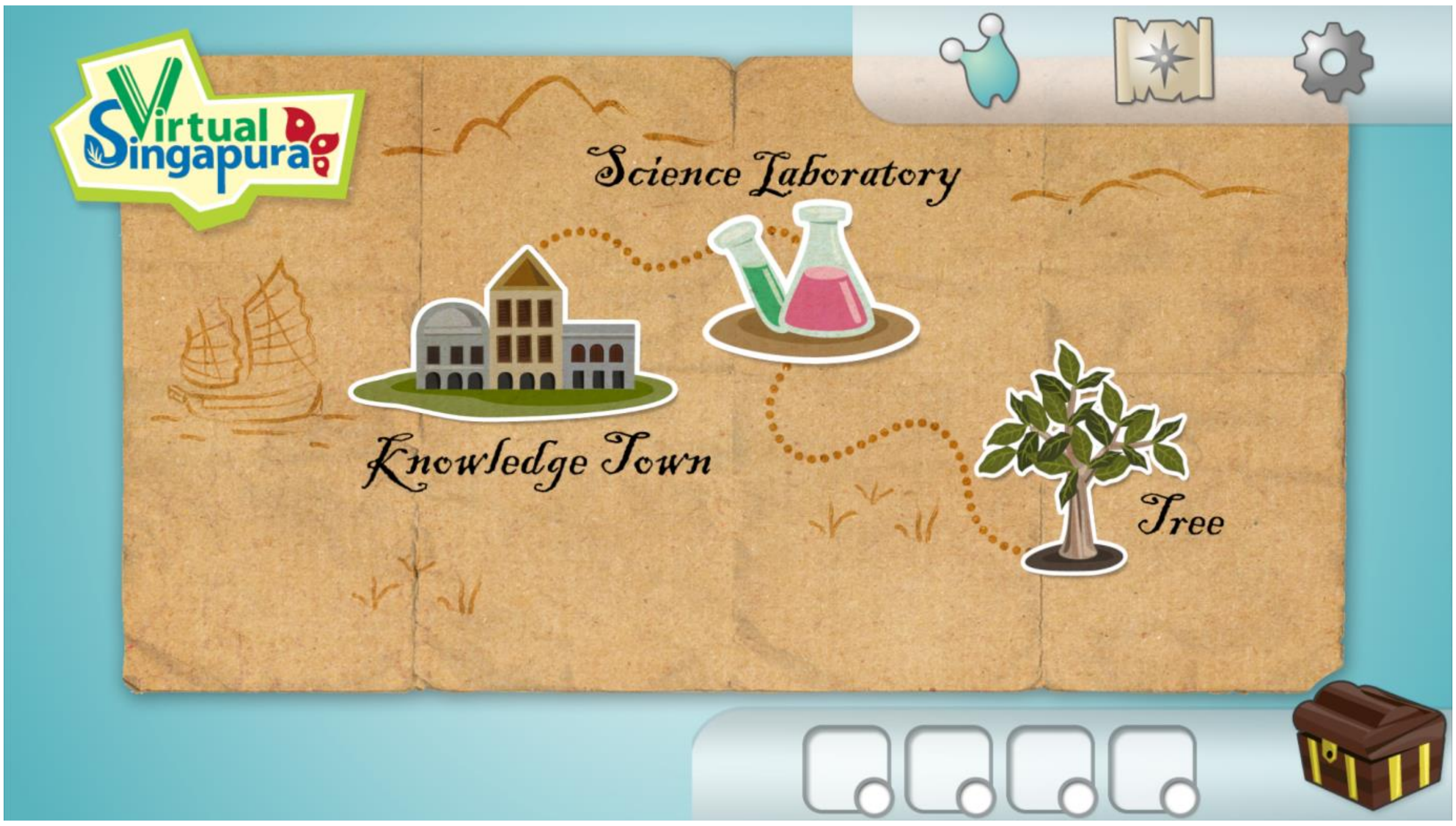

Figure 24 Main navigation in VS Saga

Unlike the previous version of VS, the VS Saga is a three part learning journey that follows the storyline of the avatar of the player who is the learner helping the town dwellers on an island by completing a series of missions. Through the completion of these missions, the learner will be able to learn science concepts. Towards the end of the mission they will be able to teach the PTA that is embodied by a water molecule. Having taught the water molecule PTA, the water molecule will be able to travel into a sick banana plant, thus saving it from dying.

The VS Saga consists of a three-part storyline, namely, the knowledge town, the science laboratory and the tree scenes. Details on each part of the storyline will be introduced in the following section.

***Knowledge Town***

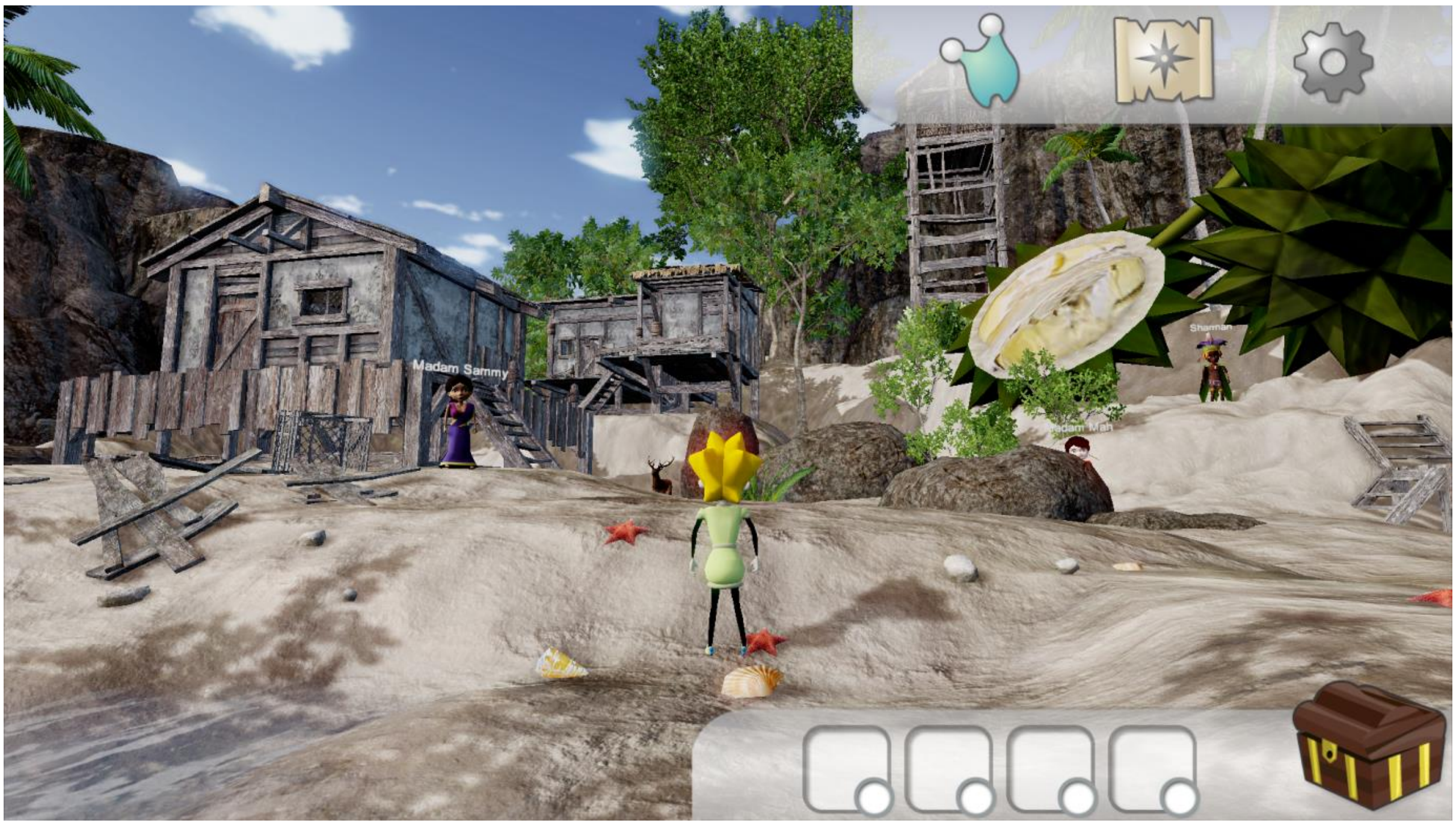

Figure 25 Knowledge town in VS Saga

The knowledge town is where the learner takes on the role as the player avatar that has just landed on the island, as shown in Fig.25. The learner will meet the water molecule PTA which then informs him or her about the need to learn more about science topics such as osmosis an diffusion by visiting exploring the island to meet other non-player characters (NPC)s. In the process of talking to different characters the learner will be able pick up information on the transportation of water molecules within a plant.

Throughout the knowledge town, there are also NPCs that are not related to learning. These NPCs are intentionally included in the scene to distract the learner players from their learning. Table 12 shows the sequence of storyline and interaction between the learner player avatar and the NPCs.

The objective of the PTA is to encourage the learner player to stay on their learning tasks and complete their mission by talking to NPCs regarding information related to their learning mission. In order to achieve to this, the PTA will have to go through its *Persuasive Reasoning* to understand the learner player's motivation and ability, thus allowing it to apply appropriate persuasion cues. The persuasion cues given by the PTA are aimed at influencing the learner player's decision to stay on track in completing their learning mission.

Table 11 Screenshot, Related Storyline and Non-Player Characters in Knowledge Town

| **Screenshot** | **Storyline and Non-Player Characters** |
|---|---|
| 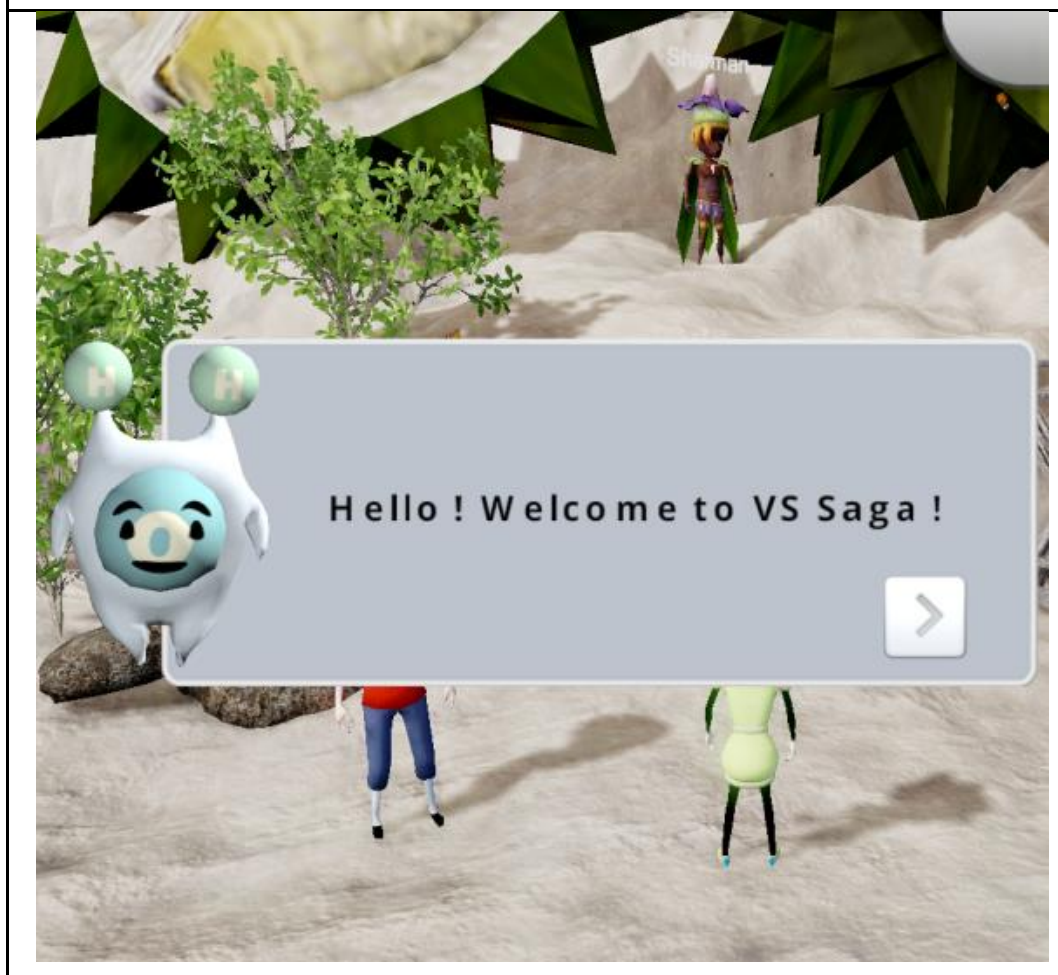  | 1. Meeting the Water Molecule PTA.<br>• Water molecule welcomes learner player.<br>• Water molecule introduces itself and the mission in the VS saga. |
| 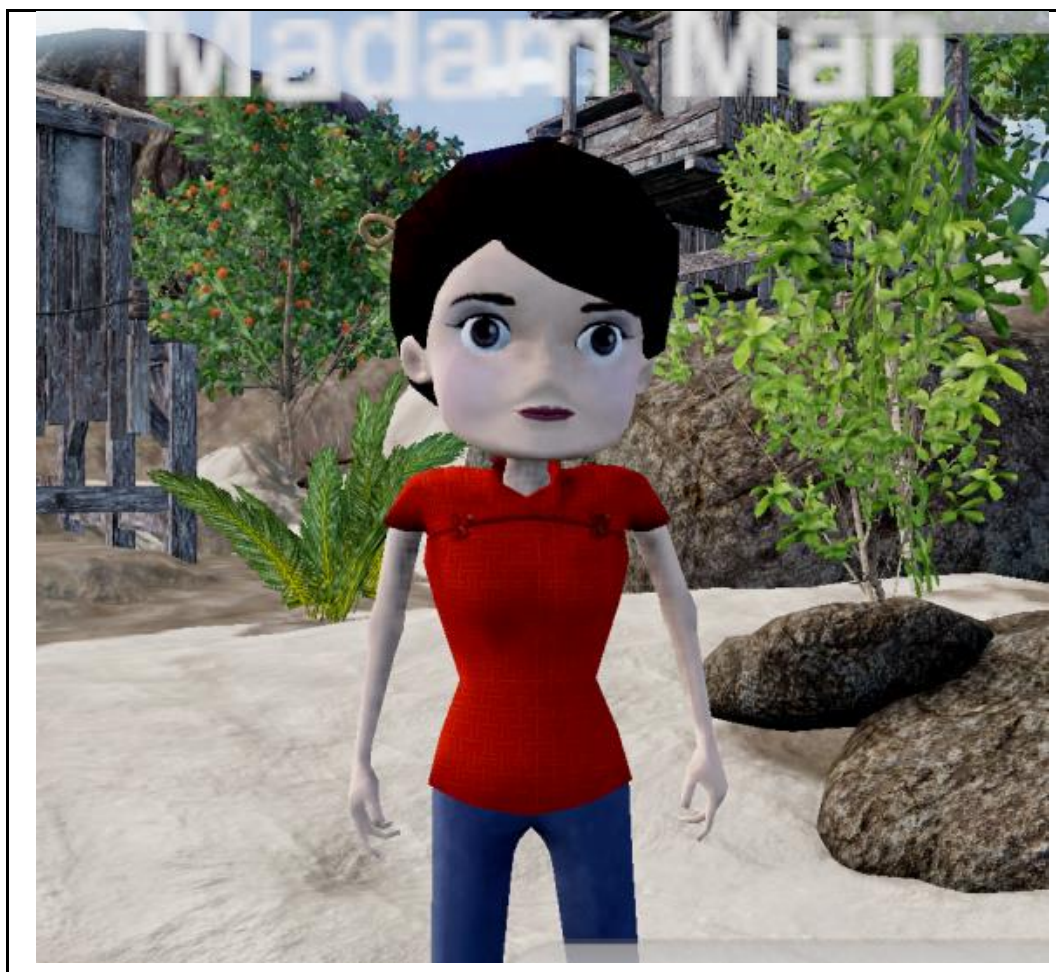<br> | 2. Talk to Madam Mah on the beach.<br>• Madam Mah welcomes the learner player.<br>• Learner player accepts the mission to meet the Sharman near the giant durian. |
| 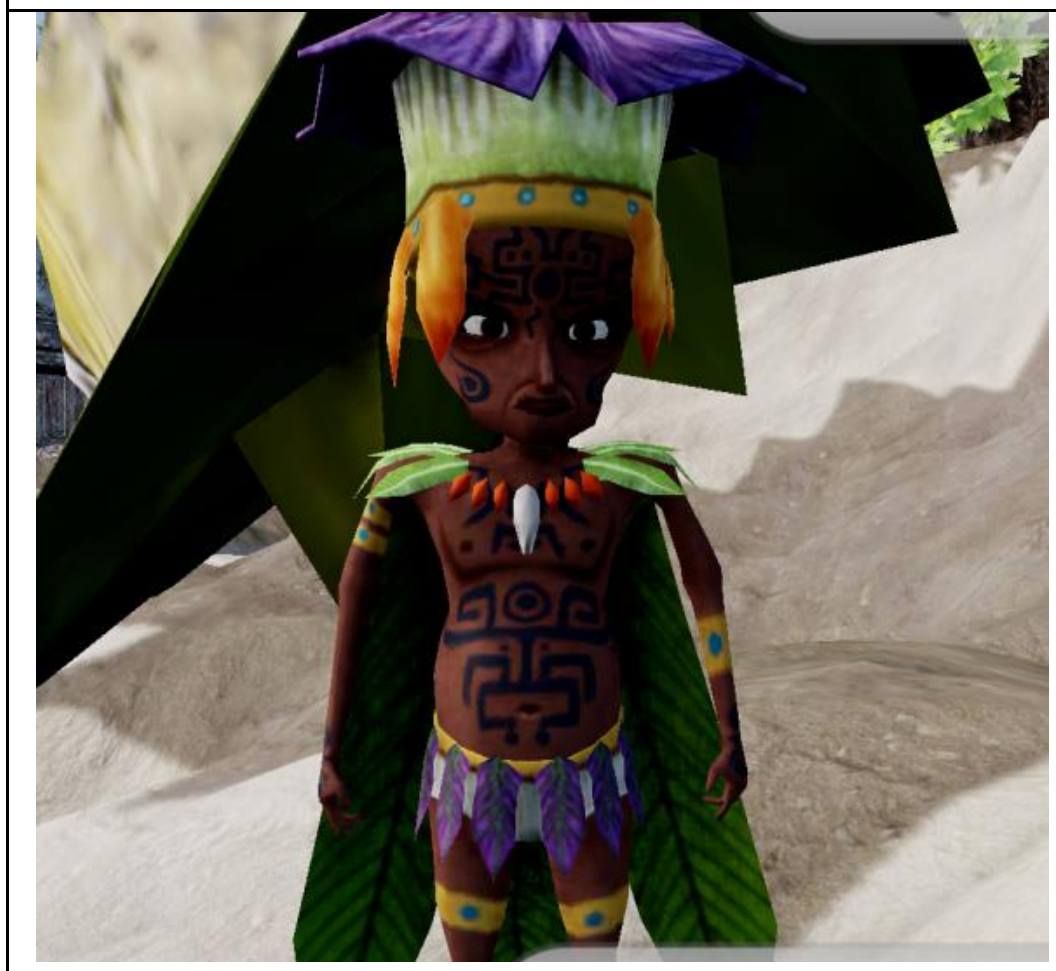 | 3. Talk to the Sharman near the giant durian.<br>• Learn the concept of diffusion from the Sharman.<br>• Learner player accepts the mission to get the perfume from Madam Sammy. Learner player refuses to learn the concept of diffusion. |
| 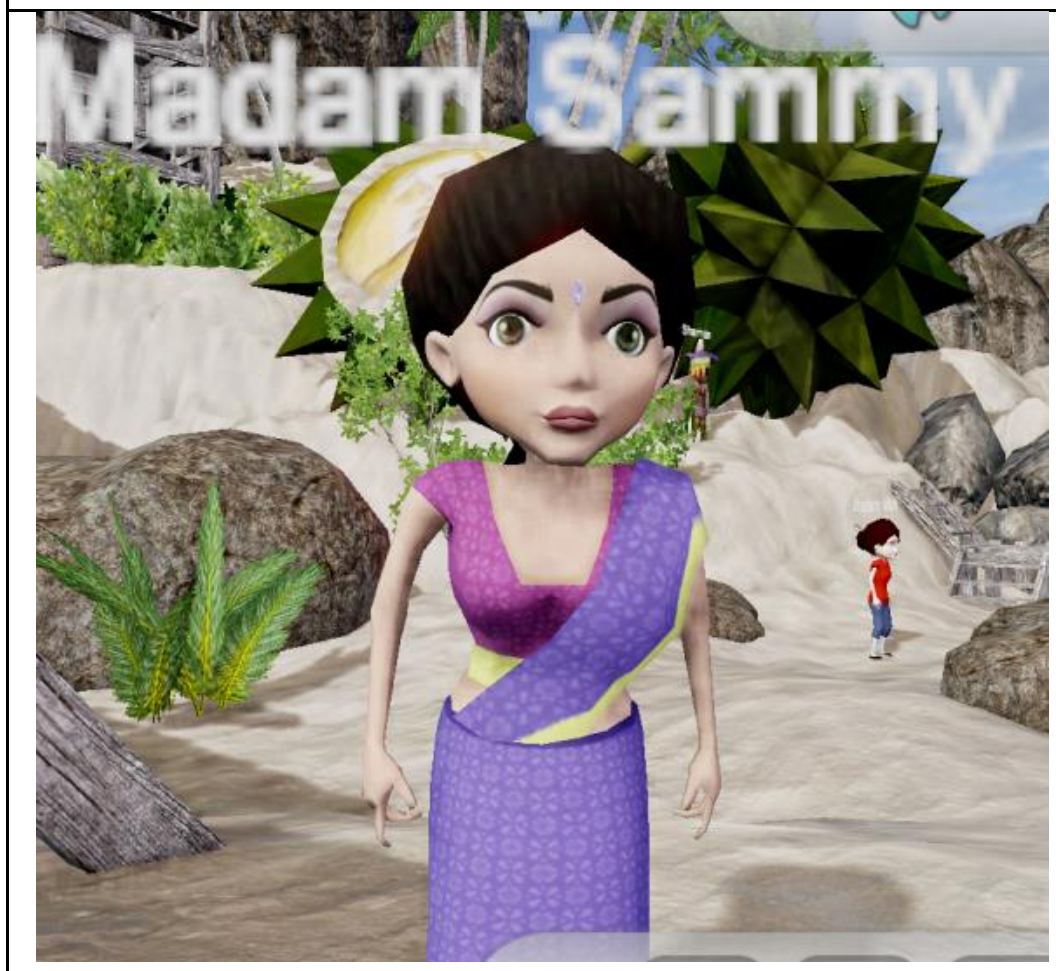<br> | 4. Talk to Madam Sammy near the stilt houses on the beach.<br>• Madam Sammy reminds the learner player to learn diffusion if not done so.<br>• Receive the potion item.<br>• Learner player receives the new mission to deliver potion to the mayor near the tree stump. |
| 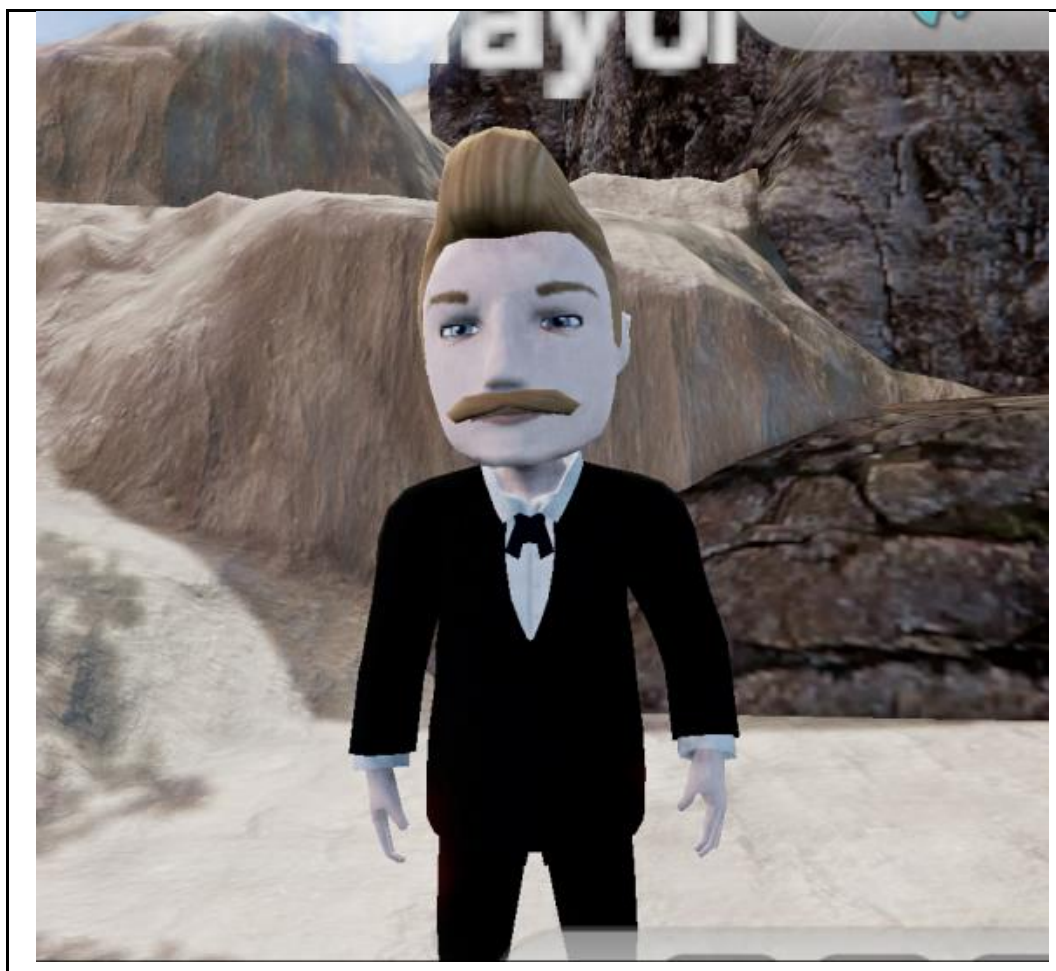 | 5. Talk to the Mayor near the coconut tree.<br>• Learn osmosis from the mayor<br>• The mayor invites the learner player to the science laboratory to practice the science concepts that they have learnt.<br>Learner player refuses to learn about osmosis. |
| 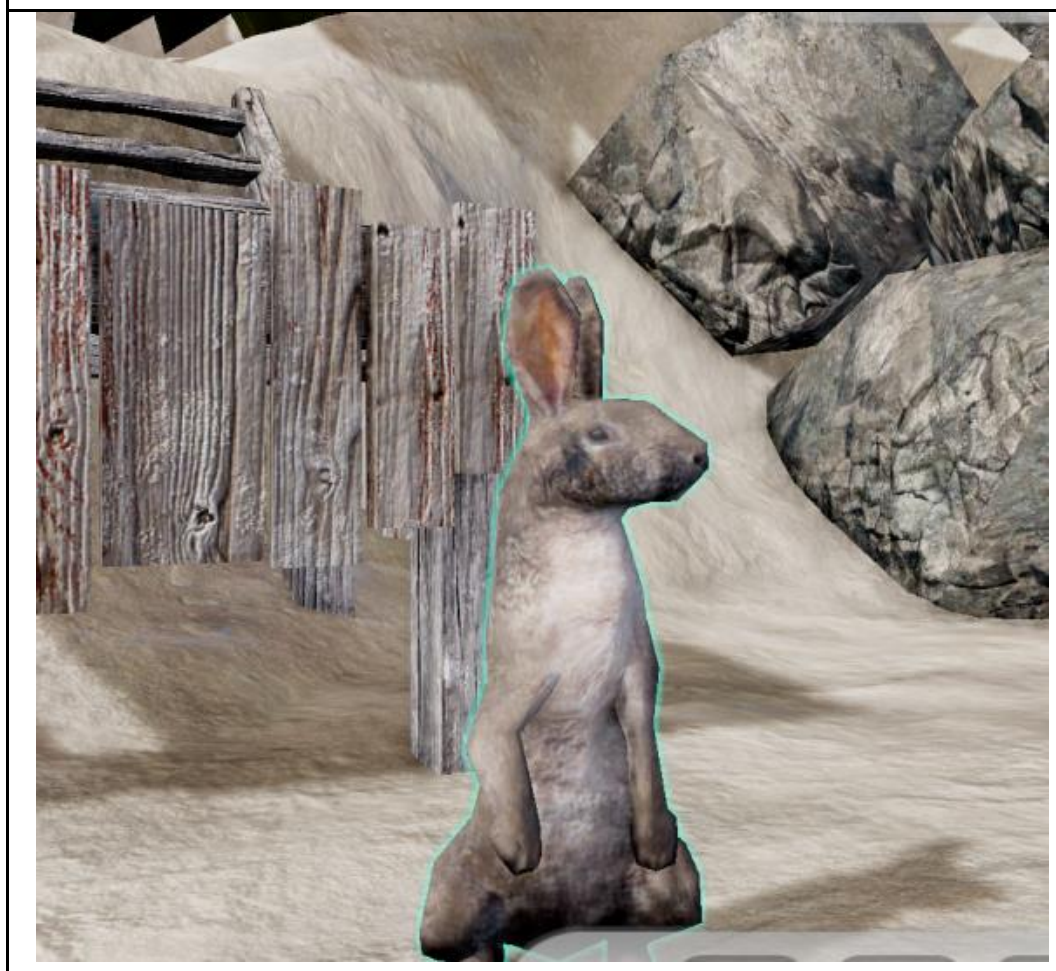 | 6. Talk to animals on the island.<br>• Animals are NPC scattered around the main island and serves as distracters. |

***Science Laboratory***

After learning the related information from the NPCs in the knowledge town, the learner player proceeds to the science laboratory as shown in Fig. 26. The science laboratory allows the learner to practice on the knowledge and information that the learner has acquired in the knowledge town.

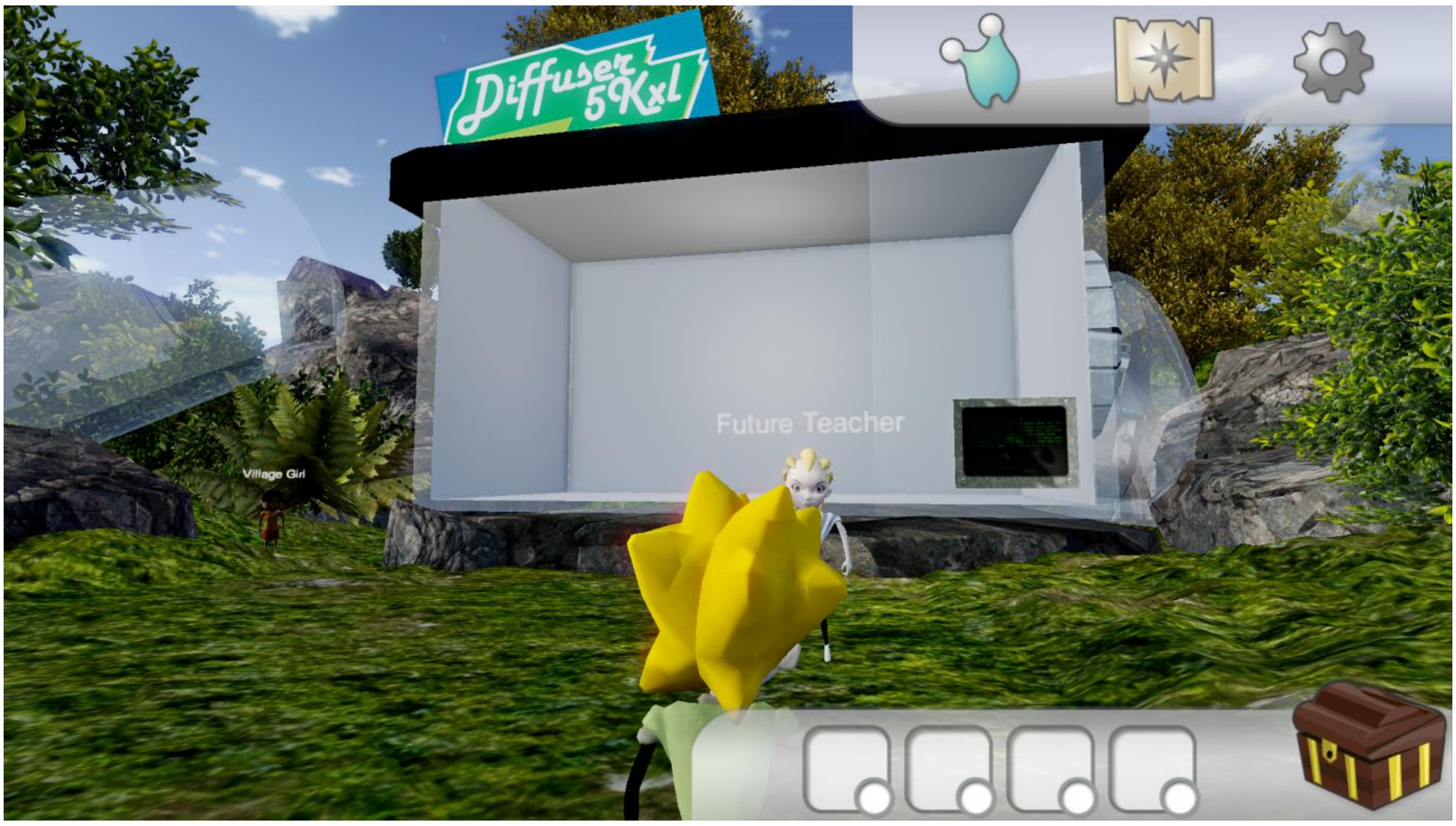

Figure 26 Science laboratory in VS Saga

The learn player can interact with new NPCs at the science laboratory. There is also a diffuser where the learner player can simulate the movement of the molecule particles during the osmosis and diffusion process through a series of experiments. This in turn helps the learner play to further retain their knowledge that they have learnt in the knowledge island scene. Table 12 shows the sequence in which the learner player avatar has to complete in this scene.

Table 12 Screenshot, Related Storyline and Non-Player Characters in Science Laboratory

| **Screenshot** | **Storyline and Non-Player Characters** |
|---|---|
| 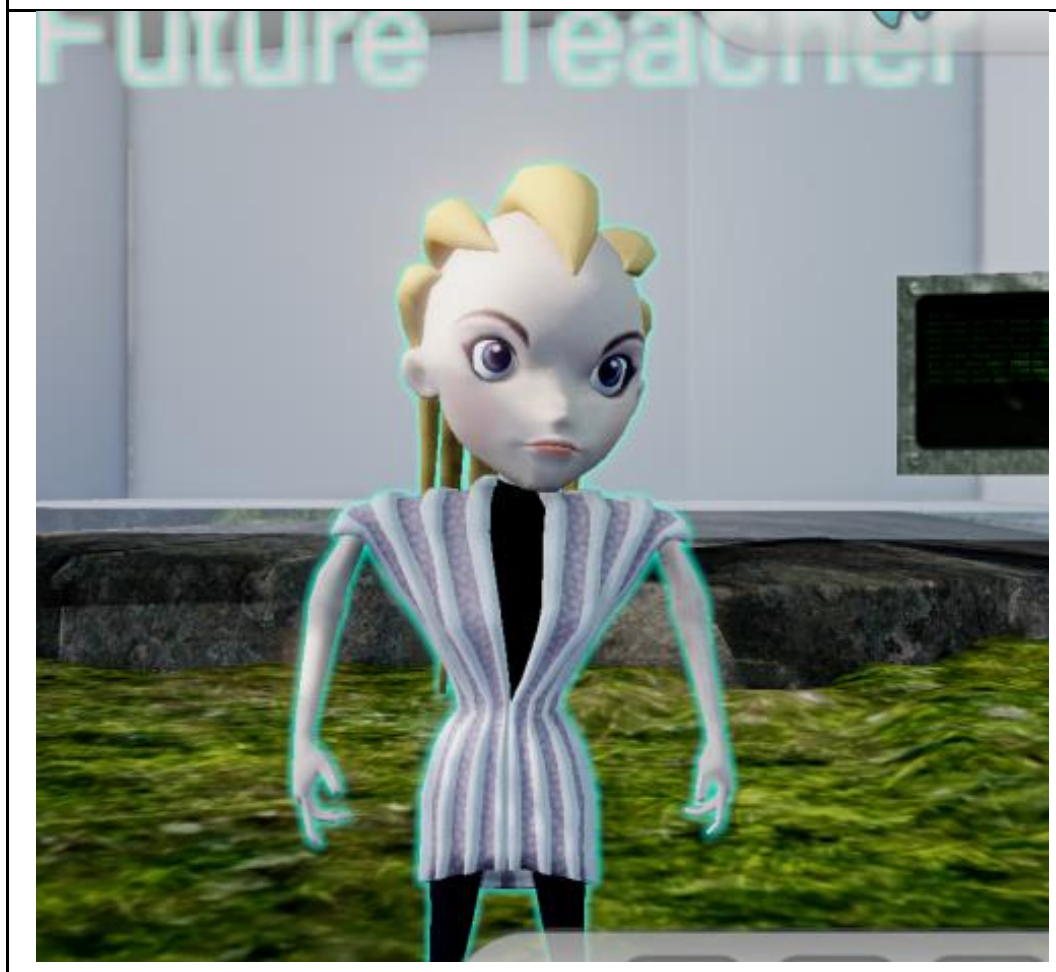 | 1. Meeting the future teacher.<br>• Learn to conduct simulation experiments.<br>• Learner player accepts task to conduct experiments.<br>• Learner player refuses to conduct experiments. |
| 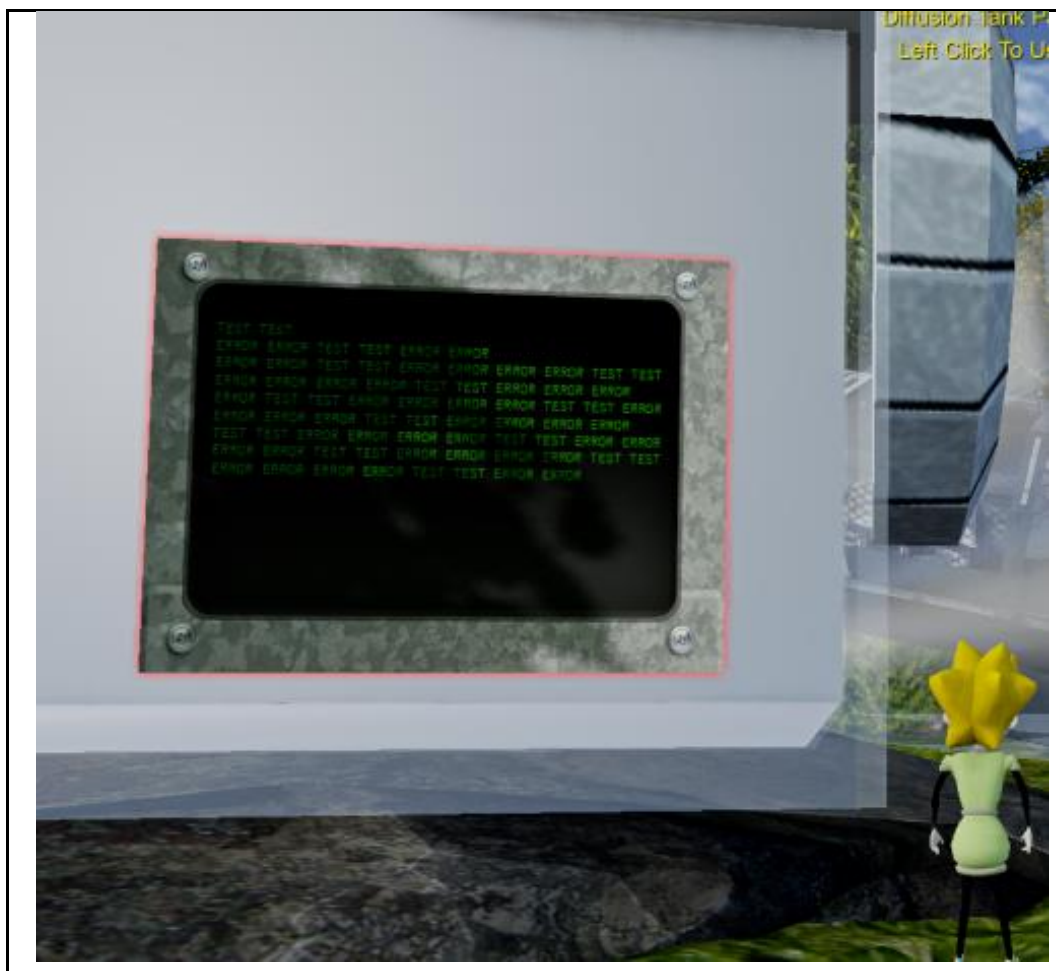 | 2. Learner player proceeds to the “diffuser 5K” simulation tank.<br>• Learner player starts the experiment by clicking on the control panel. |
| 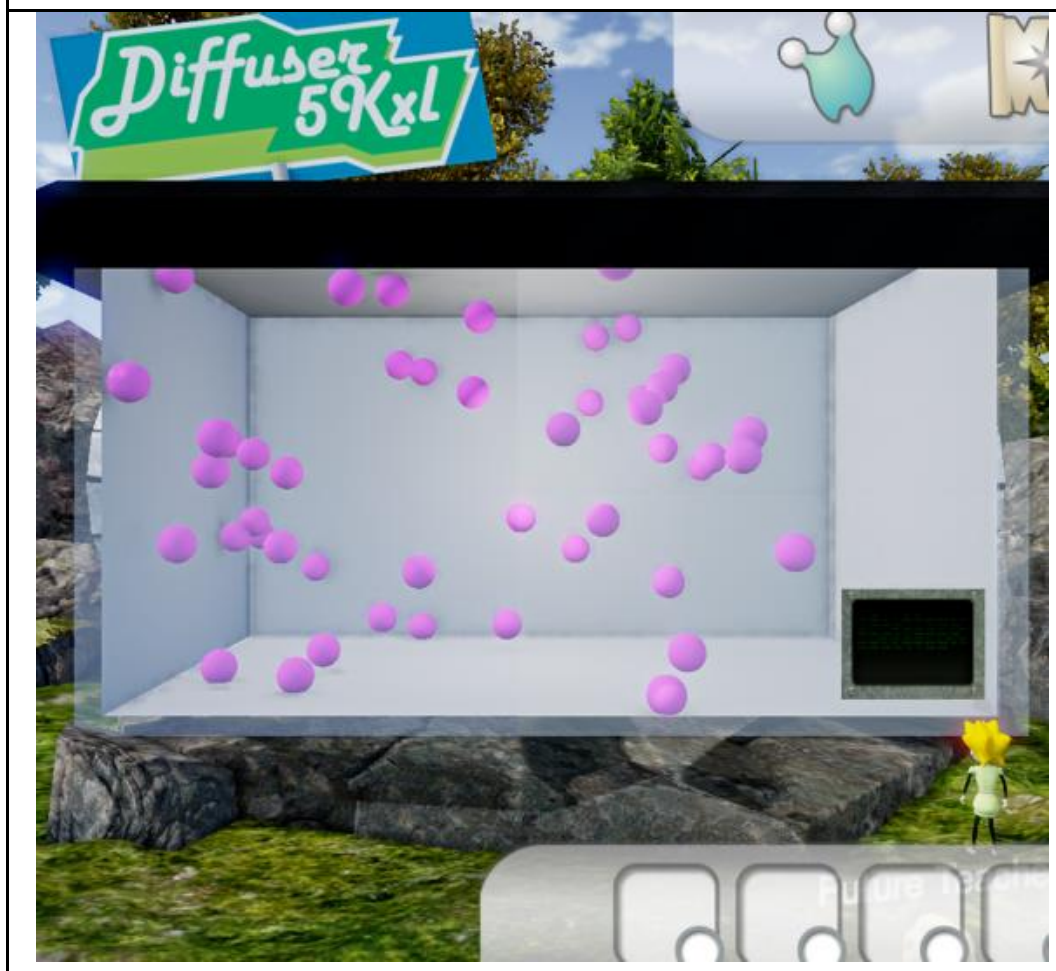  | 3. Learner player conducts the diffusion simulation experiment. |
| 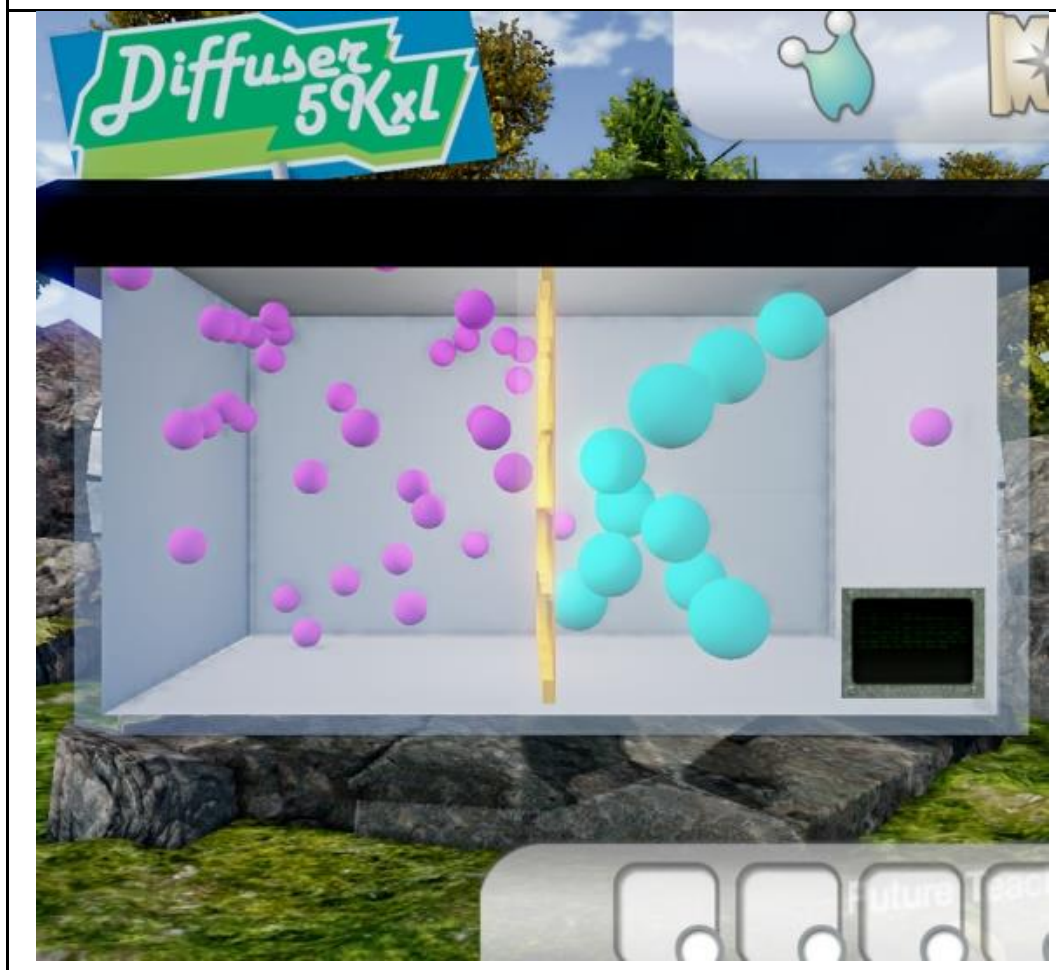  | 4. Learner player conducts the osmosis simulation experiment.<br>• Future teacher instructs the learner player to proceed to the next tree scene to teach the PTA. |
| 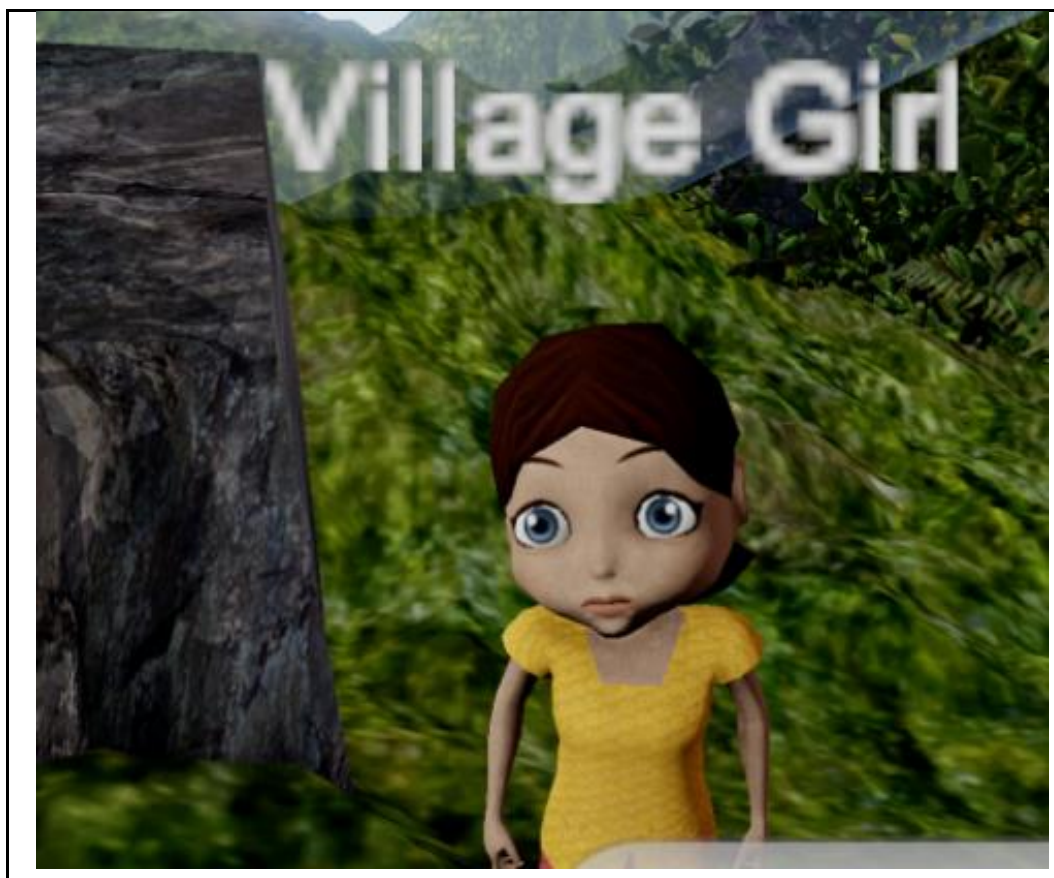  | 5. Talk to Village Girl.<br>• Village characters such as the Village Girl in the Science Laboratory are distracters. |

***Tree***

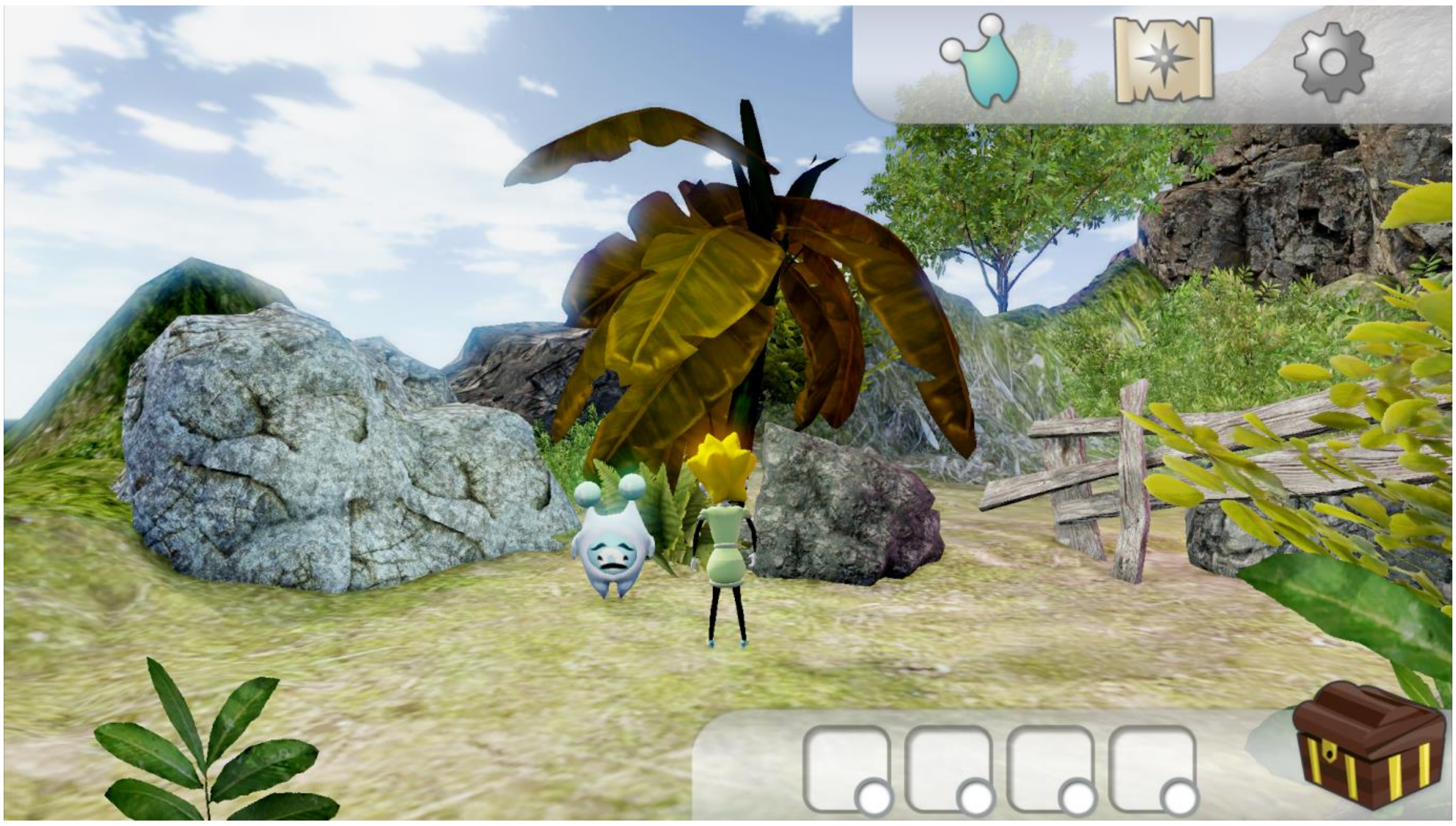

Figure 27 Save the banana plant mission in the tree scene

The last scene of the VS saga is the tree as shown in the screenshot in Fig 27. This is where the learner player teaches the water molecule PTA about the concepts of osmosis and diffusion. In this scene, learner player has to practice the knowledge he or she has learnt from the knowledge island and the science laboratory by teaching the PTA through a concept map.

Table 13 shows the screenshot, the related storyline and NPCs that interacts with the learner player in this scene.

Table 13 Screenshot, Related Storyline and Non-Player Characters in the Tree scene

<table>
<tr><th>Screenshot</th><th>Storyline and Non-Player Characters</th></tr>
<tr><td>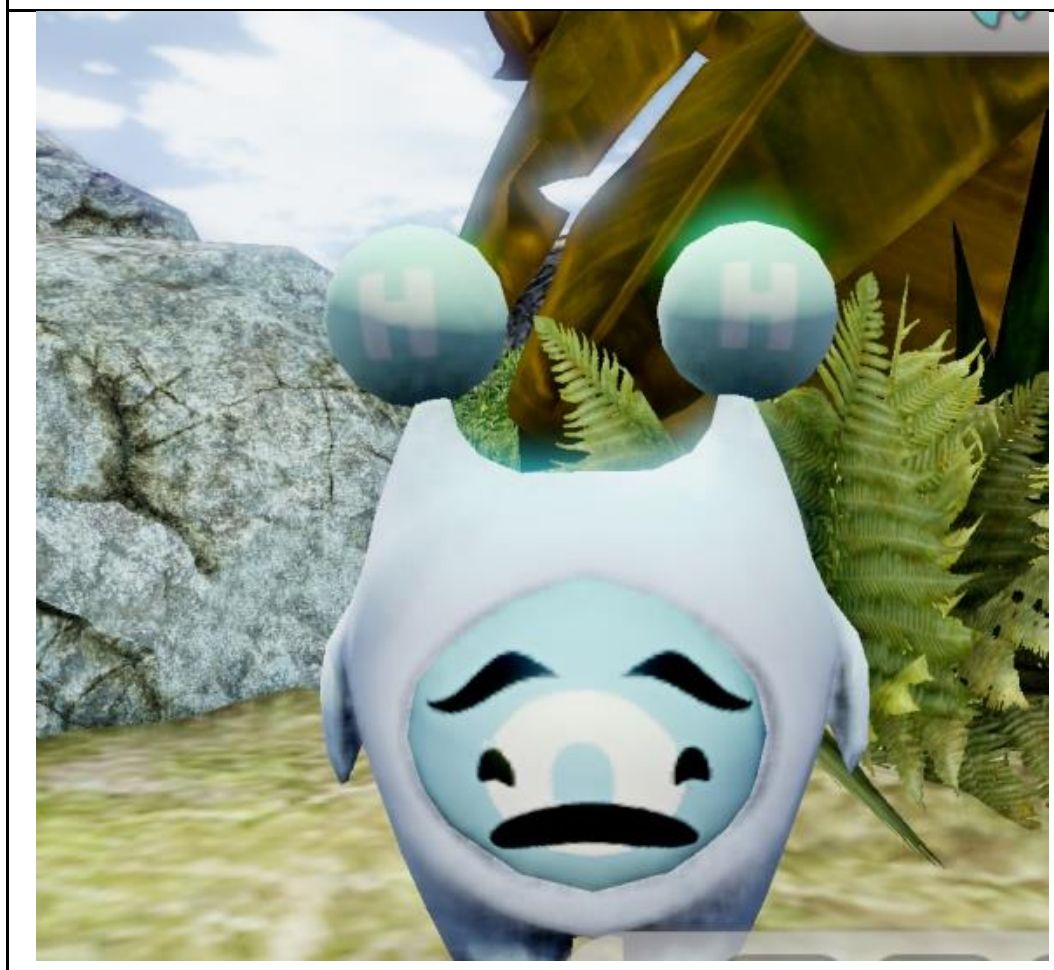</td><td>1. Learner player meets the sad PTA water molecule near a dying banana plant.</td></tr>
<tr><td>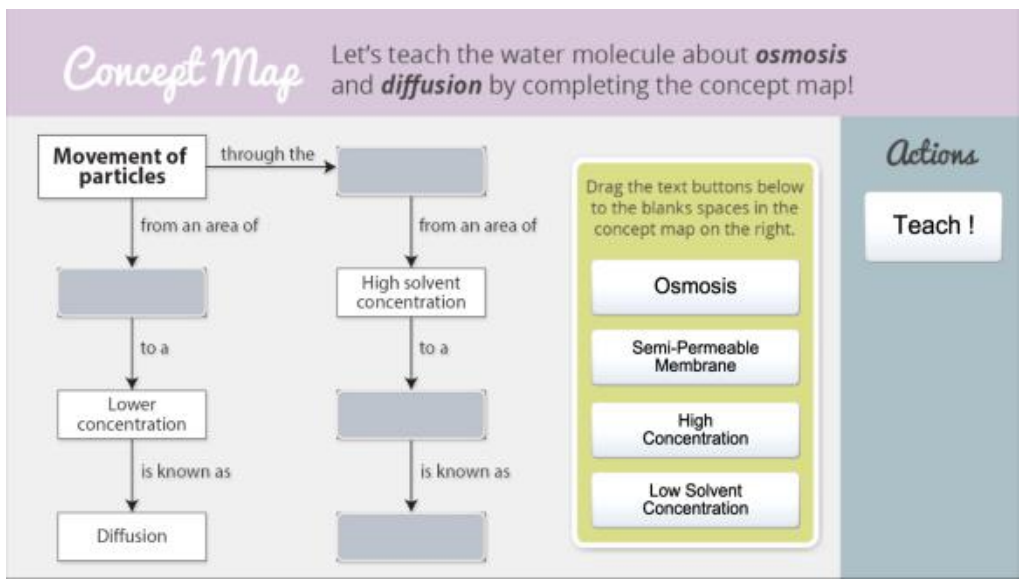
</td><td>2. Learner player is asked to teach the PTA water molecule.<br>• Learner player agrees to teach the PTA water molecule by completing the concept map.<br>• Learner player declines to teach the PTA water molecule.</td></tr>
<tr><td>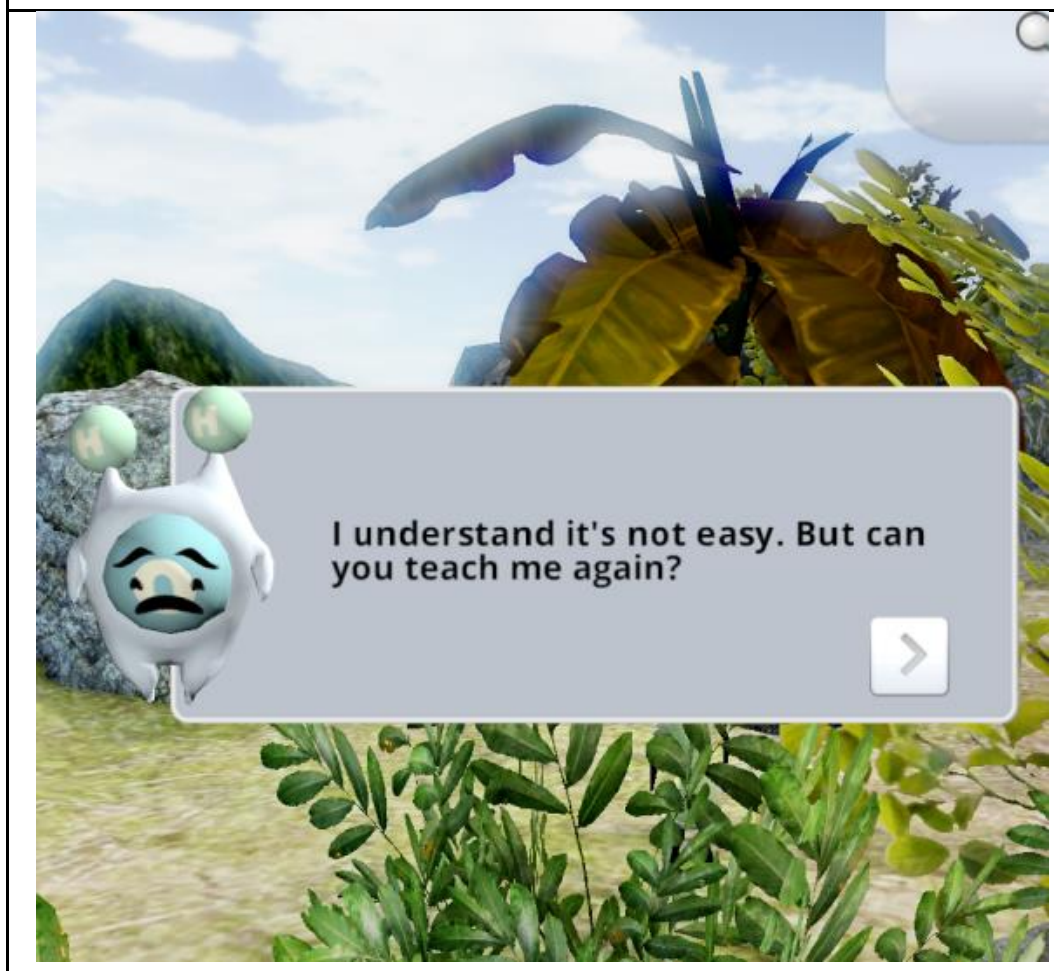
</td><td>3. The PTA reasons on the knowledge taught by the learner player.<br>• The PTA water molecule enters the banana plant through the roots and the banana plant is revitalised<br>• The PTA water molecule fails to enter the roots and ask the learner player to teach again.</td></tr>
</table>

## 3.6 PTA in VS Saga

The PTA has several controls that manage its task functionalities. In this section, the functionalities in the PTA Control, UI Control and Event Control will be discussed according to the context in VS saga.

### 3.6.1 PTA Control

***Main Routine***

When the Main Routine is executed, the PTA control activates the Event Control every 5 seconds so as to check for events. If one or more events are detected, the Main Routine decides on the events that are to be processed first in the current cycle and determines the reasoning cycle the events should go through.

***Persuasion Reasoning***

If the Persuasion Reasoning Sub-Goal Net is activated the FCM calculation will determine the ability and motivation based on the events in VS saga. The values will be compared with the pre-determined baseline values to evaluate if the motivation or the ability levels are low. If these values are lower than the baseline values persuasion cues will be activated and executed through the UI controls. This is when the PTA control calls on the UI control to display the persuasion cues in the VS saga.

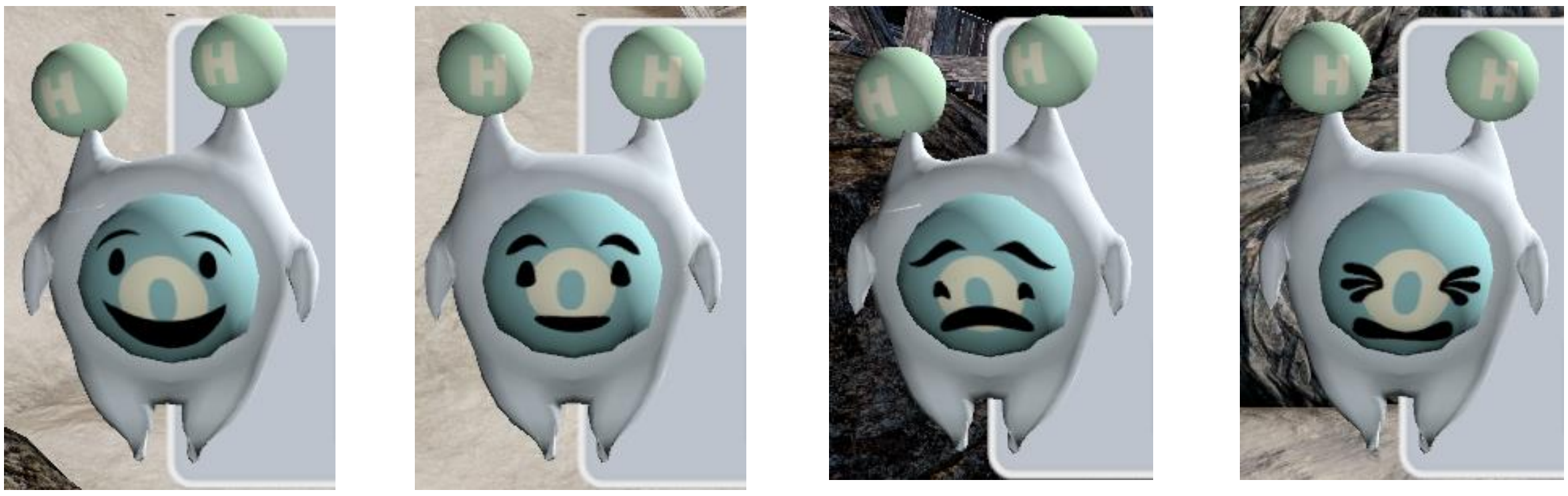

Happy Neutral Sad Shocked

Figure 28 Facial expressions and emotion persuasion cues in PTA

Persuasion cues in VS saga includes facial expressions showing affect or emotions in the PTA water molecule NPC as shown in Fig. 28. As well as dialogue speech displays such as hints from expert source or advice from attractive source that adheres to the ELM persuasion theory. An example of the PTA emotion and speech persuasion cue is shown in Fig. 29.

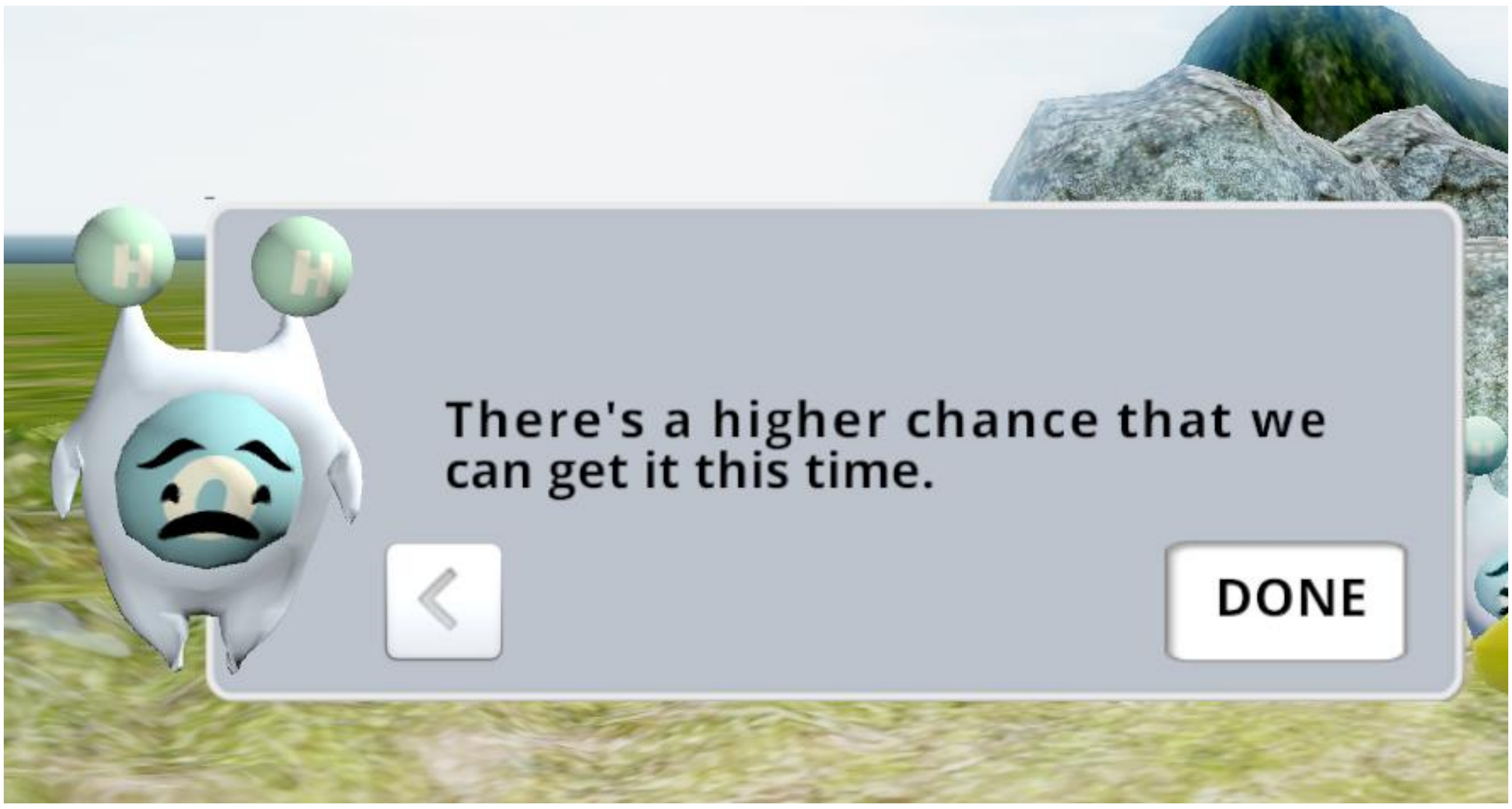

Figure 29 PTA emotion and speech persuasion cue displayed

***Teachability Reasoning***

The teaching process starts when the PTA requests the learner to teach itself through dialogue. Once the request has been accepted by the learner, the PTA control will call the UI control to active the concept map, a graphical tool that helps students organise and visualise relationships between concepts in knowledge of a subject. Fig. 30 shows the screenshot in which the PTA requests the learner for teaching.

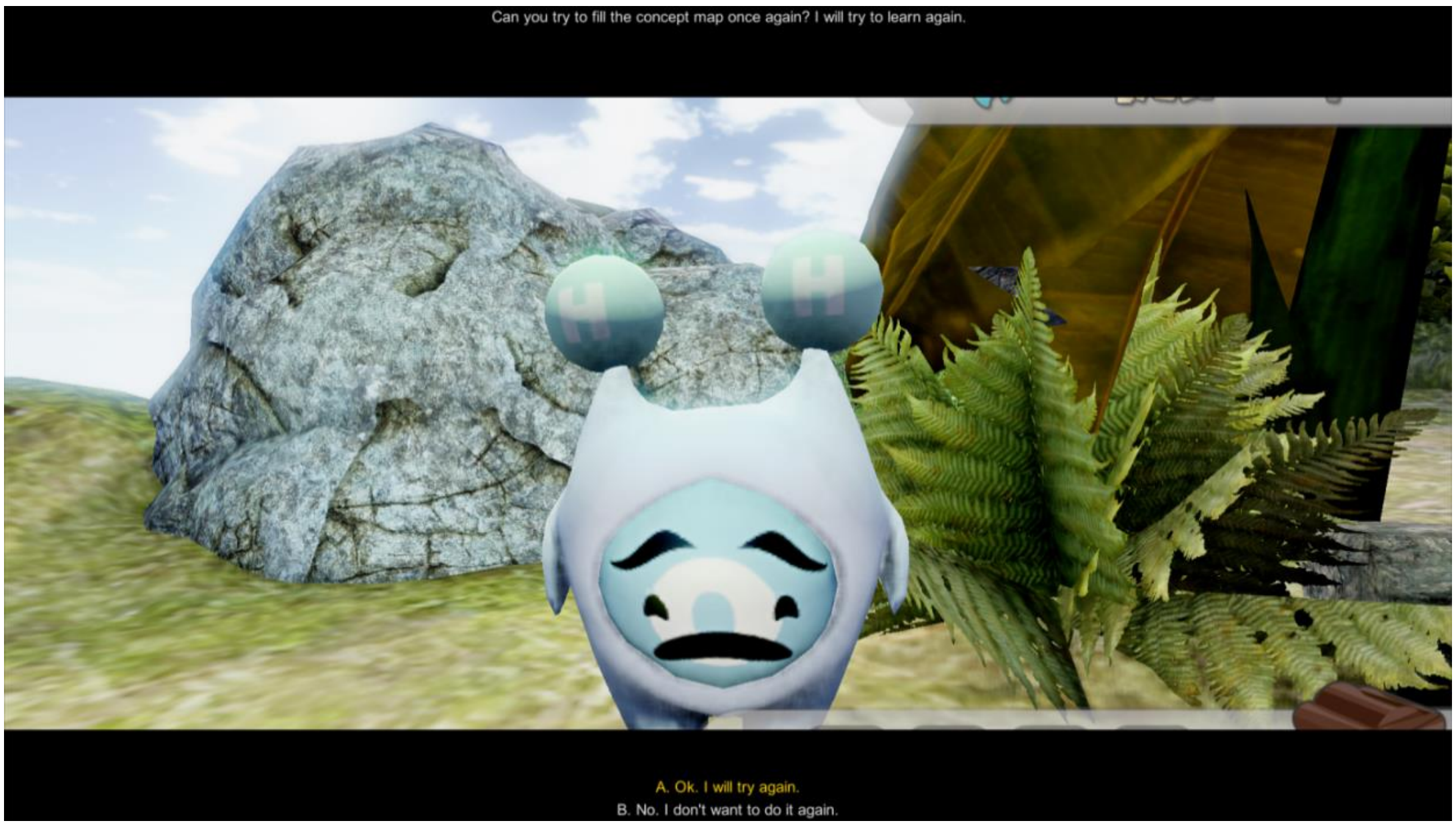

Figure 30 PTA requesting learner for teaching

The concept map will be displayed in the interface in VS saga, as shown in Fig. 31. The learner is required to complete the concept map in order to teach the PTA. If the learner rejects teaching the PTA, PTA control then generates a rejection event.

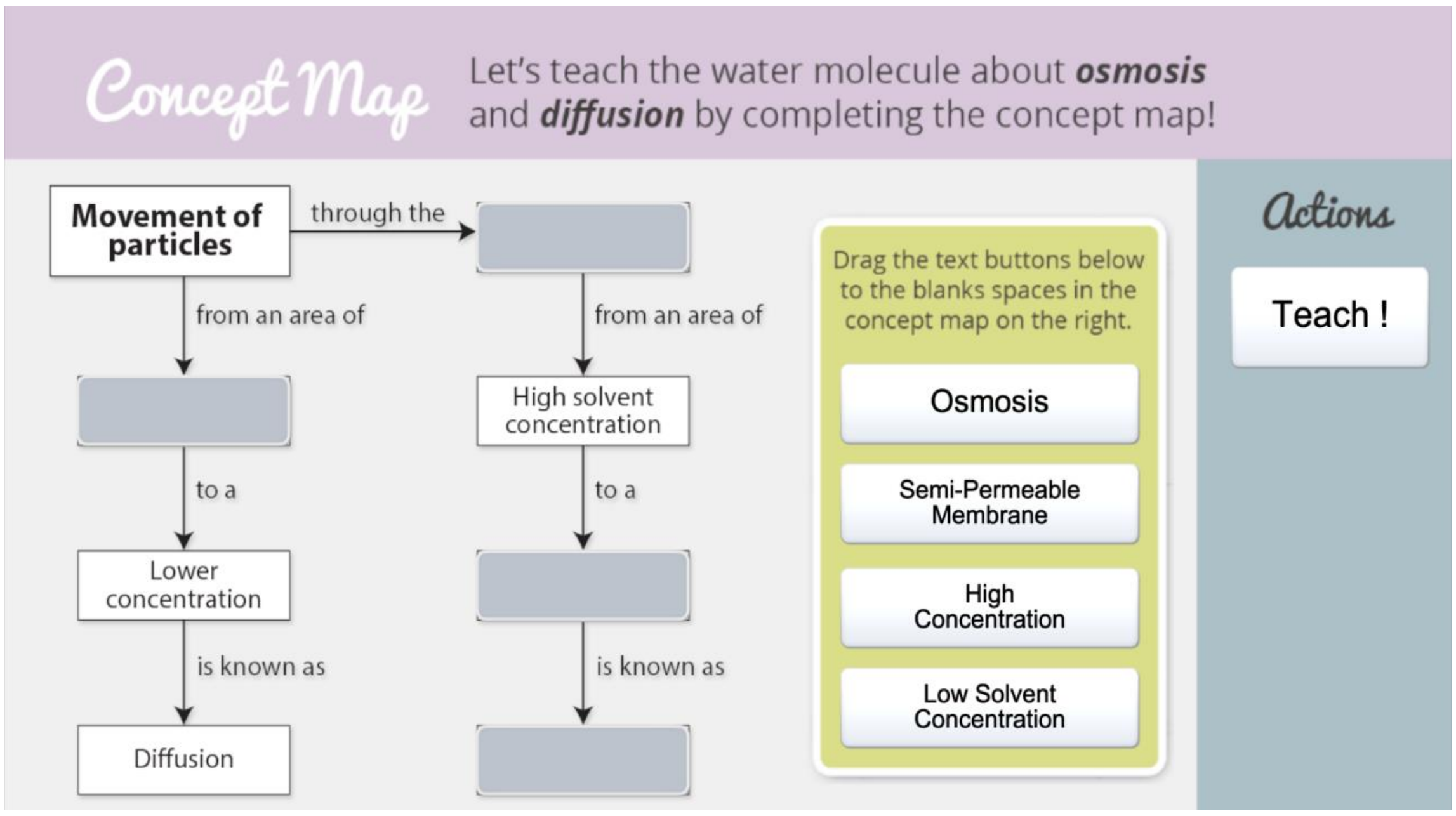

Figure 31 Concept map that require the learner to complete

***Practicability Reasoning***

In the Practicability Reasoning Sub-Goal Net, the PTA reasons on the knowledge that the learner has taught the PTA. This will determine whether the knowledge is correct. If the knowledge is correct, the PTA Control will generate an animation to show the "revival" of the dying banana plant, signifying that the PTA molecule has gained the information from the learner to enter the roots of the banana plant to transport essential water for the plant's survival. In the case where the knowledge taught is incorrect the PTA control will generate the events through the UI control to display a message to tell the learner to repeat the teaching event.

### 3.6.2 UI Control

The UI control is a component that manages the displays of and user interface (UI) in the VS saga game.

One example is the PTA panel that is an interface for the Persuasion Reasoning in the PTA to display persuasion cue. Whenever a persuasion cue is selected through Persuasion Reasoning, the PTA Control will activate the UI Control to display the information selected of the selected cue.

The concept map is where the learner player teaches the PTA. The learner player completes the teaching process by dragging the buttons selections on the left and dropping them in the blank spaces in the concept map. When they click on the "Teach!" button, the contents in the blank spaces will be saved by the UI Control as knowledge learnt by the PTA and a new event will be created for the completion of the teaching process.

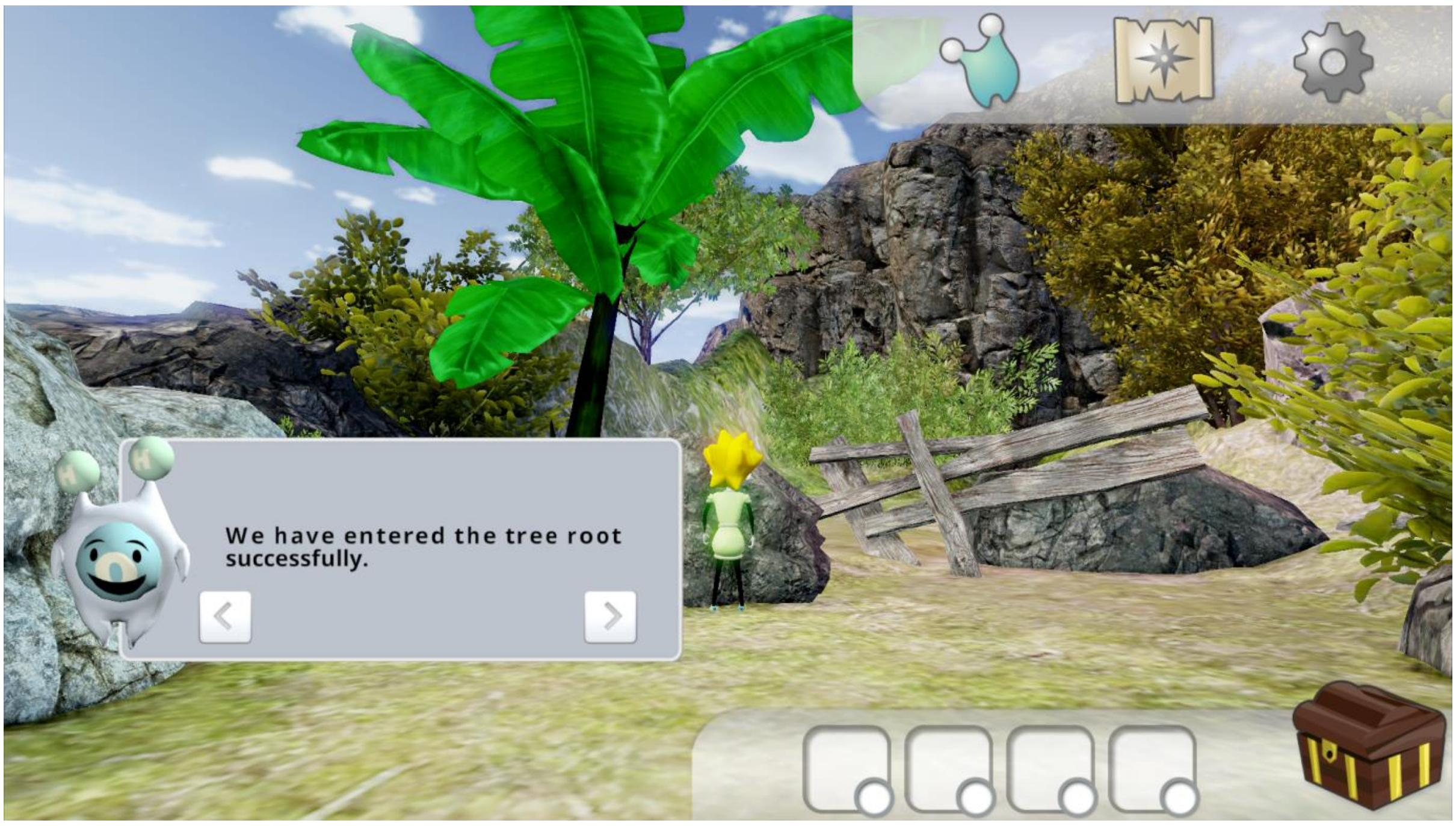

Figure 32 Banana plant after the PTA has been taught by the learner

The last example is the banana plant interface. If the learner player has successfully taught the PTA correctly, the UI Control changes the colour and rotation of the leaves to simulate an animated revival of the dying banana plant, as shown in Fig. 32.

### 3.6.3 Event Control

The Event Control in the VS saga is a supportive controller for events that support the PTA control. There are three types of events that the Event Control tracks in the VS saga, namely the Time Events, Dialogue Events and Teaching Feedback Events.

***Dialogue Events***

A dialogue event is created whenever there is a dialogue between the learner player and the NPCs. These dialogue events are created in the dialogue system in the Unity 3D game engine using the SequencerCommand script. Both learner player and the NPCs dialogues can be setup in the dialogue system. A new dialogue can be created and added to the event log with the event control.

***Time Events***

The Event Control tracks the duration in which the learner player is inactive in the Time Event. This is done with a timer that keeps track of the amount of time the learner is not engaging with learning activities by interacting with the NPCs in the VS saga. If a dialogue event is detected, the timer will be reset. In case of a time-out a time event will be generated which indicates that the learner player is inactive in the VS saga.

***Teaching Feedback Events***

The teaching feedback event provides feedback to the learner player's teaching. Two main events are tracked by the Event Control in VS Saga. Namely, the teach success event and the teach failure event.

***Events Tracked in VS Saga***

Various events are tracked in the VS saga. The table below is a list of events that are tracked in the VS saga. These events are used by the Event Control to generate responses in the PTA.

Table 14 Events Tracked in VS Saga

| **Dialogue Events** | |
|---|---|
| 1. Not learning | 9. Help Mayor NPC |
| 2. Visiting science laboratory | 10. Not teaching the water |
| 3. Learn diffusion | molecule |
| 4. Learn osmosis | 11. Teach the water molecule |
| 5. Apply diffusion | 12. Chat with animal NPC |
| 6. Apply osmosis | 13. Chat with village girl NPC |
| 7. Not conducting experiments | 14. Teachability event |
| 8. Willing to conduct experiment | |
| **Time Events** | |
| 1. Doing nothing (Time-out) | |
| **Teaching Events** | |
| 1. Teach success<br>2. Teach Failure | |

In addition to the events in the table above, the practicability event is also tracked. However, this does not belong to any of the categories in the Event Control, as the practicability event does not generate any dialogue in the VS saga.

### 3.6.4 Persuasive FCM in VS Saga

The events tracked in VS saga are categorised into the *leaf nodes* of the Persuasive FCM based on the ELM persuasion. Fig. 33 shows the complete FCM in the VS saga complete with events tracked. Note that only events that affect the motivation and ability are included in the *leaf nodes* of the Persuasive FCM.

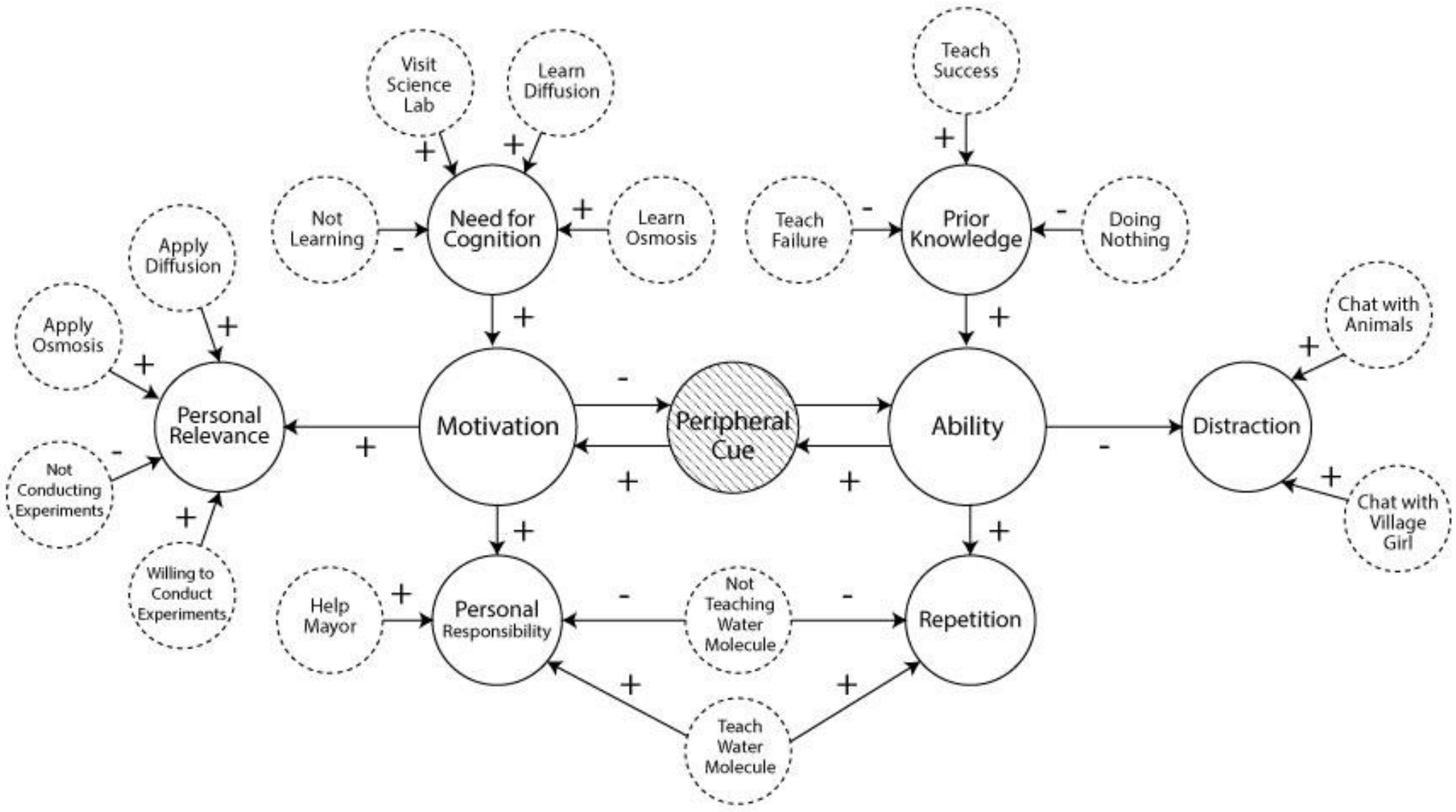

Figure 33 Complete FCM in VS saga

## 3.7 Studies during Formative Assessment of the PTA Model

Several case studies were conducted throughout the development of the PTA in VS in this section. The following are the descriptions of the case studies conducted from the year 2013 to 2015.

A preliminary study was conducted in 2013 on the VS test platform for the initial assessment of the PTA model (Lim, Ailiya, Miao, & Shen, 2013). An interview was conducted on two groups of testers who interacted with a version of VS. The control group tested on a version of the VS with a teachable agent without persuasion and compared with the treatment group who tested version of VS with the PTA prototype. The initial results showed that treatment group of testers responded that they are more willing to teach the PTA during the learning process compared to the responses from the control group.

In 2014, a separate study with a clearer model for Persuasive Goal Net allowing for tracking on the user's learning ability and motivation was conducted on a group of test users using the VS test platform with the PTA (Lim, Ailiya, Miao, & Shen, 2014). An informal interview was conducted over the advantages of PTA compared to the traditional teachable agent. The test users had indicated positive interest in learning with the PTA in the VS test platform. The test users have responded that PTA was interactive and they were willing to teach it. The PTA also appeared to be actively interacting with the test users, and provided suggestion that were related to the learning stages that they were currently in. They had also described that the PTA as "a friend in need", therefore they were more likely to teach the PTA as the test users had treated the PTA as friendly companion. It was also highlighted that PTA had to continue to be persuasive even when they were already teaching it. This indicated that the PTA had to continue to be persuasive so as to help the learners reflect on the concept that was taught to the PTA.

Table 15 is a summary of the interview and some of the suggestions that was brought up during the interview.

Table 15 Summary of Interview

| Advantages of PTA | Improvements to PTA |
|---|---|
| • Responsive especially when avatar is idling<br>• Provides appropriate responses at correct timing<br>• Provides guidance to the activities in the Virtual Learning Environment | • Needs to a wider range of emotions and more variety of actions<br>• Further integration of the PTA in actual teaching phrase. |

The case studies conducted had provided useful formative feedback during the developmental stage of the PTA leading to the improvements in VS saga. However due to the small sample size of the study group, there was no statistical significance analysis drawn from these studies. However, these small group studies provided an iterative analysis of the PTA during the design process of the VS saga.

## 3.8 Discussion

In comparison to the existing teachable agent in previous version of VS, the PTA is able to response intuitively to suit the student's learning requirements. The improved PTA is able to respond better to individual needs of the learner based on his or her ability and motivation. At the same time the PTA possessed the capability to provide strategies to encourage social interaction with the PTA, which will keep learners engaged while teaching the PTA.

The improved PTA architecture in the VS saga bridges the gap between theoretical concepts of the PTA model and the actual implementation of the PTA in a 3D game engine. The following chapter is a follow-up of the user study on intergenerational learning using the PTA VS saga, based on the research questions and hypothesis of this thesis.

# 4 Persuasive Teachable Agent for Intergenerational Learning

## 4.1 PTA for Intergenerational Learning

To our best knowledge, there is no data and studies on the intergenerational learning with teachable agents. The question is whether the PTA, a teachable agent with persuasion capabilities can benefit intergenerational learning. The study explored how different generations would respond to the PTA in the virtual learning environment.

### 4.1.1 Assessment of PTA

To assess the PTA in terms of its usage in intergenerational learning, the design of this study is based the phenomenological research. The phenomenological research method involves discovering a phenomena by gathering information through qualitative research methods, for example, by interviewing the participants, discussion and observing the participants in a situation (Lester, 1999).

The aim of phenomenological research is to describe the phenomenon in an accurate manner at the same time, refrain from any bias thus, staying with the facts (Groenewald, 2004). The advantage of the phenomenological method is that it can be applied to a single case or unforeseen or specifically selected sample participants (Lester, 1999). Phenomenological research method was adopted in this study.

### 4.1.2 Methods

The purpose of this study is to evaluate the perception and experience of intergenerational learning within groups of players interacting PTA. The participants are required to teach the PTA steps to complete the game with someone either younger or older than them. A control of paired participants will evaluate a version of VS saga with a teachable agent without persuasion and treatment group will evaluate a version of VS saga with the PTA.

In addition, there are two types of data collection in this study. The type of data collected which involves the learner player's progress in the VS saga and the assessment of the learner player's performance in the post-game questionnaire. The following are the details of the data collection procedure.

### 4.1.3 Procedure

Firstly, the participants are briefed on the purpose of the study. Followed by an invitation to complete a 20 minutes interactive session where they are required to test the VS saga. After which, they are given 15 minutes to complete a post-game questionnaire.

Once their consent is granted, the participants are required to fill up their social-demographic data which consist of their personal background, gender, age and highest education attained. The participants are thereby required to complete the rest of the post-game questionnaire.

### 4.1.4 Post-game Questionnaire

The post-game questionnaire with 32 questions is given out to each of the participant. The questions in the post-game questionnaire include 5-point likert scale (Strongly disagree = 1, disagree = 2, neutral = 3, agree = 4, strongly disagree = 5) questions, matrix questions, contingency questions, closed and open-ended questions.  Section A of the questionnaire consists of 4 questions related to the demographics and personal background of the participants.

Section B of the questionnaire consists of 8 questions that were related to the participant's learning experiences that the participants felt during the PTA. The participants were required to rank these questions according to a 5-point likert scale. In this section, questions 5 to 9 are related to the social skills that the participants have experienced or felt when they are teaching the PTA. Question 10 relates to the problem skills the participant have encountered in the game. Questions 11 and 12 are related to the people skills that the participants have felt during the interactive session in VS saga. People skills in this case refer to both psychological skills and social skills.

In Section C, there are three questions numbered 13 to 15. These questions are three 5-point likert scale questions that relate to the participant's knowledge, values and role respectively. In Section D are three questions numbered 16 to 18. These questions are related to the participant's attitude and opinion on learning with the PTA.

In Section E, there are three questions numbered 19 to 21 related to the participant's feelings towards the PTA on a 5-point likert scale rating. Note was taken that question 20 is coded on a reserve scale. Section F from 22 to 24 are three questions on a 5-point likert scale related to the participant's perceptions towards the PTA. Section G is an open-end question that requires the participants to fill in suggestions on improving the PTA to enhance their learning experience.

In the Section H of the questionnaire there are four questions, numbered 26 to 29 that relate to the participant's general intergenerational relationship background. Question 26 is related to the participant's level of interest to spend their leisure time learning with someone older or younger than them, rated on a 5-point likert scale. Question 27 is a matrix of factors that affect the participant's decision to spend time in intergenerational learning activities. Question 28 is a close-ended question that relates to the kind of activities that interest the participant in intergenerational learning. Question 29 is an open-ended question that probes on what will motivate the participants to intergenerational learning.

Section I consists of three questions numbered from 30 to 32. Question 30 and 31 are contingency questions. Question 30, relates to how the participants feel about using virtual learning environments such as VS saga to promote intergenerational learning. Question 31 relates to whether the PTA is able to aid intergenerational learning.

The last Section J concludes the questionnaire with an open-ended question that probes on the any additional comments that relates to the VS saga system or the study. The following section is a further breakdown of the categories in the data collected.

## 4.2 Data Collection

The average age of the sample (*M*=??*, SD* = ??) of a total of *N* = ?? that participated in the study. Below is the detailed breakdown of the data that are collected from the control and treatment group of participants in the post-game questionnaire. And the description of the in-game progress data which include the PTA performance, the analysis of the participant's mouse clicks and time spent in the VS saga.

### 4.2.1 Learning Experiences with PTA

In this section of the questionnaire, participants are asked to rate their learning experiences in the VS saga, according to the skills sets that are likely to gain during the test session.

***Social Skills***

The participants are required to rate 5 questions numbered from questions 5 to 9 in the post-game questionnaire in terms of the social skills and learning experiences based on a 5-point likert scale (Strongly disagree = 1, disagree = 2, neutral = 3, agree = 4, strongly disagree = 5) that participants gained in the session.

Question 5 relates to whether the participants have active listening skills that they have gained by taking up the advice from the PTA. Question 6 relates to whether the participants have taken responsibility to PTA's learning process. Question 7 probes on whether the participants felt encouraged by the PTA. Question 8 and 9 relate to whether the participants were able to focus and staying on the task assigned to them in the quests and missions that they have to complete in the VS despite the distracters that are placed in the game.

***Problem Solving Skills***

In question 10 of the post-game questionnaire, the participants are asked whether they are able to solve the problems that they had encountered in the VS saga.

***People Skills***

Questions 11 and 12 of relate to the people skills learning experience that the participants have gained during the session. Question 11 relates to whether the participants have learnt trust and respect and question 12 relates to whether the participants have gained understanding and whether the participants have empathized with the teachable agent.

### 4.2.2 Knowledge, Values and Role in Virtual Learning

Questions 13 to 15 relate to the knowledge, values and roles the participants experience in virtual learning. Question 13 explores whether they had learnt new knowledge other than information on the science concepts such as osmosis and diffusion. Question 14 relates to values that the participants gain during the session.

***Opinion on Learning with the PTA***

The opinion on learning with PTA section consists of three questions, numbered 16 to18. These questions are focused on the participant's attitudes towards interaction with their teammate and the PTA during the test session.

Question 16 relates to the whether the participants have engaged in learning during the session. Question 17 probes on whether the participants are motivated to find out more information on the science topics during the session. Question 18 asks the participants whether the participants felt that they have learnt more when teaching with the PTA with their teammate.

***Feelings towards the PTA***

In this section, there are three questions that investigate the participant's feeling towards the PTA. Question 19 probes on whether the participants have felt confident that the PTA will be able to do well. Question 20 asks whether the participants have felt nervous when they are interacting with PTA. Question 21 relates to whether the participants have felt any strong emotional feelings towards the PTA during the session.

***Perceptions towards the PTA***

The questions that probe on the participant's perception towards the PTA consist of three questions from 22 to 24.

Question 22 probes on whether the participants felt responsible for the PTA. Question 23 relates to whether the PTA responded as the participants have expected. Question 24 asks if the participants are satisfied with PTA performance on the concept map. Fig.40 shows the ratings in perceptions towards PTA.

### 4.2.3 In-game Progress

***PTA Performance***

A key indicator on how well the learner player has engaged in learning is the PTA's performance. In order to analyse this aspect in learning, the PTA learning results are tracked according to the number of attempts the learner player tries to teach the PTA. The knowledge that PTA have learnt from the learn player will also provide insights to how well that learner player has responded to the interaction with the PTA in the VS saga VLE.

***Mouse Clicks***

In addition to the post-game questionnaire and the events tracked by the PTA, the learner user's mouse clicks are recorded. This provides greater insight on the activities and richer information on what the learner user are doing in the VS saga system. The information will allow future developers of the VS saga to improve the interface design as well as aid in the evaluation and analysis of the effectiveness of the PTA in the system.

***Time Spent***

Another important factor that determines the learner player's activities in the VS saga is the time that they spent on each task assigned to them. Therefore, the time between each task to complete will be tracked.

## 4.3 Data Analysis

***Learning Experience with PTA***

Table 16 shows the comparison of the average rating and mean score of the different skills gained during the learning experience with the PTA during the session.

Table 16 Rating Comparison of Learning Experience with PTA

| | **Social Skills** | | | | | **Problem Solving Skills** | **People Skills** | |
|---|---|---|---|---|---|---|---|---|
| Questions | Qn 5 | Qn 6 | Qn 7 | Qn 8 | Qn 9 | Qn 10 | Qn 11 | Qn 12 |
| Control Group Average Rating | | | | | | | | |
| Treatment Group Average Rating | | | | | | | | |

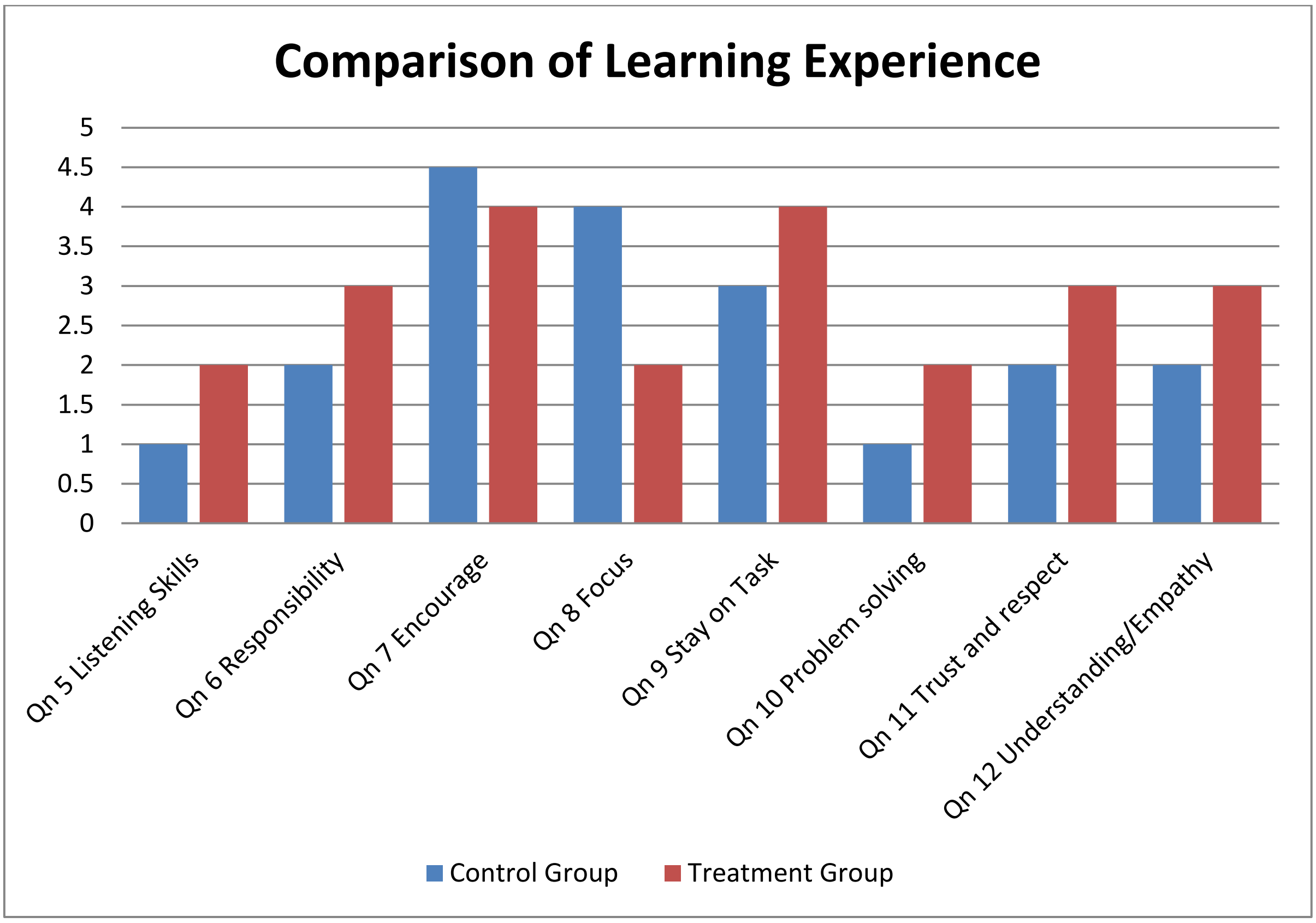

Figure 34 Comparison of learning experience average rating

### ***Knowledge, Values and Role in Virtual Learning***

Table 17 Rating Comparison on Knowledge, Values and Role in Virtual Learning

| | **Knowledge** | **Values** | **Role** |
|---|---|---|---|
| Questions | Qn 13 | Qn 14 | Qn 15 |
| Control Group Average Rating | | | |
| Treatment Group Average Rating | | | |

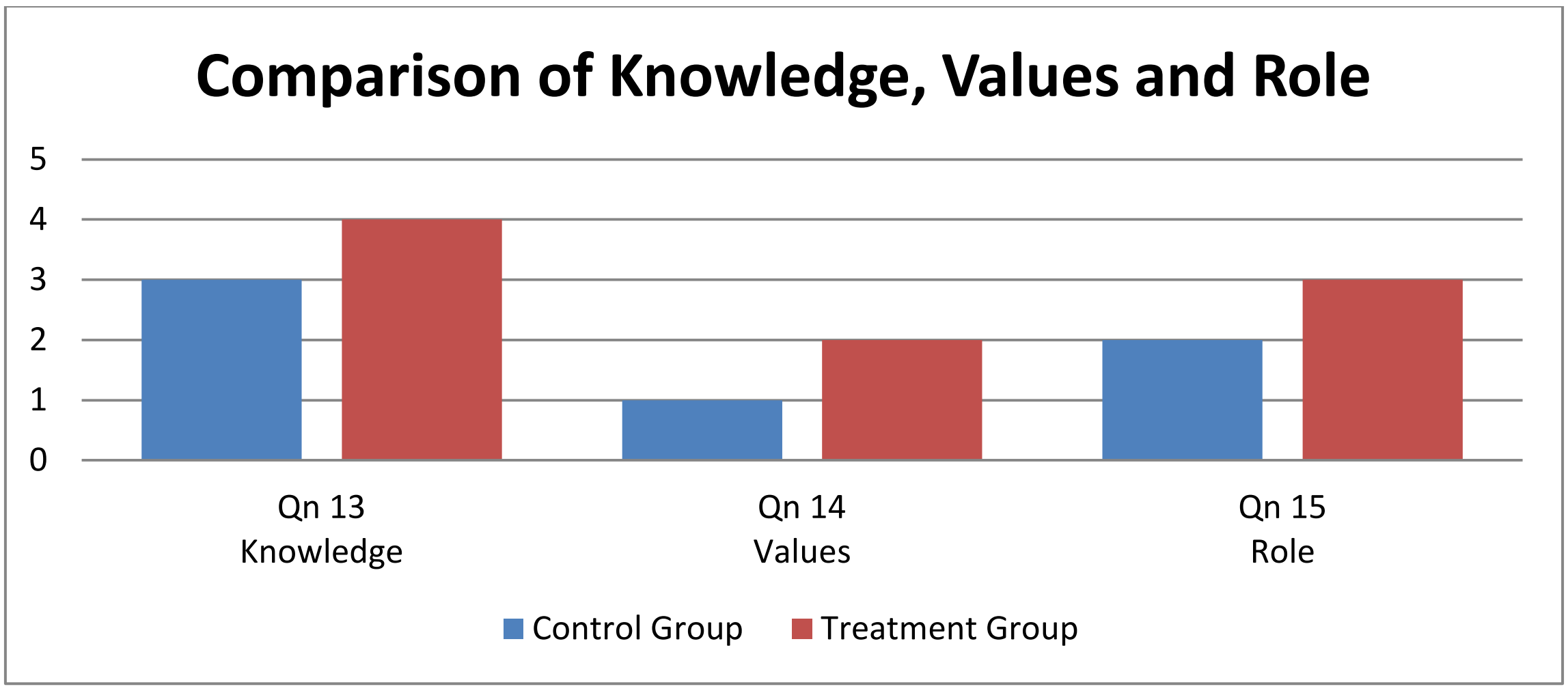

Figure 35 Comparison of knowledge, values and roles average rating

***Opinion on Intergenerational Learning***

Table 18 Rating Comparison on Opinion on Intergenerational Learning

| | **Attitude** | **Motivation** | **Learning** |
|---|---|---|---|
| Questions | Qn 16 | Qn 17 | Qn 18 |
| Control Group Average Rating | | | |
| Treatment Group Average Rating | | | |

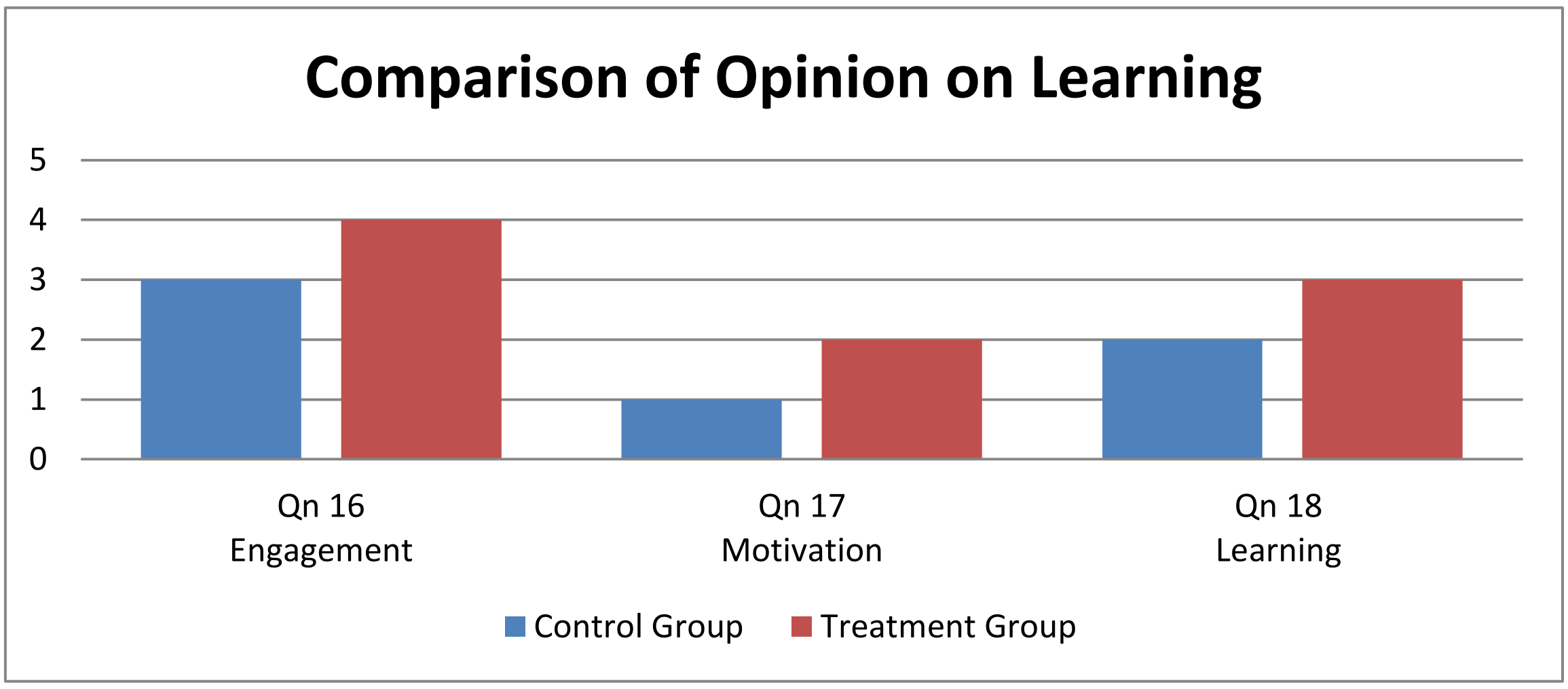

Figure 36 Comparison of attitude towards learning average rating

***Feelings towards PTA***

Table 19 Rating Comparison of Feelings towards PTA

| | **Confidence** | **Nervousness** | **Emotional** |
|---|---|---|---|

| Questions | Qn 19 | Qn 20 | Qn 21 |
|---|---|---|---|
| Control Group Average Rating | | | |
| Treatment Group Average Rating | | | |

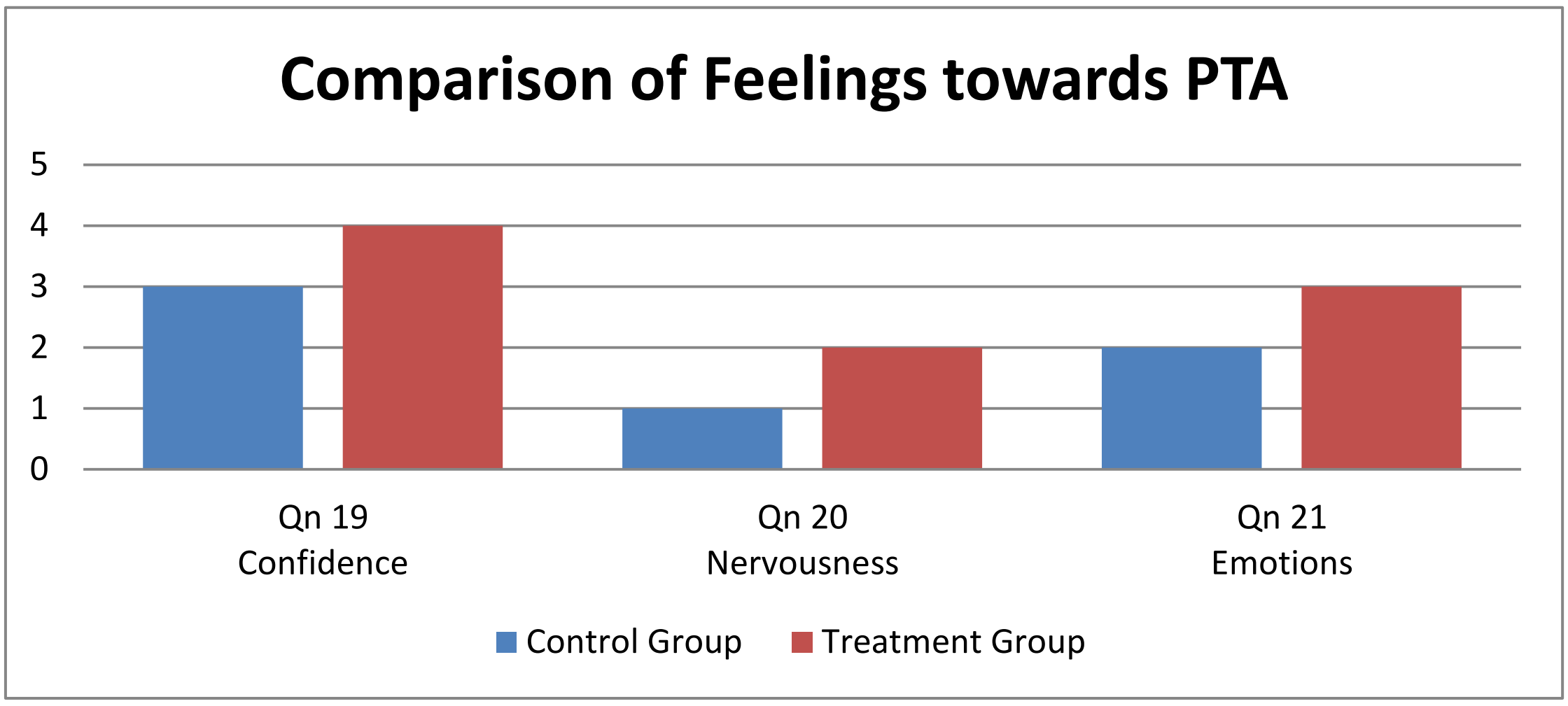

Figure 37 Comparison of feelings towards PTA average rating

***Perceptions towards PTA***

Table 20 Rating Comparison on Perceptions towards PTA

| | **Responsibility** | **Expectation** | **Performance** |
|---|---|---|---|
| Questions | Qn 22 | Qn 23 | Qn 24 |
| Control Group Average Rating | | | |
| Treatment Group Average Rating | | | |

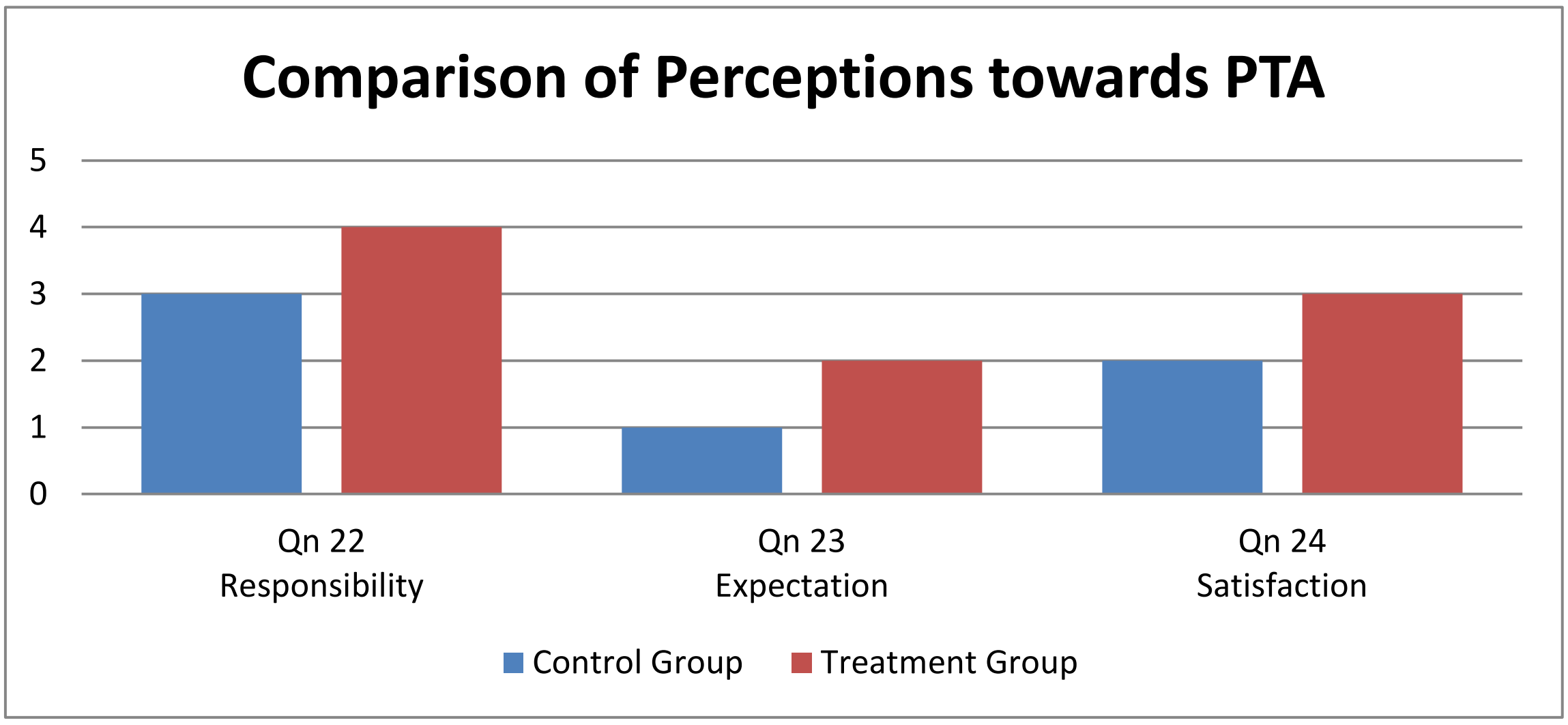

Figure 38 Comparison of perceptions towards PTA average rating

***PTA Performance***

Figure 39 Comparison with PTA and teachable agent score with age

### 4.3.1 Demographics and Intergenerational Relationships

***Level of Interest in Intergenerational Learning***

***Factors affecting decision on time spend on Intergenerational Learning***

***Activities for Intergenerational Learning***

### 4.3.2 PTA for Intergenerational Learning in Virtual Learning Environment

***Comparison of PTA and Teachable Agent for Intergenerational Learning in Virtual Environments***

### 4.3.3 Future Improvements to PTA

### 4.3.4 Summary of Findings

## 4.4 Hypotheses Conclusion

**Hypothesis 1:**

In hypothesis 1, it is predicted that the participants will benefit from events such as role taking, and other skills set such as computer skills, school related issues and academic knowledge.

**Conclusion 1:**

**Hypothesis 2:**

According to hypothesis 2, we have predicted that the participants from the older age group in an intergenerational learning setting will take a greater effort in learning with the PTA.

**Conclusion 2:**

**Hypothesis 3:**

Our hypothesis is that intergenerational learning increases the time spent teaching the PTA, and allows for greater interaction between participants from different age groups.

**Conclusion 3:**

**Hypothesis 4:**

We have predicted that the findings in this study will be useful to the future development of teachable agents, particularly in the expanding the application of teachable agents, such as informal learning and learning for different age groups.

**Conclusion 4:**

## 4.5 Discussion

# 5 Conclusion and Future Work

## 5.1 Summary of Contributions

**Contribution 1: PTA Model**

The first contribution is the formalization of the definition for the Persuasive Teachable Agent (PTA) model. The PTA model is a goal-oriented modelling approach that deliberates the teachable agent's goal to select persuasion to influence user's decision to teach the teachable agent.

**Contribution 2: Persuasive Reasoning Model**

The persuasive reasoning model is designed to assist the PTA in reasoning the methods persuade the user into teach the teachable agent. The persuasive reasoning is based on the Elaboration Likelihood Model (ELM) (Cacioppo & Petty, 1984) of persuasion. The ELM is a well-established theory of persuasion describes the process in which attitudes can be formed and altered. The ELM persuasion theory is the theoretical foundation of persuasion in the persuasive reasoning in the PTA.

To realize the persuasive reasoning in the PTA the Fuzzy Cognitive Map (FCM) is applied to the persuasive FCM computational model to compute the motivation and ability for the PTA to reason its persuasive actions. This allows the PTA to select appropriate actions to influence the user's decisions to teach the teachable agent.

**Contribution 3: PTA for Intergenerational Learning**

The PTA has been implemented to the VS Saga virtual learning environment. An improved PTA model that bridges the gap between persuasion theory and implementation as an intelligent teachable agent is attained. A formal case study on the effects of PTA on intergenerational learning has also been conducted.

## 5.2 Future Work

The research questions and hypotheses have highlighted the research direction of the PTA. The future works in this section is directed towards the implementation and evaluation of the PTA which also covers the potential of PTA research in future projects. The following are some of the possible extensive applications of PTA.

### 5.2.1 Persuasive Teachable Agent in Citizen Science Projects

The Persuasive teachable agent (PTA) is beneficial for learners in terms of improvements in learning experience. The PTA also encourages learners from different age groups to participate in learning activities. There are opportunities to look into areas of interest where learners will be interested in participating in group learning activities that are beneficial in terms of building relations across different age groups. Intergenerational learning activities create the opportunity for PTA to be applied in informal science educational projects on a large scale basis, such as a ‘citizen science’ projects.

Citizen science projects are initiatives where volunteers and enthusiasts from different backgrounds participate in scientific research, allowing the data collected to be made available for analysis on a larger scale (Duke, 2012). A citizen scientist communicates with the general public who engage the public outreach of scientific information (Clark & Illman, 2001; Duke & Emma, 2012). The participants in citizen science projects are citizen volunteers who take on the active role of data collection and analysis (Duke & Emma, 2012).

There are advantages in adopting citizen science in engaging the public in science research, especially in capitalizing the human resources available in solving scientific problems at the same time and increases the resources available and reduce the cost involved on conducting large scale by outreaching to a large number of users (Duke, 2012).

One example of a citizen science project is The Bird House Network (TBN) by Cornell Laboratory of ornithology where participants reports on observations of nest boxes setup in their neighbourhood allow for data monitoring of knowledge to be reported and collected through various media such as email, interaction by TBN staff through phone calls (Brossard, Lewenstein, & Bonney, 2005).

Other projects adopt an online approach, for example, the Foldit online game allows players to collaborate, and develop strategies to accumulate game points in different playing levels by folding proteins, (Hand, 2010). The Foldit project has benefited participants in allowing them to solve real life scientific problems. The participants in the Foldit project also experience the process of research as well as joy in finding a social community with similar interest. The emergence of ubiquitous technologies allows for collection of data to be collected on the go, mobile devices sensing enables researchers a wider form of data to be collected through logging and cataloguing measurements in everyday environments (Paulos, 2009).

Despite the apparent advantages there are issues in citizen science projects such as the management of citizen science projects which requires identifying the motivation for participation. The quality of the data collected and the effects of citizen science projects on participants (Duke & Emma, 2012).  There is research potential in the investigation dynamics of the information exchange between human and agents during citizen science projects.

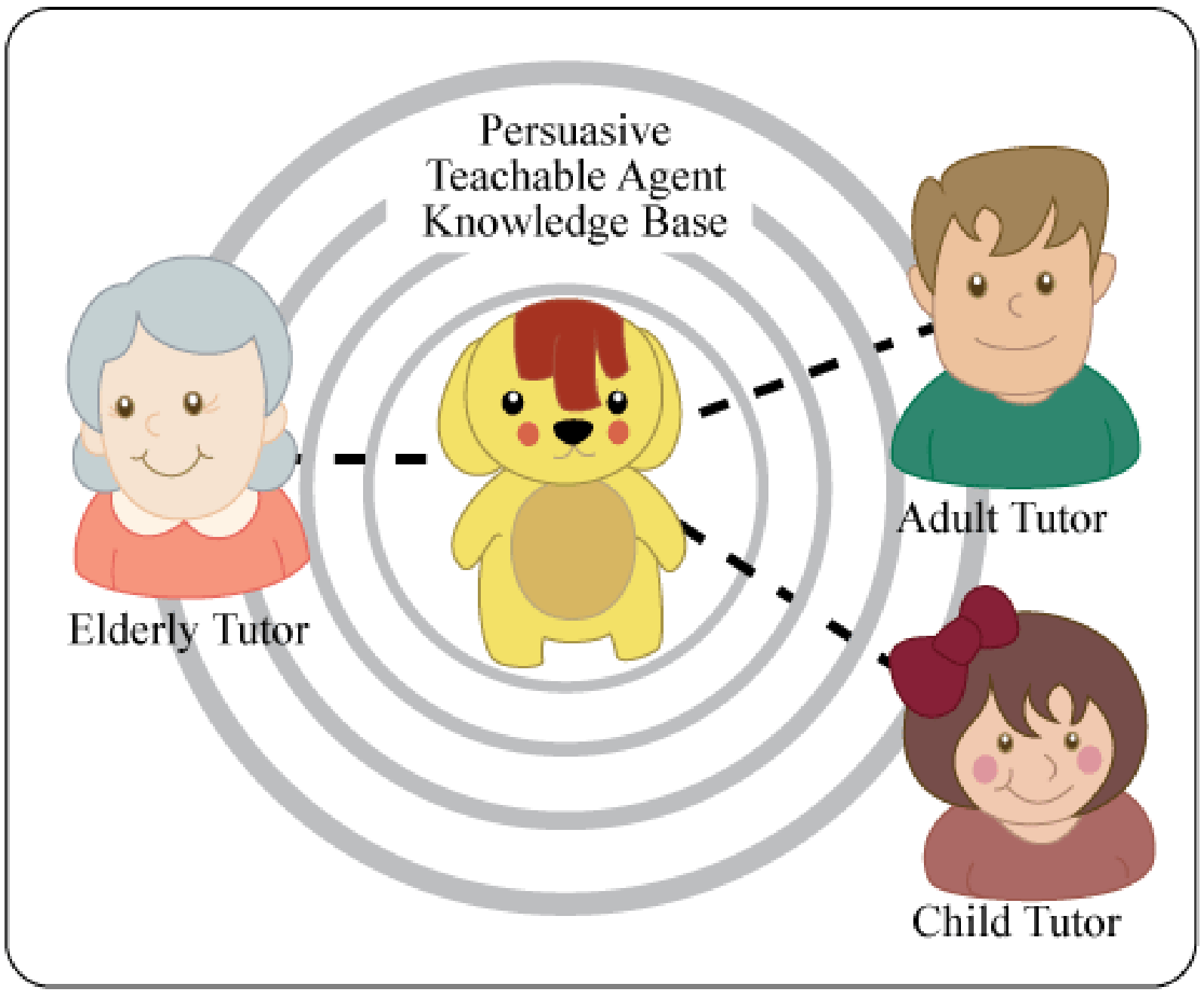

Figure 40 PTA Knowledge base increases with the user inputs.

Future research directions can also be expanded into exploring the types of information that are exchanged during the interaction with persuasive teachable agent in driving participants to contribute to citizen science and examine the potential and issues of applying PTA in intergenerational learning through citizen science projects. For example, in citizen science projects, the PTA acts as central knowledge base collecting activity logs and learning knowledge from users from different age group, at the same time analyses the information using Goal-Oriented Fuzzy Cognitive Persuasive Reasoning to encourage the learning from one another. A representation of the “ripple effect” from the exchange in information is shown in Fig. 40 where the growth of knowledge gained from the PTA affecting the different users.

### 5.2.2 Persuasive Gerontechnology

With the global trend gearing towards an aging population, there are various challenges and issues that arise with old age. Technologies are often sought after to solve problems associate with old age such as the deterioration in physical and cognitive abilities. Gerontechnology is an area of study in which technologies are used to benefit the well-being of the aging population (J.L. Fozard, 2005).

Persuasive technology on the other hand involves the use of technology to alter attitudes and behaviour without coercion (Fogg, 1999). Persuasive technology can play the part of influencing positive behaviours and attitude among the aged and aging person to enhance quality of life. However, most of the research in persuasive technologies are targeted at the younger generation, teens and preteens, and trending towards influencing the attitudes and behaviours of adults (King & Tester, 1999). Persuasive gerontechnology therefore looks at the area in which technology can provide positive influence towards attitude and behaviours to meet the ambitions, activities and wisdom of the aging society (James L Fozard & Kearns, 2006).

The PTA is a persuasive technology that can be used to enhance the life of the elderly, for example the PTA can lessen the perceived difficulty in learning a technology as the older person often perceive that mastering a new technology takes significant cognitive effort. Another form of PTA can be applied to the prevention and management of diseases where PTA can be used to encourage the elderly to adopt healthy lifestyle, through intergenerational learning activities. The PTA acts as a form persuasive gerontechnology, will encourage the elderly to lead a more active life that involve life- long learning and encourage intergeneration bonding.

### 5.2.3 User Modelling in Persuasive Teachable Agents

User modelling in human computer interaction, allow for systems to understand users in order to build systems that suit individual user's need (Fischer, 2001). Learners from different backgrounds possess different skills and personal reasons to engage in learning with teachable agents. Thus, understanding learner's ability and motivation becomes important in building adaptive systems to cater to different learners. The approach to user modelling in PTA utilizes the combination of persuasion theory and machine learning to generate users' learning behaviour and cognitive processes (Webb, Pazzani, & Billsus, 2001).

The future work in PTA will include the tracking of student learner learning activity data collected from the learning environment based on the persuasion theory such as Elaboration Likelihood Model (ELM) to evaluate the level of learning ability and motivation based on their receptiveness to either central or peripheral route of persuasion (Cacioppo & Petty, 1984). As well as analysing the learning activity data to model user's response to model persuasive strategies that cater to individual learners.

### 5.2.4 Feedback with Question Answering (QA) engine

Studies have shown that feedback support in teachable agent helped student learners to learn better, especially when they receive self-regulated learner feedback from their teachable agents (Tan & Biswas, 2006). For example, teachable agents who are able to ask questions and probe the student tutor encourage reflection and help the student tutor to structure their knowledge (Roscoe et al., 2008). Directed, corrective immediate feedback has also shown to aid in immediate learning. While guided and metacognitive feedback which comes in a form of reaffirming the knowledge that the student tutors have taught the teachable agent helps the tutors to monitor the teachable agent learning progress as well as assess the tutors own learning (Biswas & Leeawong, 2005). The body of research has shown the importance of teachable agent feedback to the student tutors.

The questioning and answering (QA) in the field of information retrieval for computer systems to search for short phrases or sentences, provides precise answers to user's query in large textual database (Prager, Chu-Carroll, Brown, & Czuba, 2006). QA systems allow for users to ask questions in natural language that they use in their daily lives and they can receive answers quickly (Hirschman & Gaizauskas, 2001).

The PTA aims to provide appropriate feedback to the student tutors with QA engine to provide learners with the information from the PTA knowledge base so as to keep them motivated in learning. The goal of the PTA feedback is to encourage student tutors to stay focus on learning tasks with the PTA. And future work can be aimed towards providing persuasive feedback using QA engine and observing the effects of persuasive feedback on learner from different age groups.

### 5.2.5 Evaluating and Assessing Persuasive Teachable Agents

The effects of persuasion on changing learning attitude with teachable agent have been positive in the preliminary experiment Future studies can be extended to include a wider range of effects on learners such as behaviour, cognition and emotional changes during interaction with PTA. The studies can also include a larger sample size of participants.

Further experiments can be conducted to evaluate favourable and unfavourable attitude change in learners and addressing the unfavourable attitude change, which is outside the area of the existing ELM persuasion model.

# Appendix A - Post-Game Questionnaire

## Intergenerational Learning with Persuasive Teachable Agent

Nanyang Technological University, School of Computer Engineering

### Information and Consent

Thank you for taking the time to participate in this research study involving the use of a virtual environment to teach lower secondary science. The purpose of this study is to collect information on your opinion of intergenerational learning with a novel artificial intelligent agent which implements persuasion in order to stimulate interest. The results of this study will be used in the writing of a thesis for the fulfillment of the requirement for a degree of doctor in philosophy in computer engineering.

As a participant, you will be asked to test two versions of an interactive virtual learning environment titled “Virtual Singapura” with a teammate. Following this you will need to answer a short written questionnaire. Participation will take approximately 20 minutes for the interactive session and 15 minutes for the survey questionnaire.

There are no known or anticipated risks associated with the participation in this study.

#### Confidentially

Data collected during this study is intended to be used solely for research purpose. Access to this data will be restricted to only the main investigator of the research thesis and its reviewers.

All information you provide is considered confidential; your name will not be included or in any way associated with the data collected in the study. Furthermore, our interest is in the average responses of the entire group of participants rather than the individual. You will not be identified individually in any way in written reports of this research, and will instead be referred to by a number within the report (e.g. Participant 1). The results of this study will also be made available to participants upon request.

#### Voluntary Participation

Participation in this study is voluntary. If you wish, you may decline to answer any questions or participate in any component of the study. You may also decide to withdraw from this study at any time and may do so without any penalty or loss of benefits to which you are entitled.

Thank you for your time and assistance in this project. If you have any further questions, I can be contacted at sufang@ntu.edu.sg or 96888523

---

**Consent Form**

I agree to participate in the study described above. I have made this decision based on the information I have read in the sections shown above. I have had the opportunity to ask and receive any additional details I wish to know about the study. I understand that my participation is voluntary and that I may withdraw from the study at any time.

Name: ______________________________________________________________________

Signature: _______________________________ Date: ________________________

## A. Personal Background

Please pick a tick in the box next to the answer of choice.

1. Gender

☐ Female ☐ Male

2. Age

☐ Below 9 ☐ 10 – 15 ☐ 11 – 16 ☐ 17 – 20 ☐ 21 – 29 ☐ 30 – 49

☐ 50 – 55 ☐ 55 and above

3. Race

☐ Chinese ☐ Malay ☐ Indian ☐ Others

4. How far did you go in school?

☐ Primary ☐ Secondary ☐ Junior College/Pre-university ☐Polytechnic

☐ University ☐ Postgraduate Studies

## B. Intergenerational Learning Experiences with Teachable Agent (water molecule)

The following are a number of statements regarding the type of skills that you have acquired during the interactive virtual learning. Please read each of the statements and indicate (circle your response from 1-5) to what extend you agree or disagree with each statement.

*Social skills*

5. I have actively listened to my teammate's advises and opinions during the session.

| Strongly Disagree | Disagree | Neutral | Agree | Strongly Agree |
|---|---|---|---|---|
| 1 | 2 | 3 | 4 | 5 |

6. I worked together with my teammate to teach the teachable agent (water molecule).

| Strongly Disagree | Disagree | Neutral | Agree | Strongly Agree |
|---|---|---|---|---|
| 1 | 2 | 3 | 4 | 5 |

7. I took turns with my teammate to teach the teachable agent (water molecule).

| Strongly Disagree | Disagree | Neutral | Agree | Strongly Agree |
|---|---|---|---|---|
| 1 | 2 | 3 | 4 | 5 |

8. I am the one who encourages my teammate.

| Strongly Disagree | Disagree | Neutral | Agree | Strongly Agree |
|---|---|---|---|---|
| 1 | 2 | 3 | 4 | 5 |

9. I have been staying on the tasks assigned to me during the session.

| Strongly Disagree | Disagree | Neutral | Agree | Strongly Agree |
|---|---|---|---|---|
| 1 | 2 | 3 | 4 | 5 |

*Problem solving skills*

10. I have learnt to solve problems that I had encountered in the game as a team.

| Strongly Disagree | Disagree | Neutral | Agree | Strongly Agree |
|---|---|---|---|---|
| 1 | 2 | 3 | 4 | 5 |

*People skills*

11. I have learnt to trust and respect my teammate during the session.

| Strongly Disagree | Disagree | Neutral | Agree | Strongly Agree |
|---|---|---|---|---|
| 1 | 2 | 3 | 4 | 5 |

12. I have more understanding towards my teammate after the session.

| Strongly Disagree | Disagree | Neutral | Agree | Strongly Agree |
|---|---|---|---|---|
| 1 | 2 | 3 | 4 | 5 |

## C. Knowledge, Values and Role in Virtual Learning

The following are a number of statements regarding the knowledge, values and role that you play during the interactive virtual learning. Please read each of the statements and indicate (circle your response from 1-5) to what extend you agree or disagree with each statement.

*Knowledge*

13. I have learnt new knowledge other than the information in the session.

| Strongly Disagree | Disagree | Neutral | Agree | Strongly Agree |
|---|---|---|---|---|
| 1 | 2 | 3 | 4 | 5 |

*Values*

14. Working together to teach the teachable agent (water molecule) with my teammate encourages me to think of others.

| Strongly Disagree | Disagree | Neutral | Agree | Strongly Agree |
|---|---|---|---|---|
| 1 | 2 | 3 | 4 | 5 |

*Roles*

15. I have taken the lead role in teaching the teachable agent (water molecule).

| Strongly Disagree | Disagree | Neutral | Agree | Strongly Agree |
|---|---|---|---|---|
| 1 | 2 | 3 | 4 | 5 |

## D. Opinion on Intergenerational Learning with the Teachable Agent (water molecule)

The following are a number of statements regarding your attitude during your interaction with the teammate and the teachable agent. Please read each of the statements and indicate (circle your response from 1-5) to what extend you agree or disagree with each statement.

*Attitude*

16. I was engaged in learning when co-teaching the teachable agent (water molecule).

| Strongly Disagree | Disagree | Neutral | Agree | Strongly Agree |
|---|---|---|---|---|
| 1 | 2 | 3 | 4 | 5 |

*Motivation*

17. I was motivated to find out more information during the interaction between my teammate and the teachable agent (water molecule).

| Strongly Disagree | Disagree | Neutral | Agree | Strongly Agree |
|---|---|---|---|---|
| 1 | 2 | 3 | 4 | 5 |

*Learning*

18. I learned more when co-teaching the teachable agent (water molecule) with a team mate than when I am learning alone.

| Strongly Disagree | Disagree | Neutral | Agree | Strongly Agree |
|---|---|---|---|---|
| 1 | 2 | 3 | 4 | 5 |

## E. Feelings towards the Teachable Agent (water molecule)

The following are a number of statements regarding your feelings towards the teachable agent. Please read each of the statements and indicate (circle your response from 1-5) to what extend you agree or disagree with each statement.

*Feelings*

19. I am confident that the teachable agent (water molecule) will do well.

| Strongly Disagree | Disagree | Neutral | Agree | Strongly Agree |
|---|---|---|---|---|
| 1 | 2 | 3 | 4 | 5 |

20. I felt nervous about the interacting with the teachable agent (water molecule) during the session.

| Strongly Disagree | Disagree | Neutral | Agree | Strongly Agree |
|---|---|---|---|---|
| 1 | 2 | 3 | 4 | 5 |

21. I have strong emotional feelings towards the teachable agent (water molecule) (happy, angry, sad etc.).

| Strongly Disagree | Disagree | Neutral | Agree | Strongly Agree |
|---|---|---|---|---|
| 1 | 2 | 3 | 4 | 5 |

## F. Perceptions towards the Teachable Agent (water molecule)

The following are a number of statements regarding your perceptions towards the teachable agent. Please read each of the statements and indicate (circle your response from 1-5) to what extend you agree or disagree with each statement.

22. I felt responsible for the teachable agent (water molecule).

| Strongly Disagree | Disagree | Neutral | Agree | Strongly Agree |
|---|---|---|---|---|
| 1 | 2 | 3 | 4 | 5 |

23. The teachable agent (water molecule) responded as I had expected.

| Strongly Disagree | Disagree | Neutral | Agree | Strongly Agree |
|---|---|---|---|---|
| 1 | 2 | 3 | 4 | 5 |

24. I am satisfied with the performance of the teachable agent (water molecule).

| Strongly Disagree | Disagree | Neutral | Agree | Strongly Agree |
|---|---|---|---|---|
| 1 | 2 | 3 | 4 | 5 |

## G. Improvements to Teachable Agent (water molecule)

25. Please tell us how the teachable agent (water molecule) can be improved to enhance your learning experience.

______________________________________________________________________

______________________________________________________________________

## H. General Intergenerational Relationship Background

Please complete the following items to the best of your ability. Please read each of the statements and indicate (circle your response: 1-5).

26. What is your level of interest to spend your leisure time learning with a person that is younger or older than you?

| Not at all Interested | A Little Interest | Neutral | Interested | Extremely Interested |
|---|---|---|---|---|
| 1 | 2 | 3 | 4 | 5 |

27. On a scale 1-5 please rate the level of factors that affect your decision to spend time learning from someone older or younger than you.

| | Does not Affect | Little Affect | Neutral | Somewhat Affects | Greatly Affect |
|---|---|---|---|---|---|
| Age difference | 1 | 2 | 3 | 4 | 5 |
| Inexperience interacting with someone from different age group | 1 | 2 | 3 | 4 | 5 |
| Lack of available time | 1 | 2 | 3 | 4 | 5 |
| Type of activities | 1 | 2 | 3 | 4 | 5 |

28. What kind of activities would you be interested in participating in an intergenerational setting (i.e. interacting with someone older or younger than you)?

☐ Sports ☐ Field Trips ☐ Games (card, puzzles, board, etc.)

☐ Volunteer work ☐ Arts/Crafts activities ☐ Outdoor recreation

☐ Conversation/talking ☐ Passive recreation (walking, watching TV, etc.) ☐ Cultural experiences (art exhibitions, museums, etc.)

☐ Others (Please specify: ______________________________ )

29. What would motivate you to learn more with some from a different age group (i.e. someone older or younger than you)?

______________________________________________________________________

______________________________________________________________________

______________________________________________________________________

______________________________________________________________________

## I. Teachable Agent for Intergenerational Learning in Virtual Environments

30. Do you feel that virtual environments such as Virtual Singapura (VS) can help in intergenerational bonding (i.e. closeness, connectedness and kinship ties among family members of over one or more generation)?

☐ Yes ☐ No

31. Do you feel that the teachable agent (water molecule) can help in intergenerational learning (i.e. learning from someone older or younger then you)?

☐ Yes ☐ No

## J. Gaming Experience in Virtual Learning Environment

The following are a number of statements regarding your gaming experience in the Virtual Learning Environment. Please read each of the statements and indicate (circle your response from 1-5) to what extend you agree or disagree with each statement.

32. I enjoyed playing the game.

| Strongly Disagree | Disagree | Neutral | Agree | Strongly Agree |
|---|---|---|---|---|
| 1 | 2 | 3 | 4 | 5 |

33. This game was fun to play.

| Strongly Disagree | Disagree | Neutral | Agree | Strongly Agree |
|---|---|---|---|---|
| 1 | 2 | 3 | 4 | 5 |

34. I thought this was a boring game.

| Strongly Disagree | Disagree | Neutral | Agree | Strongly Agree |
|---|---|---|---|---|
| 1 | 2 | 3 | 4 | 5 |

35. This game did not hold my attention at all.

| Strongly Disagree | Disagree | Neutral | Agree | Strongly Agree |
|---|---|---|---|---|
| 1 | 2 | 3 | 4 | 5 |

36. I would describe this game as interesting.

| Strongly Disagree | Disagree | Neutral | Agree | Strongly Agree |
|---|---|---|---|---|
| 1 | 2 | 3 | 4 | 5 |

37. I believe this game had some value to me.

| Strongly Disagree | Disagree | Neutral | Agree | Strongly Agree |
|---|---|---|---|---|
| 1 | 2 | 3 | 4 | 5 |

38. This game is useful for me to learn science.

| Strongly Disagree | Disagree | Neutral | Agree | Strongly Agree |
|---|---|---|---|---|
| 1 | 2 | 3 | 4 | 5 |

39. I think this game is important in helping me experience what is impossible in the real world.

| Strongly Disagree | Disagree | Neutral | Agree | Strongly Agree |
|---|---|---|---|---|
| 1 | 2 | 3 | 4 | 5 |

40. I will be willing to this game again because it has some value to me.

| Strongly Disagree | Disagree | Neutral | Agree | Strongly Agree |
|---|---|---|---|---|
| 1 | 2 | 3 | 4 | 5 |

41. I think by playing this game has helped me understand science topics.

| Strongly Disagree | Disagree | Neutral | Agree | Strongly Agree |
|---|---|---|---|---|
| 1 | 2 | 3 | 4 | 5 |

42. I believe this game is beneficial to me.

| Strongly Disagree | Disagree | Neutral | Agree | Strongly Agree |
|---|---|---|---|---|
| 1 | 2 | 3 | 4 | 5 |

43. This game is of importance to me.

| Strongly Disagree | Disagree | Neutral | Agree | Strongly Agree |
|---|---|---|---|---|
| 1 | 2 | 3 | 4 | 5 |

## K. Additional Comments

44. Please tell us know any additional comments you may have regarding the system or the study.

Thank you for your participation.